\documentclass[12pt]{article}
\usepackage{amsmath}
\usepackage{amssymb}
\usepackage{color}
\usepackage{xcolor,cancel}

\definecolor{forestgreen}{rgb}{0.13,0.54,0.13}
\newcommand{\comments}[1]{}

\numberwithin{equation}{section}

\newcommand{\rma}{{\rm a}}
\newcommand{\rme}{{\rm e}}
\newcommand{\rmd}{{\rm d}}
\newcommand{\rmi}{{\rm i}}
\newcommand{\rms}{{\rm s}}
\newcommand{\ud}[1]{\hspace{-0em}\mathrm{d}{#1}\;}
\newcommand{\udd}[2]{\hspace{-0em}\mathrm{d}{#1}\,\mathrm{d}{#2}\;}

\newcommand{\Ud}[1]{\hspace{-0ex}\mathrm{d}{#1}\;}

\newcommand{\uD}{\mathcal{D}}
\newcommand{\eff}{\rm{eff}}
\newcommand{\any}{\rm{ani}}
\newcommand{\ext}{\rm{ext}}

\begin{document}
\pagestyle{empty}

\title{Magnetization dynamics: path-integral formalism for the stochastic Landau-Lifshitz-Gilbert equation}

\author{
Camille Aron$^{1}$, 
Daniel G.\ Barci$^2$,
Leticia F. Cugliandolo$^3$, \\
Zochil Gonz\'alez Arenas$^4$, 
and Gustavo S. Lozano$^{3,5}$
\bigskip
\\
$^1${\small Department of Physics and Astronomy, Rutgers University, Piscataway, NJ 08854, USA} \\
{\small\&  Department of Electrical Engineering, Princeton University, Princeton, NJ 08544, USA}
\\
$^2${\small Departamento de F{\'\i}sica Te\'orica, Universidade do Estado do Rio de Janeiro,}
\\ {\small Rua S\~ao Francisco Xavier 524, 20550-013, Rio de Janeiro, RJ, Brazil}
\\
$^3${ \small Sorbonne Universit\'es, UPMC Univ Paris 06, UMR 7589, LPTHE, F-75005,
Paris, France}
\\
$^4$
{\small Centro Brasileiro de Pesquisas F\'{\i}sicas}\\
{\small \& National Institute of Science and Technology for Complex Systems,} \\
{\small Rua Xavier Sigaud 150, 22290-180, Rio de Janeiro, RJ, Brazil,}\\
{\small on leave from Instituto de Cibern\'etica, Matem\'atica y F\'\i sica, Cuba} \\
$^5${\small Departamento de F\'{\i}sica, FCEYN Universidad de Buenos Aires \& IFIBA CONICET,} \\
{\small Pabell\'on 1 Ciudad Universitaria, 1428 Buenos Aires, Argentina}
}

\date{}

\maketitle

\newpage
\pagestyle{plain}

\begin{abstract}
We  construct a path-integral representation of the generating functional 
for the dissipative dynamics of a classical magnetic moment as
described by the stochastic generalization of the Landau-Lifshitz-Gilbert equation proposed by Brown~\cite{Brown1963}, with the possible addition of spin-torque terms.
In the process of constructing this functional in the Cartesian coordinate system, we critically revisit this stochastic equation. We present it in a form that accommodates for any discretization scheme thanks to 
the inclusion of a drift term. The generalized equation ensures the conservation of the magnetization modulus and the 
approach to the Gibbs-Boltzmann equilibrium in the absence of non-potential
and time-de\-pen\-dent forces. The drift term vanishes only if the mid-point Stratonovich prescription is used.
We next reset the problem in the more natural spherical coordinate system. We show that 
the noise transforms non-trivially to spherical coordinates acquiring a non-vanishing mean value in this 
coordinate system, a fact that has been often overlooked in the literature. We next construct the generating functional 
formalism in this system of coordinates for any discretization prescription. The functional formalism in Cartesian or spherical coordinates 
should serve as a starting point to study different 
aspects of the out-of-equilibrium dynamics of magnets. Extensions to colored noise, 
micro-magnetism and disordered problems are straightforward.
\end{abstract}














\newpage

\tableofcontents 

\section{Introduction}

The control of magnetic materials~\cite{Hillebrands2002,Mattis1988,Mattis1985}, 
already at the heart of the information technologies developed in the second half of the twentieth century, is undergoing a second 
revolution  with the development of spintronics~\cite{Johnson1985,Brataas12}.

In ferromagnetic materials, the spin degrees of freedom carried by localized electrons tend to spontaneously long-range order 
below the Curie (second order) transition temperature. The so-called micromagnetic description consists in studying the local order 
parameter, $\mathbf{M}(\mathbf{x},t)$, the expectation value of the magnetization per unit volume averaged over, typically, 
a few lattice cells. In most relevant cases, this macroscopic composite object can be treated classically.

The spin-spin interaction is essentially due to the overlap of the electronic wave functions. Below the Curie temperature, this is the 
dominant interaction. The modulus of the local magnetization can be approximated by a temperature-dependent constant 
which takes the value of the spontaneous magnetization at the working temperature, {\it i.e.}
$|\mathbf{M}(\mathbf{x},t)|=M_\rms$. The direction of ${\mathbf M}$ is, instead, subject to the interactions
and external forces, and it is non-uniform in space and varies in time.

When a ferromagnet is used to store information, bits are encoded in the orientation of the local magnetization. Usually, the 
magnetization is manipulated by applying Oersted fields generated by electric currents. 

Very early, damping, \textit{i.e.} the transfer of spin angular momentum from the magnetization of the macrospin  
$\mathbf{M}(\mathbf{x},t)$ to the environment, became a matter of fundamental but also applied 
interest. Indeed, this is an important limiting factor in the reduction of the remagnetization rate in magnetic recording devices.
The complex environment of the macrospin  offers many different mechanisms for damping 
(spin-orbit coupling, coupling to the lattice vibrations, spin-waves, \textit{etc.}). In 1935, Landau and 
Lifshitz~\cite{Landau-Lifshitz} (LL) proposed a phenomenological classical equation to account for all these processes 
through a unique damping coefficient. 
Though the LL equation yields reasonable results for small damping coefficient, 
it was initially taken to be unphysical as it predicts an acceleration of the magnetization precession at large damping. 
Some years later, Gilbert~\cite{Gilbert55,Gilbert04} cured this problem by proposing an equation in which damping enters in a different manner. Ultimately, the two equations are equivalent after a redefinition of the parameters. 
Today, the latter equation is known as the Landau-Lifshitz-Gilbert (LLG) equation. 
Its inherent non-linearity makes it a difficult problem to solve, but also the source of very rich phenomena.
Explicit solutions can be obtained in some particular cases
but  the generic cases require numerical integration~\cite{Bertotti-etal,Cimrak2008}. 
The dynamics of magnetic moments ruled by the LLG equation have been the focus of considerable research over the last 60 years.

For sufficiently small ferromagnetic particles, thermal fluctuations can become relevant in determining the
magnetization dynamics. In 1963, Brown~\cite{Fuller63} proposed a simple extension of the LLG equation in which thermal agitation 
is introduced \textit{\`a la} Langevin, \textit{i.e.}, \textit{via} the addition of a random magnetic field the correlation of which  is proportional
to the temperature. As already shown by Brown, the stochastic magnetization dynamics can also be analyzed within the Fokker-Planck
framework, that is, \textit{via} the partial differential equation satisfied by the time-dependent probability distribution. Since then, thermal effects in the magnetization
dynamics have been studied in great detail~\cite{Bertotti-etal}. Moreover, progress 
in computational capabilities have now made possible numerical solutions of realistic situations~\cite{Cimrak2008,Aditi2012}.

A new means to manipulate the magnetization of  ferromagnetic devices
\textit{via} the spin-transfer torque effect has opened a new perspective into magnetic storage technologies.
It has also renewed the interest in the study of magnetization dynamics.
Pioneered by the theoretical work by Slonczewski~\cite{Slonczewski1996} and Berger~\cite{Berger1996}, 
 it was realized almost two decades ago
that one could also manipulate the magnetization by exchanging angular momentum with a spin-polarized current of 
electrons. This proposal rapidly found some experimental evidence~\cite{Tsoi1998,Myers1999} and, since then, the rapid advances in the field of 
spintronics have led to everyday life applications such as high-density magnetic recording devices 
(see, for instance,~\cite{Brataas12} for a short review). Although quantum in origin, the spin-transfer torque effect can be handled
phenomenologically by an appropriate generalization of the LLG equation. The LLG equation also appears in the 
context of molecular spintronics as a semi-classical limit of a full quantum problem~\cite{Brataas02,Parcollet05,Brataas08,Bode11}.

In the quest for smaller and smaller devices with higher storage density capabilities, it has recently become crucial to take 
into account the effect of thermal fluctuations on the magnetization dynamics. Indeed, smaller magnetic domains 
are more prone to thermally activated magnetization reversal because the energy barrier is proportional to the volume 
of the domain. Furthermore, as one seeks to reduce the intensity of the currents used in magnetic devices (to reduce both 
power consumption and Joule heating), thermal fluctuations limit the signal-to-noise ratio.

Most of the existing literature dealing with thermal effects in the magnetization dynamics of nanoparticles and nano-devices 
analyze the problem either by direct numerical simulation of the stochastic LLG equation or \textit{via} the study of the corresponding
Fokker-Planck equation~\cite{Coffey12}. Yet, within the study of general stochastic processes there exists also the possibility of
approaching the problem \textit{via} the path integral formulation of generating functionals~\cite{Martin73, Janssen76, deDominicis76}.
The use of generating functionals is an elegant and powerful method to derive generic properties of dynamical systems.
Compared to the other  approaches mentioned above, a path-integral 
is handy for computing moments, probability distribution functions, transition probabilities and responses. 
It is also particularly well suited when it comes to perturbation theory and renormalization group analysis, as
one can easily set up a diagrammatic  expansion. This approach has proven to be extremely useful to treat many physical systems. 
Two problems of current interest that have points in common with ours are the path-integral formulation of the 
motion of a Brownian particle confined to a finite volume~\cite{Lubensky2007} and the path-integral formulation~\cite{Andreanov05,Velenich2008,Kim2013}
of a system of interacting Langevin processes~\cite{Dean97}.

In this manuscript, we build the framework for a path-integral description of the magnetization dynamics described 
by the stochastic LLG equation. The expert reader, eager to see the actions in their various forms,
can jump to  
Eqs.~(\ref{eq:S_det_cart00}), (\ref{eq:S_diss_cart00}) and (\ref{eq:S_jac_cart00}) for the 
Landau action, and 
Eqs.~(\ref{eq:SLLGjac-det})-(\ref{eq:SLLGjac-jac}) for the Gilbert action both in Cartesian coordinates;
or to 
Eqs.~(\ref{eq:action-det-sph-L2})-(\ref{eq:esferica-jac}) for the Landau action, and  
Eqs.~(\ref{eq:action-det-sph-Gilbert-maintext})-(\ref{eq:action-jac-sph-Gilbert-maintext}) for the Gilbert 
action both in spherical coordinates.

The manuscript is organized as follows.
In Sec.~\ref{sec:landau-lifshitz-gilbert} we introduce the stochastic LLG equation in its Landau-Lifshitz and Gilbert formulations,
and we discuss the physical significance of each of its terms.

In Sec.~\ref{sec:prescriptions} we discuss the discretization of the stochastic integrals; 
the prescription used has to be carefully specified since 
it is a defining part of the dynamics in multiplicative random processes. We explain how the conservation of the 
magnetization magnitude imposes the Stratonovich discretization prescription if one persists in using the original LLG equation,
or guides one into generalizations of it if the choice is to use other discretization schemes.
This is particularly useful for numerical simulations for which the It\^o prescription has been claimed to be simpler to implement.

In Sec.~\ref{sec:functional}, we derive the generating functional of the local magnetization by using the 
Martin-Siggia-Rose-Janssen-deDominicis (MSRJD) path integral~\cite{Martin73, Janssen76, deDominicis76}, which was originally introduced in the context of 
conventional Langevin equations. Although, the MSRJD formalism is well understood for systems with additive noise or with inertia 
(see, for instance,~\cite{AronLeticia2010}), the stochastic LLG equation is a Markovian first order stochastic differential equation with
multiplicative white noise bringing in extra difficulties such as a non-trivial Jacobian~\cite{arenas2010,arenas2012a,arenas2012, Lubensky2007,Honkonen}
that contributes to the action.
 
In Sec.~\ref{sec:spherical-coordinates} we reset the general discretization scheme formulation of the stochastic LLG equation
in the spherical coordinate system.  \textit{A priori}, this is the natural coordinate system for numerical simulations 
since the non-linear constraint on the modulus of the magnetization is built in, lowering the dimensionality of the problem from $3d$ to $2d$. 
However, the random noise transforms non-trivially to the new coordinate system and, in particular, it is correlated with the polar angles, 
a fact that has often been overlooked in the literature. We therefore correct the widespread but erroneous assumption stating that the two-dimensional random field in 
spherical coordinates is a simple Gaussian white noise with vanishing mean.
We end this Section with the derivation of 
the generating functional in the spherical coordinate system.

Finally, in Sec.~\ref{sec:recap} we recap our results and we announce a number of applications
of the formalism here introduced that we plan to investigate in the future.

Technical details are reported in the Appendices.

\section{Magnetization dynamics}
\label{sec:landau-lifshitz-gilbert}

Let ${\mathbf M}({\mathbf x},t)$ be the three-dimensional vector describing the local  magnetization in a 
ferromagnet and let us assume that its modulus is constant, and equal to the spontaneous magnetization of 
the ferromagnet at the working temperature $T$: $|{\mathbf M}({\bf x},t)| \equiv M_\rms$. 
For simplicity, we also assume that the space dependence can be neglected, that is to say, we work in the macro-spin approximation, 
and we later introduce the time-dependent unit vector ${\mathbf m}(t)$ through ${\mathbf M}(t)=M_\rms {\mathbf m}(t)$.
We will not follow the historic route in the presentation but rather give a  presentation that we find more natural.

\subsection{Landau-Lifshitz-Gilbert (LLG) equation}

Whenever thermal fluctuations are negligible, the dissipative dynamics of the 
magnetization can be described  by the Landau-Lifshitz-Gilbert (LLG) equation~\cite{Gilbert55,Gilbert04}
\begin{equation}
 \rmd_t \mathbf{M} =  - \gamma_0  {\mathbf{M}} \wedge \left( {\mathbf{H}}_{\eff}
 -\frac{\eta}{M_\rms}  \rmd_t \mathbf{M} \right)
 \; ,
 \label{eq:Gilbert}
\end{equation}
where $\rmd_t$ denotes the time derivative $\rmd/\rmd t$. The modulus of the magnetization is automatically conserved thanks to the cross product denoted $\wedge$.
$\gamma_0\equiv\gamma \mu_0$ is the product of $\gamma$, the gyromagnetic ratio relating the magnetization to the angular momentum,
and $\mu_0$, the vacuum permeability constant.
The gyromagnetic factor is given by $\gamma=\mu_B g/\hbar$ (in our convention $\gamma>0$ for the electronic spin) with 
$\mu_B$  Bohr's magneton and $g$ Lande's $g$-factor.
 The first term in the {\sc rhs} describes the magnetization precession around the local effective magnetic field $\mathbf{H}_{\eff}$, while 
the second term is a phenomenological description of the dissipative mechanisms, introduced by Gilbert,
which slow down this precession and push 
${\mathbf M}$ towards the magnetic field ${\mathbf H}_{\rm eff}$ while keeping the modulus $M_\rms$ fixed.
In principle, dissipation could enter the problem with a memory kernel because the feedback of the environment
 on the magnetization involves, \textit{a priori}, retardation effects. 
In that perspective, the choice of $\eta \mathbf{M} \wedge \rmd_t \mathbf{M} /M_\rms$ for the dissipative torque  
can then be understood as the first term of a derivative expansion.
 The damping constant $\eta$ takes into account several dissipative mechanisms (spin-orbit coupling, magnon-phonon, 
 magnon-impurity, etc.). Dimensions are such that $[H]= [\gamma_0]^{-1} [t]^{-1}$ and $[\gamma_0\eta]=1$. In most relevant cases, such as magnetic recording, one has $\gamma_0\eta\ll 1$.

The local effective magnetic field can be divided into conservative and non-conservative contributions,
\begin{equation}
\mathbf{H}_{\eff}=  \mathbf{H}_{\eff}^{\rm c} + \mathbf{H}_{\eff}^{\rm nc} \;.
\end{equation}
The conservative contribution $\mathbf{H}_{\eff}^{\rm c}=-\mu^{-1}_0 \delta U/\delta {\mathbf{M}}$,  
with $U$  the energy per unit volume,
can originate from an external (possibly time-dependent) magnetic field $\mathbf{H}_{\ext}$, 
and a crystal field $\mathbf{H}_{\any}$ associated to the crystalline anisotropy,
\begin{equation}
U=- \mu_0\mathbf{M} \cdot  {\mathbf{H}}_{\ext} + V_{\any}(\mathbf{M})
\; .
\end{equation}
$V_{\any}$ is here the anisotropy potential (per unit volume).
Examples of anisotropy potentials are~\cite{Bertotti-etal}
\begin{equation}
{V}_{\any}(\mathbf{M}) = K \sum_{i\neq j} M^2_i M^2_j
\; , \qquad\qquad i,j = x,y,z
\; ,
\label{eq:cubic-anisotropy}
\end{equation}
corresponding to a cubic crystalline structure
(each of the three Cartesian axes is a minimum of this function), or
\begin{equation}
{V}_{\any}(\mathbf{M}) = K (M_\rms^2-M_{z}^2)
\; ,
\label{eq:uniaxial}
\end{equation}
associated to  the uniaxial symmetry (having a minimum along the so-called easy axis, here the $z$-axis).
 We will give a very timely example of non-con\-ser\-vative ${\mathbf H}^{\rm nc}_{\rm eff}$, the 
 spin-torque term, in Sec.~\ref{sec:spin-torque}. From the very structure of the 
 equation, one sees that only the instantaneous component of the effective field, 
 ${\mathbf H}_{\rm eff}= {\mathbf H}^\perp_{\rm eff} + {\mathbf H}^\parallel_{\rm eff}$, that is perpendicular to the 
 magnetization, \textit{i.e.} ${\mathbf H}_{\rm eff}^\perp$ such that ${\mathbf H}_{\rm eff}^\perp \cdot {\mathbf M} =0$,  
   has an effect on its dynamics. Indeed, the field appears as
 \begin{equation}
 {\mathbf M} \wedge {\mathbf H}_{\rm eff} =  {\mathbf M} \wedge {\mathbf H}^\perp_{\rm eff}
\; . 
 \end{equation}

The LLG equation can be written in the equivalent form~\cite{Landau-Lifshitz}
\begin{equation}
\rmd_t {\mathbf{M}} =  
 - \frac{\gamma_0}{1+\gamma_0^2 \eta^2}
\mathbf{M} \wedge
\left( 
 {\mathbf{H}}_{\eff}
 + \frac{\eta \gamma_0}{M_\rms} {\mathbf{M}} \wedge  {\mathbf{H}}_{\eff} 
 \right)
 \; , 
\label{eq:landau}
\end{equation}
by simply factorizing the time-derivatives on the left-hand-side of Eq.~(\ref{eq:Gilbert}).
In the rest of this manuscript, we shall refer to Eq.~(\ref{eq:Gilbert}) and Eq.~(\ref{eq:landau}) 
as the Gilbert and the Landau formulation of the same LLG equation, respectively.\footnote{We found some 
contradictions in the literature concerning the relation between the parameters in the two formulations. To be more precise, 
we find that the parameters transform as in~\cite{Bertotti-etal,Cimrak2008} but differently from what is shown in~\cite{Berkov2007}.}
(We will come back to the passage from one to the other in Sec.~\ref{sec:functional}.)

Depending on the space dimension and the interactions, the LLG can exhibit a variety of non-linear
structures (solitons, spatio-temporal patterns, etc.), see~\cite{Lakshmanan2013} for a review.
In some particular cases explicit solutions are known but the generic problem requires numerical integration~\cite{Bertotti-etal,Cimrak2008}.

\subsection{Stochastic LLG equation}

In 1963, Brown~\cite{Fuller63} showed that thermal fluctuations can be taken into account by adding a random field, ${\mathbf H}$, {\it \`a la} Langevin 
in Eq.~(\ref{eq:Gilbert}) 
\begin{equation}
 \rmd_t {\mathbf{M}} =  - \gamma_0 {\mathbf{M}} \wedge \left( \mathbf{H}_{\eff}+  \mathbf{H}  - \frac{\eta}{M_\rms}\rmd_t{\mathbf{M}} \right)
 \label{eq:sLLG1}
\end{equation}
or, equivalently, by adding this field in Eq.~(\ref{eq:landau})
\begin{equation}
 \rmd_t {\mathbf{M}} =  
 - \frac{\gamma_0}{1+\gamma_0^2\eta^2}  
 \mathbf{M} \wedge
\left[ 
  (\mathbf{H}_{\eff} + \mathbf{H})
 + \frac{\eta\gamma_0}{M_\rms}   \mathbf{M} \wedge ( \mathbf{H}_{\eff} +\mathbf{H}) 
 \right]
 \; .
\label{eq:sLL1}
\end{equation}
Assuming that the thermal environment is composed of a large number of degrees of freedom and that the 
time-scale on which they relax is the shortest one in the problem, it is natural to consider Gaussian white noise 
statistics for the stochastic field. In formal terms,
\begin{equation} \label{eq:HhatD}
\langle H_i(t) \rangle_{\mathbf H} = 0
\;, \qquad
\langle  H_i(t)  H_j(t')  \rangle_{\mathbf H}  = 2 D \delta_{ij}  \delta(t-t')
\; ,
\end{equation}
for all $i,j=x,y,z$. The parameter $D$ is such that the fluctuation-dissipation relation 
of the second kind (in Kubo's terminology~\cite{Kubo66}) is 
satisfied, that is to say, that the sLLG equation takes the magnetization to equilibrium, \textit{i.e.} with a Gibbs-Bolztmann distribution function at the temperature $T$ of the environment. As we will show explicitly in Sec.~\ref{sec:Fokker-Planck},
its relation to the parameters in the problem is 
\begin{equation}
D = \frac{\eta k_{\rm B}T}{M_\rms V \mu_0}
\; . 
\end{equation}
$V$ is the volume of the sample and $k_{\rm B}$ the Boltzmann constant.
 Dimensions are such that $[D] = [t] [H^2]$.

In both Eqs.~(\ref{eq:sLLG1}) and (\ref{eq:sLL1}), the noise couples multiplicatively to the time-dependent magnetization. 
To completely define the Markovian dynamics of such a stochastic equation with multiplicative white noise, 
one needs to specify a prescription for the way in which the noise acts at a microscopic time level.
As we will discuss in more detail in Sec.~\ref{sec:prescriptions}, written as they are, 
Eqs.~(\ref{eq:sLLG1}) and (\ref{eq:sLL1}) have to be understood as 
Stra\-to\-no\-vich equations, \textit{i.e.} with a mid-point prescription~\cite{Bertotti-etal,Langevin-Coffey,Martinez-etal}, as this is the unique 
scheme consistent with the conservation of the modulus of the magnetization. Henceforth, we will refer to these equations as 
the {\em stochastic Stratonovich} Landau-Lifshitz-Gilbert equations. We will also show in Sec.~\ref{sec:prescriptions} how to generalize these
equations to use other discretization schemes.

It can be useful to work with adimensional variables and parameters. 
If one defines $\gamma_0 M_\rms t \mapsto t$, $\mathbf{M}/M_\rms \mapsto \mathbf{m}$,
$\mathbf{H}_{\eff}/M_\rms \mapsto \mathbf{h}_{\eff}$, $\mathbf{H}/M_\rms \mapsto \mathbf{h}$, 
and $\eta\gamma_0 \mapsto \eta$ the dynamical equation becomes
\begin{equation}
 {\rm d}_t {\mathbf{m}} =-
 {\mathbf{m}} \wedge \left(\mathbf{h}_{\eff}  + {\mathbf{h}} - \eta \ {\rm d}_t{\mathbf{m}}  \right)
 \;,
 \label{eq:sLLG1-adim}
\end{equation}
or
\begin{equation}
{\rm d}_t {\mathbf{m}} =-
 \frac{1}{1+\eta^2} \ \mathbf{m} \wedge
 [(\mathbf{h}_{\eff}+ \mathbf{h}) + \eta \ {\mathbf{m}} \wedge(\mathbf{h}_{\eff}+ \mathbf{h} ) ]
 \; .
 \label{eq:sLL1-adim}
\end{equation}
The Gaussian white noise ${\mathbf h}$ has zero average and correlations characterized by a new 
diffusion constant, $\gamma_0 D/M_\rms \mapsto D$, 
\begin{equation}
\langle h_{i}(t) \rangle_{\mathbf h} = 0
\;, \qquad
\langle  h_{i}(t)  h_{j}(t')  \rangle_{\mathbf h} = 2 D \ \delta_{ij}  \delta(t-t')
\; .
\end{equation}
All variables and parameters are now adimensional.
Here again, the Gilbert and Landau stochastic Eqs.~(\ref{eq:sLLG1-adim}) and (\ref{eq:sLL1-adim}) 
have to be understood as Stra\-to\-no\-vich equations and are totally equivalent. 

\subsection{Spin-torque}
\label{sec:spin-torque}

The LLG equation~(\ref{eq:Gilbert}) or its stochastic version (\ref{eq:sLLG1}) can also be used to
describe spin-transfer torque in magnetic
multilayers~\cite{Slonczewski1996,Berger1996}. By means of this effect, a spin-polarized electric current flowing
through the multilayer can  affect the orientation of the magnetization in one of the
layers (the free moving one).
Although the origin of this effect is quantum mechanical, it can be phenomenologically studied by adding to the LLG equation an
extra ``torque" term which depends on the relative position of the layers, their polarization, the electrical current, and other properties of the 
material. This
effect has been the topic of a tremendous amount of work during the last 15 years  due to its many potential applications in the 
field of spintronics. In its
simpler form, the Landau-Lifshitz-Gilbert-Slonczewski equation reads~\cite{Stiles2006}
\begin{equation}
 \rmd_t {\mathbf{M}} =  - \gamma_0 {\mathbf{M}} \wedge \left(\mathbf{H}_{\eff} + {\mathbf H}
   - \frac{\eta}{M_\rms}  \rmd_t \mathbf{M} \right)
    - \frac{g \mu_B J {\cal P}}{2 M_\rms^2 d e}  \mathbf{M} \wedge (\mathbf{M} \wedge \mathbf{p})
    \; ,
 \label{eq:sLLG1-spintorque-dim}
\end{equation}
where we singled out from ${\mathbf H}_{\rm eff}$ the spin-torque term.
Here, $g\mu_B = \gamma\hbar$ with $\gamma$ the gyromagnetic ratio, $J$ is the current per unit area, 
${\cal P}$ represents the (dimensionless) polarization function of the fixed layer, 
${\mathbf p}$ is a unit vector in the direction of the fixed layer magnetization, 
$d$ the interlayer separation, and $e$ is the electric charge of the carriers. (Recall that in the macromagnetic approach
${\mathbf M}$ is the magnetization per unit volume.)

In dimensionless variables, this equation reads
\begin{equation}
 \rmd_t {\mathbf{m}} =  -  {\mathbf{m}} \wedge \left( \mathbf{h}_{\eff} + {\mathbf h}
  - \eta  \rmd_t \mathbf{m} \right)
   - \chi \mathbf{m} \wedge (\mathbf{m} \wedge \mathbf{p}) 
 \label{eq:sLLG1-spintorque-adim}
\end{equation}
with
\begin{equation}
 \chi=\frac{\hbar J}{2 M_\rms^2 d e \mu_0} {\cal P}
 \; , 
 \end{equation}
 a dimensionless parameter.

The last terms in Eqs.~(\ref{eq:sLLG1-spintorque-dim}) and (\ref{eq:sLLG1-spintorque-adim}) 
represent a torque that cannot be derived from a potential 
 and provides a mechanism to exchange energy between
the magnetic moment and the environment \textit{via} the current that flows in the sample. Thus,
a spin polarized current of electrons can be used  to manipulate the dynamics of the magnetic moment, its spin flip rate and 
its precession frequency~\cite{Brataas12}.

Initial studies of spin-torque assisted dynamics focused on the zero temperature limit~\cite{Stiles2006,Berkov2008}
while temperature effects were considered in \cite{Myers2002,Koch2004,Li2004,Russek2005,Apalkov2005,Xiao2005}. 
Most of this literature addresses the problem {\it via} the numerical integration of the stochastic equation~\cite{Cimrak2008} or the 
Fokker-Planck approach. A simple way to understand, at least qualitatively, thermal effects in spin-torque driven dynamics is 
by realizing that spin-torque effectively modifies the dissipation coupling. If the current is 
such that this dissipation is lowered, one can expect thermal effects to  be more relevant than in the absence of 
spin torque. Even more so, spin torque may not only reduce the effective dissipation constant but also 
make it {\em negative} meaning that the system amplifies disturbances from equilibrium~\cite{Koch2004}.

\section{Discretization prescriptions}
\label{sec:prescriptions}

In this Section, we discuss the time-discretization of the stochastic magnetization 
dynamics. Indeed, as in any stochastic differential equation, the discretization used to 
define the Wiener integral is a defining part of the model, and should be carefully taken into account to obtain sensible physical 
results~\cite{Gardiner}. As we will show, in the way the stochastic equation is currently written [either in the form in 
Eq.~(\ref{eq:sLLG1}) or in the one in Eq.~(\ref{eq:sLL1}), or in their adimensional versions], it conserves 
the magnetization modulus {\it only if} the Stratonovich prescription is used. Although this is not a new result 
(see, e.g.,~\cite{Bertotti-etal}), one still finds quite confusing statements in the literature~\cite{Martinez-etal,Berkov2002}.
In the following discussion, we first describe the discretization of generic stochastic equations and we later 
discuss the case of the stochastic LLG (sLLG) equation. We next show how Eqs.~(\ref{eq:sLLG1}) and (\ref{eq:sLL1}) have 
to be modified if one wishes to work with other stochastic 
calculi in order, notably, to ensure the conservation of the modulus of ${\mathbf M}$. 

\subsection{Rules of stochastic calculus}
\label{subsec:rules}

Let us consider a set of time-dependent random variables $x_a(t)$ satisfying a set of first order differential equations,
\begin{equation}
\rmd_t x_a(t) = f_a({\bf x}(t)) + g_{aj}({\bf x}(t)) h_j(t)
\;,
\label{eq.Langevin}
\end{equation}
where  the $x_a$, with $a=1,\dots, N$ are the components of the stochastic variable  ${\bf x}$ and $h_j$, with $j=1,\dots, M$ are the components
of  the Gaussian white-noise process ${\bf h}$ satisfying
\begin{equation}
 \left\langle h_i(t)\right\rangle_{\mathbf h}   = 0 \; , \qquad\qquad  \left\langle h_i(t) h_j(t')\right\rangle_{\mathbf h} = 2D \delta_{ij}\delta(t-t')
 \;.
\label{whitenoise}
\end{equation}
Here and in what follows, we use the Einstein summation convention over repeated indices. In Eq.~(\ref{eq.Langevin}),
$f_a({\bf x})$ and $g_{aj}({\bf x})$ are  the so-called drift term and diffusion matrix, respectively, and are arbitrary smooth
functions of ${\bf x}(t)$ (but not of its time-derivatives). Notice that, in general, the diffusion matrix 
is rectangular as the number of random variables is not necessarily equal to the number of random processes.

Due to the fact that  $h_j(t)$ has an infinite variance, Eq.~(\ref{eq.Langevin}) is ill-defined
until the product $g_{aj}({\bf x}(t)) h_j(t)$ is given a proper
microscopic meaning. This subtlety can be understood by looking at the integral
\begin{equation}
\int    g_{aj}({\bf x}(t)) h_j(t) \;  \rmd t= \int  g_{aj}({\bf x}(t))\;  \rmd W_j(t) \ ,
\end{equation}
where we have introduced the Wiener processes $W_j(t)$ as $h_j(t)=\rmd W_j(t)/\rmd t$.
The Riemann-Stieltjes integral is defined by using a set of times $t_0 < t_1 < \dots < t_N ={\cal T}$ in the 
interval $[t_0, {\cal T}]$ and constructing the sum
\begin{equation}
\int_{t_0}^{\cal T}  g_{aj}({\bf x}(t))\;  \rmd W_j(t)=
\lim _{N\to\infty} \sum_{n=0}^{N-1}  g_{aj}(\bar {\bf x}_n) \left[ W_j(t_{n+1})-W_j(t_n) \right]
\label{eq.Wiener}
\end{equation}
where $\bar x_n$ is taken in the interval $[x_n,x_{n+1}]$ and the {\sc rhs} of Eq.~(\ref{eq.Wiener}) converges in the mean-square
sense\footnote{A sequence of random variables
$X_n$ converges in the mean-square sense to another random
variable $X$ if  $\lim_{n\to\infty} \langle (X_n -
X)^2 \rangle  = 0$~\cite{Gardiner}.}.
For a smooth measure $W_j(t)$ the limit converges to a unique value, regardless of the choice of the $\bar x_n$'s.
However, a Wiener process $W_j(t)$ is not smooth;  in fact, it is nowhere differentiable.
Therefore, the value of the integral depends on the prescription used to choose the $\bar x_n$'s.
There are several ways to define this integral that can be collected in the so-called
``generalized Stratonovich prescription''~\cite{Hanggi1978}
or ``$\alpha$-prescription''~\cite{Janssen-RG}, for which one uses
\begin{eqnarray}
\bar {\bf x}_n &=& \alpha {\bf x}_{n+1} + (1-\alpha) {\bf x}_{n}
\; , 
\end{eqnarray}
with $0\le \alpha \le 1$, and
\begin{equation}
g_{aj}(\bar {\bf x}_n)=g_{aj}(\alpha {\bf x}(t_{n+1}) + (1-\alpha){\bf x}(t_n))
\; .
\label{eq.prescription}
\end{equation}
The case $\alpha=0$ corresponds to the pre-point It\^o prescription and $\alpha=1/2$ coincides with the mid-point Stratonovich one.
The post-point prescription, $\alpha=1$, is also used~\cite{Hanggi1978, Hanggi1982, Klimontovich}. 

For Markov processes with multiplicative white-noise, each choice of $\alpha$ corresponds to a different stochastic evolution.
For any physical problem, the prescription is dictated by the order of limits when sending the time scales associated to inertia and the
relaxation of the thermal bath to zero~\cite{Kupferman}. 
Once the prescription is fixed, the stochastic dynamics are unambiguously defined.

In the cases  in which the time scale associated to inertia is much smaller than the relaxation time of the bath, Eq.~(\ref{eq.Langevin}) is
given an unambiguous meaning by adding a little color to the Gaussian noise, \textit{i.e} a finite variance~\cite{Hanggi-shot}, and
by eventually taking the white-noise limit at the end of the calculations.
This regularization procedure is equivalent to the Stratonovich prescription, $\alpha=1/2$~\cite{vanKampen,Zinn-Justin}.
In the present case of magnetization dynamics, there is  {\it a priori} no term playing the role of inertia.

The rules of calculus applied to the stochastic variables also depend on the prescription.
In particular, the chain rule used to differentiate an arbitrary function
$Y({\bf x}(t))$ of a set of stochastic variables reads~\cite{arenas2010,arenas2012a,arenas2012}
\begin{equation}
\rmd_t Y({\bf x}(t))=\frac{\partial Y}{\partial x_a}\; \rmd_t x_a +(1-2\alpha)  D\; \frac{\partial^2 Y}{\partial x_a\partial x_b}\; g_{ak} g_{bk} \; .
\label{eq.chainrule}
\end{equation}
Clearly, in the Stratonovich prescription ($\alpha=1/2$), Eq.~(\ref{eq.chainrule}) is the usual chain rule of conventional calculus. For $\alpha=0$, Eq.~(\ref{eq.chainrule}) is the so-called It\^o formula.
For the $\alpha=1$ prescription, the latter differentiation rule only differs from the It\^o formula by the sign of the last term.
In fact, these two prescriptions are related by a time reversal transformation $t\to -t$ and $\alpha\to (1-\alpha)$.

\subsection{Discretization scheme for the sLLG equation}
\label{sec:sLLGprescription}

Let us start with the dimension-full Landau formulation of the sLLG equation, \textit{i.e.}, with Eq.~(\ref{eq:sLL1}) 
that we recast in the generic form of Eq.~(\ref{eq.Langevin})
\begin{equation}
{\rm d}_t M_i= g_{ij} \left(H_{{\eff}, j}+ H_j\right)
\; ,
\label{eq:sLLG-functional-int}
\end{equation}
where we introduced the $3\times 3$ diffusion matrix
\begin{equation}
g_{ij}=\frac{\gamma_0}{1+\eta^2\gamma_0^2} \ [\epsilon_{ijk} M_k + \frac{\eta\gamma_0}{M_\rms} (M_\rms^2 \delta_{ij} - M_i M_j)]
\;.
\label{Eq:gij}
\end{equation}
The Latin indices take values $i, j=x,y,z$, $\epsilon_{ijk}$ is the completely antisymmetric Levi-Civita tensor
and we are always assuming summation over repeated indices.
$g_{ij}$ can be decomposed in symmetric and antisymmetric parts, 
$g_{ij} = g_{ij}^\rms+g_{ij}^\rma$, with 
\begin{eqnarray}
&& g_{ij}^\rms = \frac{\gamma_0}{1+\eta^2\gamma_0^2} \frac{\eta\gamma_0}{M_\rms} (M_\rms^2 \delta_{ij} - M_i M_j)
\; , 
\\
&& g_{ij}^\rma = \frac{\gamma_0}{1+\eta^2\gamma_0^2} \epsilon_{ijk} M_k
\; .
\end{eqnarray}
It is simple to show  that $g$ is transverse in the sense that 
\begin{equation}
g_{ij} M_j = M_i g_{ij} = 0 \; .
\end{equation}
Equation~(\ref{eq:sLLG-functional-int}) then yields
\begin{equation}
\mathbf{M} \cdot {\rm d}_t \mathbf{M} =0\;.
\label{eq:identity00}
\end{equation}
Contrary to what it may seem, this result does not imply the conservation of the magnetization modulus, $\rmd (\mathbf{M}\cdot\mathbf{M}) /\rmd t = \rmd M_\rms^2 /\rmd t =0$, for 
all discretization schemes. Indeed, using the appropriate generalized chain rule given in Eq.~(\ref{eq.chainrule}), replacing
\begin{equation}
\label{eq.gg}
 g_{ik}g_{jk}=
 g_{ki}g_{kj} =
 \frac{\gamma_0^2}{1+\eta^2\gamma_0^2}\left(M_\rms^2 \delta_{ij}-M_i M_j \right)
 = \frac{M_\rms}{\eta} g_{ij}^\rms\;,
\end{equation}
and choosing $Y({\bf M})=\mathbf{M}\cdot \mathbf{M} = M_\rms^2$, one finds
\begin{equation}
\rmd_t M_\rms^2 =  4D (1-2\alpha) \frac{\gamma_0^2}{1+\eta^2\gamma_0^2}   M_\rms^2 
\;,
\end{equation}
where we used $M_i {\rm d}_t M_i =0$, Eq.~(\ref{eq:identity00}).
Therefore, the physical constraint that the modulus of the magnetization be conserved by the dynamics can only be satisfied when giving a 
Stratonovich prescription ($\alpha =1/2$) to Eqs.~(\ref{eq:sLLG1}) and~(\ref{eq:sLL1}). 
  
\subsection{$\alpha$-covariant expression of the sLLG} \label{sec:alpha-covariantLLG}

If for any reason one prefers to work with a prescription with $\alpha\neq 1/2$ (for instance, to perform a numerical 
simulation with an algorithm based on the It\^o calculus) while conserving the magnetization modulus, 
the equation has to be changed accordingly.
An elegant way consists in replacing the time-derivative by the {\em $\alpha$-covariant derivative}
\begin{equation}
{\rm d}_t \mapsto {\rm D}^{(\alpha)}_t= \rmd_t + 2D (1-2\alpha) \frac{\gamma_0^2}{1+\eta^2\gamma_0^2}
\;,
\label{eq.covariant}
\end{equation}
so that the $\alpha$-covariant expression of the sLLG equation, in its dimension-full Landau formulation, reads
 \begin{equation}
{\rm D}^{(\alpha)}_t M_i= g_{ij} \left( H_{{\eff},j} + H_j \right) \;.
\label{eq.LLG-covariant}
\end{equation}
The same replacement in Eq.~(\ref{eq:sLLG1}) yields the $\alpha$-covariant expression of the Gilbert formulation of the 
sLLG equation.
In both cases, this replacement is equivalent to adding a {\em spurious drift term} to the equation.
Notice that the second term in Eq.~(\ref{eq.covariant}) is odd under time reversal as it should be for a time-derivative, since the 
time-reversal transformation includes the transformation $\alpha\to (1-\alpha)$~\cite{arenas2012}.

Equation~(\ref{eq.LLG-covariant}) encodes a family of stochastic equations with different 
underlying prescriptions, \textit{i.e.} different $\alpha$. The Stratonovich equation (\ref{eq:sLLG-functional-int}) can naturally be recovered 
by setting $\alpha = 1/2$.
We emphasize that the general-$\alpha$ Eq.~(\ref{eq.LLG-covariant}) is strictly equivalent to the Stratonovich equation 
(\ref{eq:sLLG-functional-int}) and that  the choice of $\alpha$ cannot have any consequence on the physical properties of the system.
In particular, Eq.~(\ref{eq.LLG-covariant}) conserves the norm of the magnetization for any $\alpha$ since $M_i{\rm D}^{(\alpha)}_t 
M_i=0$, which implies, using the generalized chain-rule in Eq.~(\ref{eq.chainrule}), $\rmd_t M_\rms^2 =0$, and 
ensure the approach to Boltzmann equilibrium in conservative cases as we show the in the next Subsection.

The same kind of argument can be applied to the adimensional equations~(\ref{eq:sLLG1-adim}) and (\ref{eq:sLL1-adim}).

\comments{ *** Comment on adimesional equation ***
In this case, 
\begin{equation}
g_{ij}= (1+\eta^2)^{-1} [\epsilon_{ijk} m_k + \eta (m^2 \delta_{ij} - m_im_j)]
\label{eq:g-adim}
\end{equation}
and $g_{ik} g_{jk} = 
(1+\eta^2)^{-1} (m^2 \delta_{ij} - m_i m_j)$. The $\alpha$-covariant 
derivative is ${\rm D}_t^{(\alpha)}  = {\rm d}_t + 2D (1-2\alpha) (1+\eta^2)^{-1}$. 
The $\alpha$-covariant adimensional sLLG equation in its Landau form then reads
\begin{equation}
{\rm D}_t^{(\alpha)} m_i = g_{ij} ({h_{\rm eff}}_i + h_i)
\; ,
\label{eq.LLG-covariant-adim}
\end{equation}
with $g_{ij}$ given in Eq.~(\ref{eq:g-adim}). 
}

\subsection{Fokker-Planck approach}
\label{sec:Fokker-Planck}

 An alternative method to study the time evolution of a stochastic process is the 
Fokker-Planck (FP) approach, in which the system is characterized  by the  
probability of finding ${\bf M}$  (or ${\mathbf m}$)  at time $t$. 
The probability distribution $P({\bf M}, t)$ satisfies a {\it deterministic} partial differential equation,  
the solution of which completely describes the dynamics of the system. 

The FP equation associated to the sLLG equation in Gilbert's formulation and for 
Stratonovich calculus was derived by Brown~\cite{Fuller63}, see also~\cite{Bertotti-etal,GarciaPalacios98}.
We will show below that the $\alpha$-covariant stochastic equation (\ref{eq.LLG-covariant}) 
leads to a FP equation that is {\it independent} of $\alpha$.

The FP method allows one to prove that the stochastic process described by Eq.~(\ref{eq.LLG-covariant}) leads, at long times and under conservative time-independent forces, to the equilibrium Gibbs-Boltzmann probability distribution for any value of $\alpha$, provided that the noise correlation $D$ is set by an Einstein relation.
     
\subsubsection{Derivation of the Fokker-Planck equation}
     
In order to derive the Fokker-Planck equation, we begin with the identity 
\begin{equation}
 P({\bf M},t+\Delta t) = \int \ud{{\bf M}_0} P({\bf M},t+\Delta t|{\bf M}_0,t)P({\bf M}_0,t)
 \; ,
  \label{Eq.Identity}
\end{equation}
where $P({\bf M},t+\Delta t|{\bf M}_0,t)$ is the conditional probability of finding ${\bf M}$ at 
the time $t+\Delta t$, provided the system was in the state ${\bf M}_0$ at the previous time $t$ 
(note that ${\mathbf M}_0$ is not necessarily  the initial magnetization here). 
The integral in Eq. (\ref{Eq.Identity}) runs over all accessible  values of  ${\bf M}_0$.
This equation holds for any value of $\Delta t$ but we will later focus on infinitesimal time increments.

To make contact with the stochastic process in the Langevin-like description, it is convenient to define the 
conditional probability in the following way:
\begin{equation}
\label{eq.conditionalP}
 	P({\bf M},t+\Delta t|{\bf M}_0,t) = \langle \delta({\bf M} - {\bf M}(t+\Delta t)) \rangle_{\mathbf H}
	\; ,
\end{equation}
where the mean value is taken over the noise $\mathbf{H}$, and ${\bf M}(t+\Delta t)$ is determined 
by the stochastic equation (\ref{eq.LLG-covariant}) with the initial condition ${\bf M}(t)={\bf M}_0$. 
Expanding Eq.~(\ref{eq.conditionalP}) in powers of $\Delta{\bf M}={\bf M}(t+\Delta t)-{\bf M}(t)={\bf M}(t+\Delta t)-{\bf M}_0$ we 
immediately obtain 
\begin{eqnarray}
		&& P({\bf M},t+\Delta t|{\bf M}_0,t) 
		= \delta({\bf M}-{\bf M}_0) - \partial_i\delta({\bf M}-{\bf M}_0) \ \langle \Delta M_i\rangle_{\mathbf H} \nonumber \\
		 &&
		 \qquad\qquad\qquad\qquad\qquad+ \frac{1}{2}
		\partial_{i}\partial_{j}\delta({\bf M}-{\bf M}_0) \ \langle \Delta M_i\Delta M_j\rangle_{\mathbf H} +  \ldots 
		\qquad
		\label{eq.delta_Cond}
\end{eqnarray}
where $\partial_i \equiv \partial/\partial{M}_i$ and the ellipsis indicate terms involving higher order  correlations.

The idea is to compute the correlations $\langle \Delta M_i\rangle_{\mathbf H}$ and $\langle \Delta M_i\Delta M_j\rangle_{\mathbf H}$ 
to leading order in $\Delta t$ and then take the limit 
$\Delta t\to 0$. To do this, we integrate the sLLG equation (\ref{eq.LLG-covariant}) 
in the interval $(t, t+\Delta t)$  obtaining 
\begin{equation}
	\Delta M_i = f_i ({\bf M}_0)\Delta t + g_{ij}[{\bf M}_0+ \alpha \Delta {\bf M}(t + \Delta t)]
	\int_{t}^{t+\Delta t} \ud{t'} H_j(t')\;,
	\label{eq.deltam}
\end{equation}
where we have used the $\alpha$-discretization procedure to define the last Wiener integral  as explained in Sec.~\ref{subsec:rules},
$W_j(t+\Delta t) - W_j(t) = \int_t^{t+\Delta t} dt' H_j(t')$, 
and  
\begin{eqnarray}
\label{eq.f}
f_i ( {\bf M}_0)&=&g_{ij}({\mathbf M}_0) H_{{\eff},j} 
- 2D \frac{\gamma_0^2}{1+\eta^2\gamma_0^2}\; (1-2\alpha) {M_0}_i\;, \\   
\label{eq.gij}
g_{ij}( {\bf M}_0)&=&\frac{\gamma_0}{1+\eta^2\gamma_0^2} \ \left[\epsilon_{ijk} {M_0}_k + \frac{\eta\gamma_0}{M_\rms}  
(M_\rms^2 \delta_{ij} - {M_0}_i {M_0}_j)\right]\;,
\end{eqnarray}
see Eq.~(\ref{Eq:gij}).
Solving Eq.~(\ref{eq.deltam}) to order $\Delta t$ (by expanding $g_{ij}$ 
in powers of $\Delta{\bf M}$ and solving iteratively), and computing the mean values using $\langle H_i(t)H_j(t') \rangle_{\mathbf H}=2D\delta_{ij}\delta(t-t')$, we finally obtain 
\begin{eqnarray}
 \langle \Delta M_i\rangle_{\mathbf H} &=& f_i({\bf M}_0)\Delta t + 2D \alpha g_{k\ell}({\bf M}_0)\partial_k g_{i\ell}({\bf M}_0) \Delta t\,, \label{prom_deltax} \\
 \langle \Delta M_i\Delta M_j\rangle_{\mathbf H} &=& 2 D g_{ik}({\bf M}_0)g_{jk}({\bf M}_0)\Delta t \,. 
 \label{prom_deltaxdeltay}
\end{eqnarray}
Interestingly enough, the mean value as well as  the two point correlation are 
of order $\Delta t$. Higher momenta of the distribution such as 
$\langle \Delta M_i\Delta M_j\Delta M_k\rangle_{\mathbf H}$ are of higher order in $\Delta t$ 
and do not contribute to the expansion 
in Eq.~(\ref{eq.Pm}) for sufficiently small $\Delta t$. It is important to note that these results depend on ${\mathbf M}_0$.

Replacing now Eq.~(\ref{eq.delta_Cond}) into Eq.~(\ref{Eq.Identity}) and integrating over 
${\bf M}_0$ we have
\begin{eqnarray}
		&&
		P({\bf M},t+\Delta t)-P({\bf M},t) 
		=  - \partial_i \ [   \langle \Delta M_i\rangle_{\mathbf H} P({\bf M},t) ] \nonumber \\
		& & 
			\qquad\qquad\qquad\qquad\qquad\qquad
			+ \frac{1}{2}
		\partial_{i}\partial_{j} \ [\langle \Delta M_i\Delta M_j\rangle_{\mathbf H} P({\bf M},t) ]
		+ \ldots
		\qquad\qquad
		\label{eq.Pm}
\end{eqnarray}
Using the averages found in Eqs.~(\ref{prom_deltax}) and (\ref{prom_deltaxdeltay}) and  
performing the continuum limit $\Delta t \to 0$ we finally find the desired partial differential equation for $P({\bf M},t)$, 
\begin{eqnarray}
	\label{eq.FP0}
	&&
	\partial_t P({\bf M},t) 
		=  
		-\partial_i \left[ \left( f_i({\bf M})+ 2D \alpha g_{k\ell}({\bf M})\partial_k g_{i\ell}({\bf M}) \right)
		   P({\bf M},t)  \right]  \nonumber \\
		& & 
		\qquad\qquad\qquad 
		+ \partial_i\partial_j \left[
		D g_{ik}({\bf M})g_{jk}({\bf M})
		\ P({\bf M},t) \right]
\;.
\label{eq:FP}
\end{eqnarray}

It is instructive to rewrite Eq.~(\ref{eq:FP}) in the form of a continuity equation, 
\begin{equation}
\partial_tP({\bf M},t) +\boldsymbol{\nabla}\cdot \mathbf{J}({\bf M},t) = 0\;,
\end{equation}
where the components of the current probability $\mathbf{J}({\bf M},t)$ are given by
\begin{eqnarray}
J_i&=& \left[ f_i + D (2\alpha-1) g_{k\ell}\partial_k g_{i\ell}  - D g_{ik}\partial_j g_{jk} -  
D g_{ik}g_{jk}\partial_j \right] P \;.
\end{eqnarray}
The two following properties of the diffusion matrix $g$ 
\begin{eqnarray}
g_{k\ell}\partial_k g_{i\ell} &=& -\frac{2 \gamma_0^2}{1+\eta^2 \gamma_0^2} {M_0}_i \;,\\
g_{ik}\partial_j g_{jk}&=&0 \; ,
\end{eqnarray}
and the explicit form of $f_i$ given in Eq.~(\ref{eq.f})
allow us to arrive at the simpler expression, 
\begin{eqnarray}
J_i= \left(  g_{ij} H_{{\rm eff},j} - 
D g_{ik}g_{jk}\partial_j \right)  P\;.
\end{eqnarray}
Thus, the Fokker-Planck equation, related with the stochastic process governed by 
the $\alpha$-covariant sLLG equation (\ref{eq.LLG-covariant}), is given by
\begin{equation}\label{eq:FP333}
\partial_tP({\bf M},t) =\partial_i
\left\{
\left[ g_{ij} H_{{\rm eff},j} -
D g_{ik}g_{jk}\partial_j \right] P({\bf M},t)
 \right\}
 \;.
\end{equation}
Note that, as anticipated, this differential equation is $\alpha$-independent. Thus, the $\alpha$-covariant sLLG equation 
(\ref{eq.LLG-covariant}) leads to a unique time  evolution for the magnetization probability for any value of the parameter 
$\alpha$. Also, from $g_{ij} M_j=0$, it is immediate to check that the current probability is transverse,  
$ {\bf J} \cdot {\bf M}=0$, meaning that there is no dynamics in the direction of the magnetization and, consequently, 
the time evolution preserves the magnetization modulus.

\subsubsection{Approach to equilibrium}

With the Fokker-Planck equation defined in Eq.~(\ref{eq:FP333}), we can study the asymptotic probability distribution.  
Any stationary state at long times, 
\begin{equation}
P^{\rm st}({\bf M})=\lim_{t\to\infty} P({\bf M},t)\;,
\end{equation}  
has an associated stationary current that satisfies $\partial_i J^{\rm st}_i=0$.  
However, solutions to this equation do not necessarily represent equilibrium distributions 
as they could be non-equilibrium steady states with non-vanishing probability current.
Indeed, the equilibrium distribution is defined as a stationary solution of the Fokker-Planck equation with zero current,
${\mathbf J}_{\rm eq}={\mathbf 0}$, and 
it is expected to be reached asymptotically in the absence of explicit time dependent  or non-potential forces. 

Considering the {\it Ansatz} $P_{\rm eq}({\bf M})= 
N \exp(-{\cal V}[{\bf M}])$ for the equilibrium probability, where $N$ is a normalization constant, the condition $J^{\rm eq}_i=0$ implies
\begin{equation}
g_{ij} H_{{\rm eff},j} +
D g_{ik}g_{jk}\partial_j {\cal V} =0\;.
\end{equation}
In order to solve it for ${\cal V}({\bf M})$, we assume that the effective magnetic field can be obtained from a potential energy density 
$H_{{\rm eff},i}=H^c_{{\rm eff},i} = -\mu_0^{-1} \partial_i U[{\bf M}]$. Then, 
\begin{equation}
-\mu_0^{-1} g_{ij} \partial_j U + 
D g_{ik}g_{jk}\partial_j {\cal V} =0\;.
\end{equation}
We can further simplify this equation by noting that the antisymmetric part contributes with 
a  topological divergence-less term since  $\partial_ i \left(g^\rma_{ij} \partial_j U\right)= 0$ for symmetry reasons.  Also, from 
Eq.~(\ref{eq.gg}),   $g^\rms_{ij}=(\eta/M_\rms) g_{ik}g_{jk}$. Then the  null stationary current condition  takes the very simple form, 
\begin{equation}
 g^\rms_{ij} \partial_j \left( \frac{U}{\mu_0} - 
\frac{DM_\rms}{\eta} {\cal V}\right) =0\;,
\end{equation}
with obvious solution ${\cal V}=\eta/(DM_\rms\mu_0) \ U$. The equilibrium solution of the Fokker-Planck equation is of the Gibbs-Boltzmann type 
$P_{\rm GB}=N\exp(-\beta U V)$ provided we choose
\begin{equation}
D=\eta k_{\rm B}T/(M_\rms V \mu_0)
\label{eq:D}
\end{equation}
which is the Einstein relation or, 
more generally, a consequence of the fluctuation-dissipation theorem of the second kind in Kubo's terminology~\cite{Kubo66}.

\section{Generating functional}
\label{sec:functional}

It is a well established fact that stochastic process can be analyzed with
the help of path integrals. Janssen~\cite{Janssen76} introduced such a description for
stochastic processes, a very convenient formalism to manipulate correlation and response
functions,  that is close in spirit to the operator approach of Martin-Siggia-Rose~\cite{Martin73}.
The formalism was later applied to quenched disorder by De~Dominicis~\cite{deDominicis76,deDominicis78} and is often referred as the MSRJD formalism.
Over the years, there have been numerous attempts to generalize the work of Janssen to
the case of multiplicative noise, {\it i.e.}, the case in which the diffusion
matrix depends on the stochastic variable. The literature on this problem is
rather extensive and in many cases also confusing. No
attempt will be made here to review this literature (the interested reader can look at~\cite{Tirapegui82} for a thorough 
description of path integral methods in the stochastic context). Instead, we will
focus on the specific problem at hand, the stochastic dynamics
of the magnetization in the Cartesian and spherical coordinate systems.

Let us consider that the system is prepared at an initial time, $t_0$, and subsequently let evolve until the final time of the experiment, 
$\mathcal{T}$. It is common to consider the limit $\mathcal{T}\to \infty$, but we prefer to keep the final time arbitrary. 
In the rest of this manuscript, we will encounter many time-integrals of the form $\int_{t_0}^{\cal T} \Ud{t} \ldots $ and,
for the sake of simplicity, we will most frequently simply denote them  $\int\ud{t} \ldots$

The generating functional of observables averaged over thermal histories is defined as 
\begin{align}
 \mathcal{Z}[\boldsymbol{\lambda}] = \langle \ \langle \
 \exp \int_{t_0}^{\cal T} \Ud{t} \boldsymbol{\lambda}(t) \cdot \mathbf{M}_{\mathbf H}(t)  \ \rangle_\rmi \ \rangle_\mathbf{H}  \;,
 \label{eq:uno}
\end{align}
where $ \langle \cdots \rangle_\rmi = \int \rmd \mathbf{M}_0 \cdots P_\rmi [\mathbf{M}_0, {\mathbf H}_{\rm eff}(t_0)] $
is the average over the initial conditions, ${\mathbf M}_0 \equiv {\mathbf M}(t_0)$ with $|\mathbf{M}_0| = M_\rms$,
distributed with the probability distribution function $P_\rmi[\mathbf{M}_0, {\mathbf H}_{\rm eff}(t_0)]$ which is normalized to unity.
The average over the realizations of the thermal noise, which are distributed according to the probability distribution 
functional $P_{\rm n}[\mathbf{H}]$, is denoted
\begin{align}
 \langle \cdots \rangle_\mathbf{H} \equiv \int\uD{[\mathbf{H}]} \cdots P_{\rm n}[\mathbf{H}]
 \; .
\end{align}
$P_{\rm n}[\mathbf{H}]$ is also normalized to unity.
In the rest of the manuscript, we use the notation
\begin{align}
\langle \cdots \rangle \equiv \langle\langle \cdots \rangle_\rmi  \rangle_\mathbf{H} \;.
\end{align}
$\boldsymbol{\lambda}$ is a vector source that couples linearly to the fluctuating magnetization configuration 
$\mathbf{M}_{\mathbf H}(t)$ which is the unique solution to the sLLG
equation for a given initial condition $\mathbf{M}_0$
and a given history of the noise $[\mathbf{H}]$. By construction ${\cal Z}[\boldsymbol{\lambda} = \mathbf{0}]=1$.

In the rest of this Section, we construct the MSRJD representation of the generating functional $\mathcal{Z}[\boldsymbol{\lambda}]$
according to the following steps. The first one is to exchange the dependence on the unique solution $[\mathbf{M}_{\mathbf{H}}]$ with a sum over all trajectories 
by imposing the equation of motion written in the form $\mbox{\textbf{Eq}}[\mathbf{M},\mathbf{H}]=\mathbf{0}$
 \textit{via} a delta functional:
\begin{eqnarray}
&&
{\cal Z}[\boldsymbol{\lambda}] = \int {\cal D}[\mathbf{H}] \ P_{\rm n}[\mathbf{H}] \
 \int {\cal D}[\mathbf{M}]\ 
 P_{\rm i}[\mathbf{M}_0, {\mathbf H}_{\rm eff}(t_0)] 
\nonumber\\
&&
\qquad\;\;
\times \
\delta\Big{[} \mbox{\textbf{Eq}}[\mathbf{M},\mathbf{H}] \Big{]}
\ |{\cal J}[\mathbf{M},\mathbf{H}]| \
\exp \int_{ t_0}^{\cal T} \! \ud{t} {\boldsymbol{\lambda}}(t) \cdot \mathbf{M}(t) 
\;. \label{eq:generatingZ}
\end{eqnarray}
The measure ${\cal D}[\mathbf{M}] $, defined precisely in App.~\ref{app:discrete}, 
has to be understood as the sum, at each time step, over vectors $\mathbf{M}(t)$ in  the entire $\mathbb{R}^3$ space. In particular, it includes the integration over the initial conditions at time $t_0$.
As discussed in Sec.~\ref{sec:sLLGprescription}, the constraint $|\mathbf{M}(t)| = cst$ is encoded in the equation of 
motion (see later Sec.~\ref{sec:conservation_modulus}).
Notice that, at the level of the path integral, 
this allows one to consider $M_x$, $M_y$ and $M_z$ as unconstrained variables, 
\textit{i.e.} $|\mathbf{M}(t)| \neq \mbox{cst}$. 
The Jacobian ${\cal J}[\mathbf{M},\mathbf{H}]$ ensures that the {\sc rhs} of Eq.~(\ref{eq:generatingZ}) does not depend 
on the particular formulation of $\mbox{\textbf{Eq}}[\mathbf{M},\mathbf{H}]$:
\begin{eqnarray}
\mathcal{J}[\mathbf{M},\mathbf{H}] \equiv
\det_{ij;uv}
\begin{array}{c}
\displaystyle
\frac{\delta {\rm{Eq}}_{\it i}[\mathbf{M},\mathbf{H}](u)}{\delta M_{j}(v)}
\end{array}
\end{eqnarray}	
with the coordinate indices $i,j=x,y,z$ and the times $u,v$.
(If one thinks in
terms of discrete time, the equation is imposed by a product of $\delta$-functions starting 
at time indexed $n=1$ and ending at time indexed $n=N$.
We next exponentiate the functional Dirac delta with the help of a Lagrange
multiplier  $[\rmi\hat{\mathbf{M}}]$. 
Afterwards, we average over the initial conditions $\mathbf{M}_0$ and the noise realizations $[\mathbf{H}]$. Finally, 
we obtain a path integral representation of $\mathcal{Z}[\boldsymbol{\lambda}]$ in which the trajectories are weighted 
by the exponential of an action functional $S[\mathbf{M},\rmi\hat{\mathbf{M}}]$:
\begin{equation}
{\cal Z}[{\boldsymbol{\lambda}}] =
\int {\cal D}[{\mathbf M}] {\cal D}[\rmi \hat {\mathbf M}] \ 
\exp\left[ 
S[{\mathbf M}, \rmi \hat {\mathbf M}] + \int \rmd t \ {\boldsymbol{\lambda}}(t) \cdot {\mathbf M}(t)
\right]
\; . 
\end{equation}

\subsection{Landau formulation}
\label{sec:MSRDJ}

In Sec.~\ref{sec:landau-lifshitz-gilbert}, we presented two formulations of the same 
sLLG equation (Gilbert and Landau) with generic discretization prescriptions. 
These different starting points lead to distinct formulations of the generating functional that describe the same physics. 
In the presentation below, we choose to start with the dimension-full 
Landau formulation of the $\alpha$-covariant sLLG equation, \textit{i.e.} with Eq.~(\ref{eq.LLG-covariant})
\begin{eqnarray}
{\rm{Eq}}_{Li}[\mathbf{M},\mathbf{H}] &\equiv& {\rm D}_t^{(\alpha)} M_i - g_{ij}( H_{{\eff},j}+H_j ) = 0
\; , 
\label{eq:def-Eq}
\end{eqnarray}
and we construct a formalism that is valid in any $\alpha$-prescription. The subscript $L$ stands for Landau formulation
here and in the rest of this Section. We collected in the
magnetic field ${\mathbf H}_{\rm eff}$ all contributions from conservative as well as
non-conservative origin, including the possible spin-torque terms.
We discuss the generating functional  obtained when starting from the Gilbert formulation, 
and its equivalence to the Landau formulation, in Sec.~\ref{sec:Landau2Gilbert} and App.~\ref{app:path-integral:Gilbert}.
In Sec.~\ref{sec:observables} we recall how these generating functionals enable one to compute all cumulants and linear 
responses of the magnetization.

The operator in the determinant can be worked out explicitly and put into the form
\begin{equation}
\frac{\delta {\mbox{Eq}}_{Li}(u)}{\delta M_j(v)}
=
\delta_{ij} {\rm d}_u \delta(u-v) + A_{ij}(v) \delta(u-v)\;,
\end{equation}
with
\begin{eqnarray}
A_{ij}
=
2
D(1-2\alpha) \frac{\gamma_0^2}{1+\eta^2\gamma_0^2}
 \delta_{ij}
-
\frac{\partial g_{ik}}{\partial M_j}  ( H_{{\eff},k}+H_k )
-
  g_{ik} \frac{\partial  H_{{\eff},k}}{ \partial M_j}
\end{eqnarray}
where we assumed that ${\mathbf H}_{\rm eff}$ is ultra-local in time in the sense that it involves only the 
magnetization evaluated at time $u$ but no time-derivatives, {\it i.e.} the effective magnetic 
field can be of the form ${\mathbf H}_{\rm eff}({\mathbf M}(u), u)$.
From now on we will use the notation $\partial_i = \partial /\partial M_i$ as in Sec.~\ref{sec:Fokker-Planck}.
Using that the inverse of 
 $\delta_{ik} {\rm d}_u \delta(u-w)$ is proportional to
 the Heaviside function $ \delta_{ik}\Theta(w-v)$, the Jacobian becomes
\begin{align}
  \mathcal{J}_L[\mathbf{M}, \mathbf{H}]
& =
 \; \det_{ik;uw}  \left[  \delta_{ik}(u) {\rm d}_u\delta(u-w)\right]
 \nonumber\\
&
 \qquad\qquad \times
 \det_{kj;wv}   \left[
 \delta(w-v) \delta_{kj}
   + \Theta(w-v)  A_{kj}(v) \right]
   \; .
   \label{sec:discussRVsA}
  \end{align} 
We treat the second determinant in the way described in App.~\ref{sec:determinants}.
Notice that $C_{kj}(w,v) = \Theta(w-v)  A_{kj}(v)$ is causal. Usually, the expansion stops at the first order due to the 
causality of the operator $C$. However, when there is a white-noise dependence in this operator, 
as it is the case here, one has to be careful and consider the possible contribution of the second order
term, $C^2$~\cite{Arnold2000,Lubensky2007}. This is explained in App.~\ref{sec:determinants}. In the Cartesian framework, the contribution of the second order term is 
an irrelevant constant and only the first term has a non-trivial dependence on the magnetization field.
We obtain
 \begin{align} \label{eq:jacjacjac}
   \mathcal{J}_L[{\mathbf M}, {\mathbf H}]
 \propto \; &
\exp\ [ \alpha \int{} \ud{t} A_{ii}(t) ]
\;,
 \end{align}
where the factor $\alpha$ comes   $\Theta(0)=\alpha$ in the $\alpha$-prescription. Indeed, 
when working with continuous-time notations, it can be shown~\cite{Gardiner} that the $\alpha$ discretization prescription introduced in 
Sec.~\ref{sec:prescriptions} translates into the prescriptions $\int_{t_1}^{t_2}\Ud{t} G(t) \delta(t-t_1) = (1-\alpha) G(t_1)$ and $\int_{t_1}^{t_2}\Ud{t} G(t) \delta(t-t_2) = \alpha G(t_2)$ 
for any $G(t)$ that is a causal functional of the random field. In particular, for $G(t)=1$, this can be conveniently collected into $\Theta(0) \equiv \alpha$.
The integrand in Eq.~(\ref{eq:jacjacjac}) is
\begin{displaymath}
 A_{ii}(t)  =  
 2
 D(1-2\alpha) \frac{\gamma_0^2}{1+\eta^2\gamma_0^2}
  \delta_{ii}
+
\frac{2\eta\gamma_0^2}{1+\eta^2\gamma_0^2}
\frac{M_i}{M_\rms} (H_{{\eff},i}+H_i )
 -
  g_{ik} \partial_i H_{{\eff},k}
\;.
\end{displaymath}
Finally, dropping all terms that are constant in the overall normalization, we obtain
 \begin{eqnarray}
   \mathcal{J}_L[{\mathbf M}, {\mathbf H}]
& \!\!  \!\!
\propto \!\! \!\! &
 \exp \left\{ \frac{\alpha\gamma_0^2}{1+\eta^2\gamma_0^2}
 \frac{1}{M_\rms}
 \int{} \ud{t}  \Big{[}
2 \eta \ \mathbf{M} \cdot  {\mathbf{H}}_{{\eff}}
\right.
\nonumber\\
&&
\qquad\qquad 
+ \eta
(M_i M_j -\delta_{ij} M_\rms^2)\partial_j H_{{\eff},i}
\nonumber\\
&&
\qquad\quad\;\;\;
\left.
+ M_\rms \epsilon_{ijk} M_k \partial_j H_{{\rm eff}, i}^{\rm nc} + 2 \eta \ \mathbf{M} \cdot  {\mathbf{H}}
 \Big{]}\!  \right\}
 \; .
 \label{eq:Jcart-L}
 \end{eqnarray}

Coming back to the generating functional ${\cal Z}[\boldsymbol{\lambda}]$, we now exponentiate 
the delta functional that imposes the sLLG equation as
\begin{align}
\delta \Big{[} \mbox{\textbf{Eq}}_L[\mathbf{M},\mathbf{H}] \Big{]} \propto
\int {\cal D}[\hat{\mathbf{M}}] \
\exp  \int{} \ud{t} \ \rmi\hat{\mathbf{M}} \cdot\mbox{\textbf{Eq}}_L[\mathbf{M},\mathbf{H}]  
\; ,
\end{align}
in which $\hat{\mathbf{M}}$ is integrated over the entire $\mathbb{R}^3$ space at each time slice
and has dimension $[\rmi\hat {\mathbf M}] = [{\mathbf M}]^{-1}$.
The integration over the Gaussian white noise $\mathbf{H}$ 
yields
\begin{eqnarray}
&&
 \int {\cal D}{[\mathbf{H}}]
 \
 \displaystyle
 \exp \left\{
 \int{} \ud{t}
 \left(
 {\frac{ 2\alpha \eta\gamma_0^2}{1+\eta^2\gamma_0^2}  M_j H_j
-\rmi\hat{M}_i  g_{ij} H_j
- \frac{1}{2} \frac{H_j H_j}{2D}}
\right) \right\}
\nonumber\\
&&
\qquad\qquad
\propto \exp \left\{ D \int{} \ud{t}  \ g_{ji} g_{ki} \rmi\hat{M}_j \rmi\hat{M}_k  \right\}
\; ,
\end{eqnarray}
 where the cross term vanishes thanks to  the property $g_{ij} M_j =0$. 
We will also drop a factor that depends only on $M_\rms^2$ in the overall normalization.
Accordingly, the effect of the random field contribution coming from the Jacobian disappears. 
Altogether, we recast the generating functional in the form
\begin{equation}
{\cal Z}[\boldsymbol{\lambda}] =
\int {\cal D}[\mathbf{M}] {\cal D}[\hat{\mathbf{M}}]
\ \exp \left\{ S_L+\int{} \! \ud{t}  \boldsymbol{\lambda} \cdot \mathbf{M} \right\}
\; .
\end{equation}
The action $S_L$ is a functional of the histories of the magnetization $[\mathbf{M}]$ 
and the auxiliary field $[\rmi\hat{\mathbf{M}}]$. It reads
\begin{align}
 S_L =&\; \;  \ln P_\rmi[\mathbf{M}_0, \mathbf{H}_{\rm eff}(t_0)] 
 \nonumber \\
& +  \int{} \ud{t} \rmi\hat M_i \left( {\rm D}^{(\alpha)}_t M_i - g_{ij} H_{{\eff}, j} \right)
 + D \int{} \ud{t}  g_{ji} g_{ki} \rmi\hat{M}_j \rmi\hat{M}_k  
 \nonumber \\
& + \frac{\alpha\gamma_0}{1+\eta^2 \gamma_0^2 } \frac{1}{M_\rms}  \int{} \ud{t}  \Big{[}
2 \eta\gamma_0 \ \mathbf{M} \cdot  {\mathbf{H}}_{{\eff}}
+ \eta\gamma_0 (M_i M_j -\delta_{ij} M_\rms^2)\partial_j H_{{\rm eff}, i} 
\nonumber\\
& \qquad\qquad\qquad\qquad\qquad
+ M_\rms \epsilon_{ijk} M_k \partial_j H_{{\rm eff}, i}^{\rm nc}
 \Big{]}\!
\; . 
\label{eq:generating-functional0}
\end{align}
Notice that we identified $\mathbf{M}_0$ with $\mathbf{M}(t_0)$ and we included the integral over the initial conditions, 
$\int\rmd \mathbf{M}_0$, into the functional integral $\int\uD[\mathbf{M}]$ and their probability distribution, $P_\rmi(\mathbf{M}_0),$ into the action functional.

One identifies the contribution of the deterministic dynamics, $\widetilde S_{L,{\rm det}}$, the dissipative and 
thermal effects, $\widetilde S_{L,{\rm diss}}$, and the Jacobian
\begin{equation}
S_L = \widetilde S_{L,{\rm det}} + \widetilde S_{L,{\rm diss}} + S_{L,{\rm jac}}\;.
\label{eq:SLandau}
\end{equation}
Using the decomposition of $g_{ij}$ in symmetric and antisymmetric parts, 
$g_{ij} = g_{ij}^\rms+g_{ij}^\rma$,  
already used in Sec.~\ref{sec:Fokker-Planck},
and renaming indices conveniently
one has
\begin{align}
\widetilde  S_{L,{\rm det}} =
& \;\; \ln P_\rmi[\mathbf{M}(t_0), \mathbf{H}_{\rm eff}(t_0)]  
 + \int \Ud{t}
 \rmi\hat{M_i}  \cdot \left( {\rm D}_t^{(\alpha)} M_i -g^\rma_{ij} {H_{\rm eff}}_j \right)
 \; , 
 \label{eq:S_det_cart00} 
 \\
\widetilde S_{L,{\rm diss}} =& \;\;
 \int \Ud{t}
\ g_{ij}^\rms \ \rmi\hat{M_i} \left( 
\frac{DM_\rms}{\eta} \
\rmi\hat{M_j} 
- 
{H_{\rm eff}}_j
\right)
 \;,
 \label{eq:S_diss_cart00}
 \\
S_{L,{\rm jac}}=& \;\;
 \frac{\alpha\gamma_0}{1+\eta^2\gamma_0^2}
 \frac{1}{M_\rms}
 \int \Ud{t} \Big{[}
2 \eta\gamma_0 \ \mathbf{M} \cdot  {\mathbf{H}}_{{\eff}}
+ \eta\gamma_0
(M_i M_j -\delta_{ij} M_\rms^2)\partial_j H_{{\rm eff}, i}
\nonumber\\
&
\qquad\qquad\qquad\qquad\;\;\;
+ M_\rms \epsilon_{ijk} M_k \partial_j H_{{\rm eff}, i}^{\rm nc}
 \Big{]}\!
\; .
 \label{eq:S_jac_cart000}
\end{align}
Remembering that $D$ is proportional to $\eta$ [see Eq.~(\ref{eq:D})], one sees that $\widetilde S_{L,{\rm diss}}$ vanishes  at $\eta=0$ while the remaining deterministic and Jacobian parts, $\widetilde S_{L,{\rm det}}+S_{L,{\rm jac}}$,  
yield what one would have obtained starting from the equation without dissipation. This writing of the Landau action,
shows that $g^\rms$ contains the information on the dissipative aspects of the dynamics.

Note that $g^\rms$ does not have an inverse. This is related to the ``gauge invariance" of the action, 
that retains the same form under the parallel translation $\rmi \hat {\mathbf M} \to \rmi \hat {\mathbf M} + a {\mathbf M}$,
with $a$ generic. 

\subsubsection{Rewriting of the Jacobian}
\label{sec:rewriting-Jacobian}

As mentioned in the introduction, one should distinguish the parallel and perpendicular components of the
effective field ${\mathbf H}_{\rm eff}={\mathbf H}^\perp_{\rm eff} + {\mathbf H}^\parallel_{\rm eff}$. We will 
write them as
\begin{equation}
H_{{\rm eff},i} = H_{{\rm eff}, i}^\parallel +  H_{{\rm eff}, i}^\perp 
=
f M_i + \epsilon_{ilk} M_l T_k  
\; , 
\end{equation}
where $f$ and ${\mathbf T}$ can be functions of  ${\mathbf M}$. This separation is different from the separation in conservative and non-conservative contributions. 
Notice that ${\mathbf T}$ is not defined uniquely as any translation parallel to ${\mathbf M}$ leaves this relation 
unchanged.
Using this separation, one  proves that the Jacobian contribution to the action, $S_{L,{\rm jac}}$,  is independent 
of the parallel component of the effective field, more precisely, it is independent of $f$. This arises 
due to the cancellation of the first term with one stemming from the second one.

The Jacobian contribution to the action then reads
\begin{eqnarray}
&& S_{L,{\rm jac}} = \;\;
 \frac{\alpha\gamma_0}{1+\eta^2\gamma_0^2}
 \frac{1}{M_\rms}
 \int \Ud{t} \Big{[}
\eta\gamma_0
(M_i M_j -\delta_{ij} M_\rms^2)\partial_j H^\perp_{{\eff}_i}
\nonumber\\
&& 
\qquad\qquad\qquad\qquad
+ M_\rms \epsilon_{ijk} M_k \partial_j H^\perp_{{\eff}_i}
 \Big{]} 
\; .
 \label{eq:S_jac_cart00}
\end{eqnarray}
As expected, we conclude that the full action does not depend on ${\mathbf H}_{\rm eff}^\parallel$.

\subsubsection{Conservation of the modulus}
\label{sec:conservation_modulus}

The decomposition of the auxiliary field  
 $ \rmi\hat{\mathbf{M}} = \rmi \hat {\mathbf M}_\parallel + \rmi \hat {\mathbf M}_\perp$ into a sum of parallel and perpendicular components 
to  the magnetization ${\mathbf M}$ (\textit{i.e.} $\rmi\hat{\mathbf{M}}_\perp \cdot {\mathbf M} = 0$ and 
$\rmi\hat{\mathbf{M}}_\parallel \wedge {\mathbf M} = {\mathbf 0}$) will allow us to show that the modulus 
$M_\rms$ is conserved by the dynamics and to derive the Gilbert formulation of the 
action functional (see Sec.~\ref{sec:Landau2Gilbert} for the latter). 

As $M_i g_{ij} = 0$, then 
 \begin{equation}
 \rmi \hat M_i g_{ij} = 
 \frac{\gamma_0}{1+\eta^2\gamma_0^2} [ \epsilon_{jkl} M_k \rmi \hat M_l + \eta\gamma_0 M_\rms \rmi \hat M_j ]
 \; .
 \end{equation}
Therefore, 
we find $ g_{ji} g_{ki} \rmi\hat{M}_j \rmi\hat{M}_k = 
M_\rms^2 \gamma_0^2/ (1+\eta^2\gamma_0^2) \
\rmi\hat{\mathbf{M}}_\perp \cdot  \rmi\hat{\mathbf{M}}_\perp$. This property allows us to rewrite the 
action in (\ref{eq:generating-functional0}) in an equivalent form:
\begin{equation}
S_L= {S}_{L,{\rm det}} + {S}_{L,{\rm diss}} + S_{L,{\rm jac}}\;,
\end{equation}
with
\begin{align}
  {S}_{L,{\rm det}} =
& \ln P_\rmi[\mathbf{M}(t_0), \mathbf{H}_{\rm eff}(t_0)]  
 + \int \Ud{t}
 \rmi\hat{\mathbf{M}}_\parallel  \cdot {\rm D}_t^{(\alpha)} \mathbf{M} 
\nonumber \\
&+ \frac{1}{1+\eta^2\gamma_0^2}
 \int \Ud{t}
 \rmi\hat{\mathbf{M}}_\perp \cdot 
 \Big{[} \rmd_t \mathbf{M} - 
\eta \gamma_0^2 M_\rms  \ \mathbf{H}_{\eff}
 \nonumber\\
 & 
 \qquad\qquad\qquad\;
 - 
 \left( \eta\gamma_0 M_\rms^{-1} \ \rmd_t \mathbf{M}  \wedge  \mathbf{M} 
 + \gamma_0 \ {\mathbf M} \wedge \mathbf{H}_{\eff}  \right) \Big{]}
 \;, 
 \label{eq:S_det_cart0} \\
{S}_{L,{\rm diss}} =&
\frac{1}{1+\eta^2\gamma_0^2 }  
\int \Ud{t}
\rmi\hat{\mathbf{M}}_\perp \cdot 
\Big{[}
D \gamma_0^2 \, M_\rms^2 \ 
\rmi\hat{\mathbf{M}}_\perp
\nonumber\\
& 
\qquad\qquad\qquad +
\left( \eta\gamma_0 M_\rms^{-1} \ \rmd_t \mathbf{M} \wedge \mathbf{M} +  \eta^2\gamma_0^2 \rmd_t \mathbf{M}\right)
\Big{]}
 \; .
 \label{eq:S_jac_diss0}
\end{align}
The Jacobian contribution, $S_{L,{\rm jac}}$, is again given by Eq.~(\ref{eq:S_jac_cart00}) as it does not involve $\rmi\hat{\mathbf M}$.
We added and subtracted a term in what we called deterministic and dissipative contributions for later convenience.
The dissipative part, ${S}_{L,{\rm diss}}$, only regroups terms that involve the interactions 
with the environment such as thermal effects and the dissipative torque.  It does not 
depend on the deterministic forces acting on the problem, grouped in ${\mathbf H}_{\rm eff}$, that 
appear only in ${S}_{L,{\rm det}}$.
At $\eta=0$, ${S}_{L,{\rm diss}}$ vanishes  while the remaining deterministic part, 
${S}_{L,{\rm det}}$,  again yields what one would have obtained starting from the equation without dissipation.
This cutting up will take a clear meaning in the Gilbert formulation of the generating 
functional (see Sec.~\ref{sec:Landau2Gilbert}).

The sector of the formalism that involves  $\rmi\hat{{\mathbf M}}_\parallel$, the component 
of $\rmi\hat{\mathbf{M}}$ which is parallel to  $\mathbf{M}$,  encodes the conservation of the modulus of the magnetization.
Indeed, the only term involving $\rmi{\hat{ \mathbf{M}}}_\parallel$ in the action functional given in 
Eqs.~(\ref{eq:S_det_cart0}) and (\ref{eq:S_jac_diss0}) is
\begin{align}
 \int{} \ud{t}  \rmi\hat{{\mathbf M}}_\parallel  \cdot  {\rm D}_t^{(\alpha)} {\mathbf M}
  =     \int{} \ud{t}  \rmi\hat{M}_\parallel  \ \frac{\mathbf{M}}{M_\rms} \cdot {\rm D}_t^{(\alpha)} {\mathbf M} 
  \;.
\end{align}
If one were to integrate over $\hat M_\parallel$, this would yield a delta functional
\begin{align}
 \delta\left[ \mathbf{M} \cdot {\rm D}_t^{(\alpha)} {\mathbf M} \right]\;,
\end{align}
that imposes the constraint
${\mathbf M} \cdot {\rm D}_t^{(\alpha)} {\mathbf M} = 0$
at all times, simply expressing the conservation of the modulus, $\rmd_t M_\rms^2 = 0$, as we explained in Sec.~\ref{sec:alpha-covariantLLG}.

\subsection{Gilbert formulation}
\label{sec:Landau2Gilbert}

As we stressed in Sec.~\ref{sec:landau-lifshitz-gilbert}, the Gilbert and Landau 
formulations of the sLLG equation,  Eqs.~(\ref{eq:sLLG1}) and (\ref{eq:sLL1}), or their adimensional 
form in Eqs.~(\ref{eq:sLLG1-adim}) and (\ref{eq:sLL1-adim}),
are strictly equivalent. In Sec.~\ref{sec:MSRDJ}, we constructed a prescription-covariant functional formalism starting from 
the $\alpha$-covariant expression of the Landau formulation of the sLLG equation, namely Eq.~(\ref{eq.LLG-covariant}).
Starting from the $\alpha$-covariant expression of Eq.~(\ref{eq:sLLG1-adim}) 
 and following a similar route 
(see App.~\ref{app:path-integral:Gilbert}), 
one can construct another action 
functional corresponding to the Gilbert formulation of the problem.
The ensuing Gilbert action functional  reads (the subscript $G$ stands for Gilbert formulation):
\begin{align}
 S_G = S_{G,{\rm det}} + S_{G,{\rm diss}} + S_{G,{\rm jac}}
 \label{eq:total-action-Gilbert}
\end{align}
with
\begin{eqnarray}
 S_{G,{\rm det}} \! &\! = \! & \!
 \ln P_\rmi[\mathbf{M}(t_0), \mathbf{H}_{\rm eff}(t_0)] 
 +  \int\ud{t}  \rmi\hat{\mathbf{M}}_\parallel \cdot {\rm D}_t^{(\alpha)} {\mathbf M}
 \nonumber \\
 & & +
   \int\Ud{t} 
   \rmi\hat{\mathbf{M}}_\perp  \cdot  
   \left( \frac{1}{ M_\rms^2} {\rm D}^{(\alpha)}_t \mathbf{M} \wedge \mathbf{M} 
   + \gamma_0 \mathbf{H}_{\eff} \right)
 \; ,
 \;\;\;\;\;  \label{eq:SLLGjac-det} 
 \\
 S_{G,{\rm diss}} \! &\! = \! & \!
 \int \Ud{t}
  \gamma_0\rmi\hat{\mathbf{M}}_\perp  \cdot   \left( D \gamma_0 \rmi\hat{\mathbf{M}}_\perp
  - \frac{\eta}{M_\rms} \ {\rm D}^{(\alpha)}_t \mathbf{M} \right)
 \; . 
\label{eq:SLLGjac-diss}
\\
S_{G,{\rm jac}}
  \! &\! = \! & \!
 \frac{\alpha\gamma_0}{1+\eta^2\gamma_0^2}
 \frac{1}{M_\rms}
 \int \Ud{t} \Big{[}
\eta\gamma_0
(M_i M_j -\delta_{ij} M_\rms^2)\partial_j H^\perp_{{\eff}_i}
\nonumber\\
&& 
\qquad\qquad\qquad\qquad
+ M_\rms \epsilon_{ijk} M_k \partial_j H^\perp_{{\eff}_i}
 \Big{]} 
\; .
 \label{eq:SLLGjac-jac}
\end{eqnarray}
The Jacobian contribution, $S_{G,{\rm jac}}$, is identical to Eq.~(\ref{eq:S_jac_cart00}), 
$S_{G,{\rm jac}}=S_{L, {\rm jac}}$.
The equivalence between the Landau and the Gilbert formulations simply 
corresponds to a transformation of the auxiliary field $\rmi\hat{\mathbf{M}}$.
One passes from the Landau action functional given in Eqs.~(\ref{eq:S_det_cart0}) and (\ref{eq:S_jac_diss0}) 
to the Gilbert formulation in Eqs.~(\ref{eq:SLLGjac-det}) and (\ref{eq:SLLGjac-diss})
\textit{via} the following change
\begin{align}\label{eq:Trans_L2G}
 \rmi\hat{\mathbf{M}}_\perp \mapsto
  -\frac{1}{M_\rms} \left( \frac{1}{M_\rms} \rmi\hat{\mathbf{M}}_\perp \wedge\mathbf{M} + 
  \eta\gamma_0 \ \rmi\hat{\mathbf{M}}_\perp \right)\;,
\end{align}
which is a linear transformation with a constant Jacobian that can be dropped into the overall normalization.
Conversely, one passes from the Gilbert action functional to the Landau formulation 
\textit{via} the inverse transformation
\begin{align}\label{eq:Trans_G2L}
 \rmi\hat{\mathbf{M}}_\perp \mapsto 
 \frac{1}{1+\eta^2\gamma_0^2}  \left[ \rmi\hat{\mathbf{M}}_\perp \wedge\mathbf{M} - 
 \eta\gamma_0 M_\rms \  \rmi\hat{\mathbf{M}}_\perp \right]\;.
\end{align}

\subsection{Observables}
\label{sec:observables}

The generating functional can be used to evaluate the average of functions of the magnetization, in particular
its $n$-times correlation functions, cumulants, and linear responses, by taking variations with respect to 
sources conveniently introduced in the action through linear couplings to the magnetization and the 
auxiliary field. We list a number of these observables below.
 
The averaged magnetization is given by
\begin{align}
 \langle M_i(t)\rangle = 
 \frac{1}{\mathcal{Z}[\boldsymbol{\lambda}=0]} \frac{\delta \mathcal{Z}[\boldsymbol{\lambda}]}{\delta \lambda_i(t)}\Big{\rvert}_{\boldsymbol{\lambda}=0}
 \; . 
\end{align}

The two-time correlations can be obtained from the variation of the ${\mathcal Z}$ with respect to two sources:
\begin{align}
 \langle M_i(t)\, M_j(t')\rangle = \frac{1}{\mathcal{Z}[\boldsymbol{\lambda}=0]} \frac{\delta \mathcal{Z}[\boldsymbol{\lambda}]}{\delta \lambda_i(t) \, \delta \lambda_j(t')}\Big{\rvert}_{\boldsymbol{\lambda}=0}
 \; . 
\end{align}
Similarly, one derives the $n$-times correlation functions by taking variations with respect to 
$n$ factors $\lambda_i$ evaluated at different times.

 The cumulant generating functional is defined as 
 $W[\boldsymbol{\lambda}] \equiv \ln \mathcal{Z}[\boldsymbol{\lambda}]$. For instance, one generates the 
 second order cumulant as
 $$\left. \frac{\delta^2 W[\boldsymbol{\lambda}]}{\delta \lambda_i(t) \delta \lambda_j(t')} \right|_{{\mathbf \lambda} = {\mathbf 0}}
 = 
 \langle M_i(t) M_j(t') \rangle - \langle M_i(t)\rangle \langle M_j(t') \rangle \; . $$

The linear response is the result of 
the effect of a linear perturbation of the local effective magnetic field, 
$\mathbf{H}_{\rm eff} \mapsto \mathbf{H}_{\rm eff} + \widetilde{\mathbf{H}}$, performed at a time $t'$ on the observable of choice. 
The equation of motion and the dynamic action do not depend on $H_{\rm eff}^\parallel$. Therefore, the only 
variation that may have an effect on the averaged observables is the one on $H_{\rm eff}^\perp$.
The linear response of the magnetization component $M_i$ measured at a later time $t$, in the Landau formulation, is
\begin{align}
R_{ij}(t,t') = 
&
\left. 
\frac{\delta \langle M_i(t)\rangle_{S_L} }{\delta {\widetilde{H}_j(t')}}
\right|_{\widetilde{\mathbf H}={\mathbf 0}} 
\!\!\! 
=  
\left.
\frac{\delta \langle M_i(t)\rangle_{S_L} }{\delta {\widetilde{H}^\perp_j(t')}}
\right|_{\widetilde{\mathbf H}^\perp={\mathbf 0}} 
\nonumber\\
= &
\;\;
\langle M_i(t)  \rmi\hat M_k(t') g_{kj}[{\mathbf M}(t')] \rangle_{S_L}\;.
\nonumber
\end{align}
The average has to be taken with the weight $\exp S_L$
with the Landau formulation of the action functional, $S_L$, given in Eqs.~(\ref{eq:S_jac_cart00}), (\ref{eq:S_det_cart0}) and (\ref{eq:S_jac_diss0}).
The presence of the auxiliary field $\rmi\hat{\mathbf{M}}$ in the expression of the response 
is the reason why it is often referred to as the ``response field''.
This ``classical Kubo formula'' can be generalized to the response of any observables $A$:
\begin{align}
R_{A\,j}(t,t')    
 \equiv &
\left. \frac{\delta \langle A(t) \rangle_{S_L}}{\delta \widetilde{H}_j(t')}  \right|_{\widetilde{\mathbf{H}}=0} 
= 
\left. \frac{\delta \langle A(t) \rangle_{S_L}}{\delta \widetilde{H}^\perp_j(t')}  \right|_{\widetilde{\mathbf{H}}=0} 
\;\;
\nonumber\\
= & \;\;
\langle A(t)  \,  \rmi\hat M_k(t') g_{kj}[{\mathbf M}(t')] \rangle_{S_L} 
\;.
\end{align}
The trivial case in which $A$ is set to be a constant, gives $R_{A\,j}(t,t') = 0$ for all $t$ and $t'$, yielding the identity
\begin{align} \label{eq:KuboID0}
\langle  \rmi\hat{M}_k(t) g_{kj}[{\mathbf M}(t')] \rangle_{S_L} = 
0
\; . 
\end{align}

Within the Gilbert formulation of the $\alpha$-covariant generating functional,
one can also compute the linear response function.
Another classical Kubo formula expressing the linear response as a two-time correlator 
is found
\begin{align}
R_{ij}(t,t')  
&
=   
\left. \frac{\delta \langle M_i(t) \rangle_{S_G}}{\delta \widetilde{H}_j(t')}  
\right|_{\widetilde{\mathbf{H}}=0} 
= 
\left. \frac{\delta \langle M_i(t) \rangle_{S_G}}{\delta \widetilde{H}^\perp_j(t')}  
\right|_{\widetilde{\mathbf{H}}^\perp=0} 
\nonumber\\
& 
= \;\;  \langle M_i(t) \,\gamma_0 \rmi\hat{M}_{\perp\,j}(t') \rangle_{S_G} 
\; , 
\end{align}
where the averages are weighted by $\exp S_G$ 
given in Eqs.~(\ref{eq:total-action-Gilbert}) with the contributions~(\ref{eq:SLLGjac-det}), (\ref{eq:SLLGjac-diss}) and (\ref{eq:S_jac_cart00}).
Applied now  to any observable $A$ this relation reads
\begin{align}
R_{A\,j}(t,t')    
= &
\;\;  \langle A(t)  \, \gamma_0 \rmi\hat{M}_{\perp\,j}(t') \rangle_{S_G}
\; . 
\end{align}
The trivial case in which $A$ is set to be a constant, gives 
$R_{A\,j}(t,t') = 0$ for all $t$ and $t'$, yielding the identity
\begin{align} \label{eq:KuboID0G}
\langle  \rmi\hat{\mathbf{M}}_\perp(t) \rangle_{S_G} = 
0
\; . 
\end{align}

\subsection{Equilibrium dynamics}
\label{sec:equilibrium_cond}

The magnetization undergoes equilibrium dynamics if it is prepared in and let evolve under equilibrium conditions. More specifically,
 initial conditions at temperature $T$ have to be drawn from a Gibbs-Boltzmann distribution in a potential $U$ (per unit volume),
the system has to evolve with the same
(time-independent) potential $U$  with no additional non-potential fields, ${\mathbf H}_{\rm eff}^{\mathrm{nc}}={\mathbf 0}$,
and it has to be in contact with a thermal bath in equilibrium at the same temperature.

Thermal initial conditions in the potential $U$ correspond to the Gibbs-Boltzmann probability distribution
\begin{equation}
P_\rmi[\mathbf{M}_0, {\mathbf H}_{\eff}(t_0)] = \rme^{\displaystyle -\beta V U[\mathbf{M}_0, \mathbf{H}_{\ext}(t_0)]-\ln Z[\mathbf{H}_{\ext}(t_0)]}
\end{equation}
with $Z[\mathbf{H}_{\ext}(t_0)] \equiv \int\ud{\mathbf{M}_0} \rme^{\displaystyle -\beta V U[\mathbf{M}_0, \mathbf{H}_{\ext}(t_0)]}$.
The deterministic contribution to the Landau action functional in Eqs.~(\ref{eq:S_det_cart00})-(\ref{eq:S_jac_cart00}) reads
\begin{align}
  S_{L,{\rm det}} =
& \;
\ln P_\rmi[\mathbf{M}_0, \mathbf{H}_{\rm eff}(t_0)]  
 + \int \Ud{t}
 \rmi\hat{M_i}  \left( {\rm D}_t^{(\alpha)} M_i -g^\rma_{ij} H^c_{{\rm eff},j} \right)
 \; ,
 \end{align}
with ${\mathbf H}^c_{\rm eff} = -\mu_0^{-1} \partial U/\partial {\mathbf M}$.   
The Jacobian can be expressed in terms of $U$ by using $H^c_{{\rm eff},i} = -\mu_0^{-1} \partial_i U$ in Eq.~(\ref{eq:S_jac_cart000})
or $H^c_{{\rm eff},i} = -\mu_0^{-1} P_{ij} \partial_j U$ in Eq.~(\ref{eq:S_jac_cart00}) with $P_{ij}$ the projector onto the direction 
perpendicular to the magnetization, $P_{ij} = \delta_{ij} - M_i M_j /M_\rms^2$.
The dissipative part, $S_{L,{\rm diss}}$, remains unchanged.

\section{Spherical coordinate formalism}
\label{sec:spherical-coordinates}

As the modulus of the magnetization $\mathbf{M}$ is constant, the vector rotates on a sphere of radius $M_\rms$, and it is natural 
to work in a spherical coordinates system.
In this Section, we present an equivalent functional formalism for the dynamics of the magnetization, that uses a system of spherical coordinates. 
In Sec.~\ref{sec:sph_LLG_cov}
we present the sLLG equation in spherical coordinates and in any discretization 
prescription.
We stress in Sec.~\ref{sec:sph_random_field} that  the statistics of the random field are not as trivial as they are in a Cartesian description. 
Although the non-trivial character of the noise has been correctly treated in some references (see, for instance, \cite{Fuller63} and Sec.~4.4.5 in~\cite{Gardiner}), 
this subtlety has led to mistaken statements~\cite{Langevin-Coffey} and 
omission or lack of clarity in the literature~\cite{Bertotti-etal,Martinez-etal}. We hope to clarify this matter  once and for all here.
It is also important to note that the transformation from Cartesian to spherical coordinates is non-linear and one cannot naively apply it to the 
generating functional. We construct the corresponding path integral formalism starting from the 
dynamic equations in the spherical coordinate system in Sec.~\ref{sec:contruct_spherical}.
The resulting action functional is given by the sum of the terms in Eqs.~(\ref{eq:action-det-sph-L2}), 
(\ref{eq:action-diss-sph-L}) and (\ref{eq:esferica-jac}).

We introduce the usual coordinates $M_\rms$, $\theta$ and $\phi$ 
where $M_\rms$ is the radial component, 
$\theta$ the polar angle, and $\phi$ the azimuthal angle (see App.~\ref{app:conventions} for more details on 
the conventions used). 
The vector $\mathbf{M}$ defines the usual local basis ($\mathbf{e}_{M_\rms}, \mathbf{e}_\theta, \mathbf{e}_\phi$) 
with 
\begin{eqnarray}
\mathbf{M}(M_\rms,\theta,\phi) &\equiv&
M_\rms\, \mathbf{e}_{M_\rms}(\theta,\phi) 
\nonumber\\
&=& 
M_\rms 
\left( 
\sin\theta \cos\phi \ {\mathbf e}_x + \sin \theta \cos \phi \ {\mathbf e}_y + \cos \theta \ {\mathbf e}_z 
\right)
\; . 
\end{eqnarray}
Here and after, Greek indices such as $\mu$ or $\nu$ label the spherical coordinates $M_\rms, \ \theta, \ \phi$, and 
Latin indices continue to label the Cartesian coordinates $x, \ y, \ z$. We collect the spherical coordinates in 
a vector $\Omega_\mu$.
The rotation matrix is called $R$ and we give its explicit form in App.~\ref{app:conventions}.

\subsection{$\alpha$-covariant sLLG equation}
\label{sec:sph_LLG_cov}

Similarly to what was done in the Cartesian coordinate system, the Stra\-to\-no\-vich sLLG equation in spherical coordinates should be modified 
 to work in a generic $\alpha$-prescription, while maintaining the physics unchanged.
We wish to find the equation satisfied by the spherical coordinates $M_\rms, \theta, \phi$
knowing that the Cartesian components of the magnetization vector ${\mathbf M}$ satisfy the sLLG given 
 in Eq.~(\ref{eq.LLG-covariant}) in the Landau formulation,
\begin{align}
 \mathbf{Eq}_L[\mathbf{M},\mathbf{h}] \equiv {\rm D}^{(\alpha)}_t \mathbf{M} - g \left( \mathbf{H}_{\eff} + \mathbf{H} \right) = {\mathbf 0}
 \label{eq:LLG-recall}
 \, ,
\end{align}
and that the Cartesian chain rule in Eq.~(\ref{eq.chainrule}) applied to our problem  reads
\begin{align} \label{eq:cr11}
\rmd_t & =  {\rmd_t} M_i  \ \partial_i + 
 \frac{  D(1-2\alpha) \gamma_0^2}{1+\eta^2\gamma_0^2}  \left(M_\rms^2 \delta_{ij} - M_i  M_j \right) \, \partial_i \partial_j \; .
\end{align}

We start by re-writing the chain rule in spherical coordinates. 
We first work out the first term in Eq.~(\ref{eq:cr11}) as
\begin{align}
 {\rmd_t} M_i  \ \partial_i  & =
    -\frac{  2D(1-2\alpha) \gamma_0^2}{1+\eta^2\gamma_0^2} M_\rms \partial_{M_\rms}
 \nonumber\\
 &
 \;\;\;\;\;
 + \frac{1}{M_\rms}
 \left[
 g_{\theta \nu_\perp} ({H_{\rm eff}}_{\nu_\perp}+H_{\nu_\perp})  \ \partial_\theta 
 + \frac{1}{\sin\theta} g_{\phi \nu_\perp} ({H_{\rm eff}}_{\nu_\perp}+H_{\nu_\perp}) \  \partial_\phi
 \right]
 \nonumber
\end{align}
(see App.~\ref{app:metric}).
The relevant elements of $g_{\mu\nu}$ are
\begin{eqnarray}
g_{\theta\theta} = g_{\phi\phi} = \eta\gamma_0 g_{\theta\phi} = - \eta\gamma_0 g_{\phi\theta} =  \frac{\eta\gamma_0^2 M_\rms}{1+\eta^2\gamma_0^2}
\label{eq:g-elements-spherical}
\end{eqnarray}
while $g_{M_\rms\mu}=g_{\mu M_\rms}=0$.
In order to treat the second term of Eq.~(\ref{eq:cr11}) we notice that
\begin{align}
 (M_\rms^2 \delta_{ij}  - M_i M_j  )  \partial_i \partial_j
 &=  M_\rms^2 \nabla^2 -  M_i M_j  \,  \partial_i \partial_j   \;, \nonumber
\end{align}
where $\nabla^2$ is the Laplacian operator that in spherical coordinates reads
\begin{align}
 \nabla^2  = \frac{1}{M^2_\rms} \left( M_\rms^2 \partial^2_{M_\rms} + 2 M_\rms \partial_{M_\rms}    +  \cot\theta\partial_\theta +  \partial^2_\theta + \frac{1}{\sin^2\theta}\partial^2_\phi \right) \;.
\end{align}
We also have
$  M_i M_j  \,  \partial_i \partial_j    
  =  M^2_\rms \ \partial_{M_\rms}^2  
  $.
Therefore, the second term in Eq.~(\ref{eq:cr11}) becomes
\begin{align}
  \frac{  D(1-2\alpha) \gamma_0^2}{1+\eta^2\gamma_0^2} M_\rms^2 \left(  \nabla^2 - \partial_{M_\rms}^2 \right)\;.
\end{align}
Altogether we have
\begin{align} \label{eq:cr22-maintext}
 \rmd_t  & =  
 \frac{1}{M_\rms}
 \left[ 
 g_{\theta \nu_\perp} H_{\nu_\perp} \partial_\theta 
 + \frac{1}{\sin\theta} g_{\phi \nu_\perp} ({H_{\rm eff}}_{\nu_\perp}+H_{\nu_\perp}) \ \partial_\phi
 \right]
 \nonumber\\
 &
 \;\;\;\;\;
 + \frac{  D(1-2\alpha) \gamma_0^2}{1+\eta^2\gamma_0^2}
 \left( \cot\theta\partial_\theta +  \partial^2_\theta + \frac{1}{\sin^2\theta}\partial^2_\phi \right)\;.
\end{align}
We now apply the differential operator (\ref{eq:cr22-maintext}) to $M_\rms$, $\theta$ and $\phi$, respectively, to  obtain
the equations of motion in spherical coordinates
\begin{align}
\rmd_t M_\rms &= 0\;, \\
 \rmd_t \theta  &= \frac{1}{M_\rms} g_{\theta \nu_\perp} ({H_{\rm eff}}_{\nu_\perp}+H_{\nu_\perp})  + \frac{  D(1-2\alpha) \gamma_0^2}{1+\eta^2\gamma_0^2} \cot\theta \;,
  \\
\rmd_t \phi &= \frac{1}{M_\rms \sin\theta} g_{\phi \nu_\perp} ({H_{\rm eff}}_{\nu_\perp}+H_{\nu_\perp}) \; ,
\end{align}
and we use these identities in Eq.~(\ref{eq:cr22-maintext}) to re-write the time-differential operator in 
a form that is explicitly independent of the external and random fields
\begin{align}
\rmd_t    &=   \  \rmd_t \Omega_{\mu_\perp} \  \partial_{\Omega_{\mu_\perp}}
  + \ \frac{  D(1-2\alpha) \gamma_0^2}{1+\eta^2\gamma_0^2}\left(  \partial^2_\theta + \frac{1}{\sin^2\theta} \ \partial^2_\phi \right)  \;.
\end{align}
This is the chain-rule in spherical coordinates.

Introducing the covariant derivatives
\begin{align}
{\rm D}_t^{(\alpha)} M_\rms &\equiv \rmd_t M_\rms \; , 
\label{eq:covDMs}\\
{\rm D}^{(\alpha)}_t(\theta) &\equiv \rmd_t \theta-\frac{D(1-2\alpha)\gamma_0^2}{(1+\eta^2\gamma_0^2)} \cot\theta \; , \label{eq:covDtheta} \\
{\rm D}^{(\alpha)}_t(\phi) &\equiv \rmd_t \phi \; ,
\label{eq:covDphi}
\end{align}
we now recast Eqs.~(\ref{eq:covDMs})-(\ref{eq:covDphi}) as
\begin{align}
\mbox{Eq}^{\rm sph}_{L,M_\rms}[M_\rms,\theta,\phi] &\equiv  {\rm D}_t^{(\alpha)} M_\rms = 0 \; ,
\label{eq:ralpha} \\
\mbox{Eq}^{\rm sph}_{L,\theta}[M_\rms,\theta,\phi] &\equiv  M_\rms \ {\rm D}^{(\alpha)}_t(\theta) - g_{\theta\nu_\perp} ({H_{\rm eff}}_{\nu_\perp}+H_{\nu_\perp})  = 0 \; ,
 \label{eq:thetaalpha} \\
\mbox{Eq}^{\rm sph}_{L,\phi}[M_\rms,\theta,\phi] &\equiv  M_\rms \sin\theta\,  {\rm D}^{(\alpha)}_t(\phi) - g_{\phi\nu_\perp}  ({H_{\rm eff}}_{\nu_\perp}+H_{\nu_\perp}) =0\;.
 \label{eq:phialpha} 
\end{align}
Equation~(\ref{eq:ralpha}) encodes the conservation of the modulus.
Using the explicit form of $g_{\mu\nu}$ given in Eq.~(\ref{eq:g-elements-spherical}), Eqs.~(\ref{eq:thetaalpha}) and (\ref{eq:phialpha}) become
\begin{eqnarray}
&& {\rm D}_t^{(\alpha)}(\theta) -  
\frac{\gamma_0}{1+\eta^2\gamma_0^2} 
\left[ H_{{\rm eff},\phi} + H_\phi + \eta\gamma_0 ( H_{{\rm eff},\theta} + H_\theta) \right]  
= 0
\;, 
\label{eq:spherical-Landau1}
\\
&&
\sin\theta \ {\rm D}_t^{(\alpha)}(\phi) - \frac{\gamma_0}{1+\eta^2\gamma_0^2} 
\left[ 
\eta\gamma_0 ( H_{{\rm eff},\phi} + H_\phi ) 
- ( H_{{\rm eff},\theta} + H_\theta) 
\right] 
= 0
\; . 
\qquad
\label{eq:spherical-Landau2}
\end{eqnarray}

Had we started from the Gilbert formulation of the sLLG equation in Cartesian coordinates, we would have naturally obtained
\begin{align}
{\rm D}^{(\alpha)}_t(\theta) + \eta\gamma_0 \ \sin \theta \ {\rm D}^{(\alpha)}_t(\phi) -  \gamma_0 \left( H_{\eff, \phi} + H_\phi \right) = 0 \; ,
 \label{eq:thetaalpha-Gilbert} \\
-  \sin\theta\,  {\rm D}^{(\alpha)}_t(\phi) + \eta\gamma_0 \ 
 {\rm D}^{(\alpha)}_t(\theta) -\gamma_0  \left( H_{\eff, \theta} + H_\theta \right) =0 \;.
  \label{eq:phialpha-Gilbert} 
\end{align}
In this form, the random field $H_\theta, H_\phi$ may be  erroneously interpreted as being additive,
 and that all discretization prescriptions are equivalent in spherical coordinates. This is not the case as the 
 time-derivative of $\phi$ in Eqs.~(\ref{eq:phialpha}) and (\ref{eq:phialpha-Gilbert}) are multiplied by a function of $\theta$.
 Moreover, we will see in Sec.~\ref{sec:sph_random_field} that, in the local coordinate system, the random field 
 has a non-trivial distribution that depends on the discretization.

Summarizing, we have shown how to write the sLLG equation in spherical coordinates in a generic $\alpha$-prescription.
For each prescription the stochastic equations are {\em different} and one can simply encode the dependence on $\alpha$ by 
introducing $\alpha$-covariant time-derivatives. When treated with 
the correct rules of stochastic calculus, all equations yield the same physical results.

An interesting observation is that in the case of a planar ferromagnet, \textit{i.e.} when the magnetization is bound to live on the plane (moving on a circle), 
the equations no longer have any explicit dependence on $\alpha$.
In this case, one can project the equations on the $x,y$ plane by setting $\theta=\pi/2$ and the $\alpha$-covariant time-derivatives 
reduce to the usual time-derivatives. This means that the sLLG equation is the same for all discretization schemes. 
We will see in Sec.~\ref{sec:sph_random_field} that in the case $\theta=\pi/2$, the noise in polar coordinate is a usual Gaussian white noise with vanishing mean.
In other words, the stochastic evolution in two dimensions is driven by an  effective {\em additive} noise, despite its original multiplicative
character. However, this property only holds in two dimensions and it is not true in general.

\subsection{Random field statistics}\label{sec:sph_random_field}

The probability distribution function of the random noise in the Cartesian coordinate system 
is Gaussian with zero average. This statistics does not directly translate into another coordinate
system. The distribution in the transformed system has to be carefully studied if one wishes 
to use the rotated components of the random field. As we found some misconceptions
in the literature regarding this point, in this
Subsection we derive this distribution in spherical coordinates. The reader who is just interested in 
the generating functional construction can jump over this Subsection and go directly to Sec.~\ref{sec:contruct_spherical}.

The equations of motion do not involve the radial component of the random field, $H_{M_\rms}$.
We are then naturally interested in deriving the probability distributions of the orthoradial components, $P^{\rm sph}_{\rm n}[H_\theta,H_\phi]$.
Let us start with the statistics of the random field in the Cartesian basis.
The probability distribution of histories for such an isotropic Gaussian white noise is given by
\begin{equation}
 P_\mathrm{n}[H_x,H_y,H_z] \propto \exp\left\{-\frac{1}{4D}
 \int{} \ud{t}  \left[ H_x(t)^2+  H_y(t)^2 +  H_z(t)^2 \right]  \right\}\;.
\end{equation}
The rotation to the spherical coordinate system, 
\begin{align}
 P^{\rm sph}_{\rm n}[H_{M_\rms}, H_\theta, H_\phi] = |\mathcal{J}^{\rm rot}| \ P_\mathrm{n}[R^{-1}_{x\mu} H_\mu,R^{-1}_{y\mu} H_\mu,R^{-1}_{z\mu} H_\mu]\;,
\end{align}
with 
\begin{equation}
R^{-1}_{i\mu}(\mathbf{M}) =  R^{-1}_{i \mu}(\theta,\phi) =
\!\!
\begin{array}{cc}
\left[
\begin{array}{ccc}
\sin\theta\cos\phi &  \cos\theta\cos\phi & -\sin\phi \\
\sin\theta\sin\phi &  \cos\theta\sin\phi & \cos\phi \\
\cos\theta &  -\sin\theta & 0
\end{array}
\right]
\; ,
\end{array}
\label{eq:inverse-rotation-matrix0}
\end{equation}
involves the Jacobian
\begin{eqnarray}
\mathcal{J}^{\rm rot}
\equiv
\mathrm{det}_{i\nu,tt'} 
\frac{\delta R^{-1}_{i\mu}(t) H_\mu(t)}{\delta H_\nu(t')} 
\; . 
\label{eq:Jacobian-rot}
\end{eqnarray}
After a series of transformations detailed in App.~\ref{app:rotation-noise} we set the calculation of the 
determinant in a form that allows us to use the identity (\ref{eq:identity})
with a causal $C_{\mu\nu}(w,v)$. In the present case, the noise dependence in the operator $C$ 
requires to keep the second-order contribution in the expansion, but all higher order terms vanish~\cite{Arnold2000,Lubensky2007}. 
We therefore use Eq.~(\ref{eq:identity-random-field}).
After a lengthy computation detailed in App.~\ref{app:rotation-noise} the Jacobian ${\cal J}^{\rm rot}$ is found to be
\begin{align}
\ln\mathcal{J}^{\rm rot} 
&= 
 \int\Ud{t} R_{\mu j}(t)  \frac{\partial R^{-1}_{j\rho}(t)}{\partial \Omega_{\nu_\perp}}  \frac{\delta \Omega_{\nu_\perp}(t)}{\delta H_{\mu}(t)} H_\rho(t) 
\nonumber\\
 & -\frac12 \iint\udd{t}{t'} 
 R_{\mu_\perp j}(t) \frac{\partial R^{-1}_{j\rho}(t)}{\partial \Omega_{\tau_\perp}} \frac{\delta \Omega_{\tau_\perp}(t)}{\delta H_{\nu_\perp}(t')} 
  R_{\nu_\perp k}(t') \frac{\partial R^{-1}_{k\sigma}(t')}{\partial \Omega_{\kappa_\perp}} \frac{\delta \Omega_{\kappa_\perp}(t')}{\delta H_{\mu_\perp}(t)} 
 \nonumber\\
 & \qquad\qquad\qquad 
 \times 
 H_\rho(t) H_\sigma(t')
 \; . 
 \nonumber
\end{align}
The ``responses" $ {\delta \Omega_{\tau_\perp}(t)}/{\delta H_{\nu_\perp}(t')}$ are causal, making 
the integrand in the last term vanish for all $t'\neq t$. However, as it involves two random field factors (which are delta correlated) it may still yield a non-trivial contribution at $t=t'$.
We will see that in cases in which the ``equal-time responses''
$ {\delta \Omega_{\tau_\perp}(t)}/{\delta H_{\nu_\perp}(t)}$
vanish (as in, {\it e.g.}, non-Markovian processes) the Jacobian turns out to be trivial and equals one, 
$ \mathcal{J}^{\rm rot} = 1$. This is the case for the It\^o convention.
However,  the sLLGs in spherical coordinates and generic discretization prescription
yield finite and non-vanishing equal-time responses and, hence, a non-trivial ${\mathcal J}^{\rm rot}$.

Using a more  compact notation, the probability distribution function reads
\begin{eqnarray*}
&&
\ln P^{\rm sph}_{\rm n}[H_\mu] =  -\frac{1}{4D} \int {\rm d} t \ H^2_\mu(t) + \int\ud{t}  L_\rho(t) H_\rho(t)
\nonumber\\
&&
\;\;\;\qquad - \frac{1}{2} \iint\udd{t}{t'}
Q_{\rho\sigma}(t,t') 
H_\rho(t) H_\sigma(t')
\; ,
\end{eqnarray*}
with 
\begin{eqnarray}
L_\rho(t) &\equiv&  R_{\mu j}(t)   \frac{\partial R^{-1}_{j\rho}(t)}{\partial \Omega_{\tau_\perp}}  \frac{\delta \Omega_{\tau_\perp}(t)}{\delta H_{\mu}(t)} \; ,  \\
Q_{\rho\sigma}(t,t') &\equiv& R_{\mu j}(t) \frac{\partial R^{-1}_{j\rho}(t)}{\partial \Omega_{\tau_\perp}} \frac{\delta \Omega_{\tau_\perp}(t)}{\delta H_{\nu}(t')} R_{\nu k}(t') \frac{\partial R^{-1}_{k\sigma}(t')}{\partial \Omega_{\kappa_\perp}} \frac{\delta \Omega_{\kappa_\perp}(t')}{\delta H_{\mu}(t)}
\; . 
\end{eqnarray}
  
The responses can be computed by first formally recasting the solutions of the equations of 
motion~(\ref{eq:thetaalpha-Gilbert}) and (\ref{eq:phialpha-Gilbert}) into
\begin{align}
 \theta(t) &= \theta_0 + \int_{t_0}^t\Ud{t'} \!\! \ldots +  \frac{\gamma_0}{1+\eta^2\gamma_0^2} \int_{t_0}^t\Ud{t'} \left[ H_\phi(t') + \eta\gamma_0 H_\theta(t') \right]\,, \label{eq:integre_theta} \\
 \phi(t) &= \phi_0 + \int_{t_0}^t\Ud{t'} \!\! \ldots + \frac{\gamma_0}{1+\eta^2\gamma_0^2} \int_{t_0}^t\Ud{t'} \frac{1}{\sin\theta(t')}  \left[ \eta\gamma_0 H_\phi(t') -  H_\theta(t') \right]\,, \label{eq:integre_phi}
\end{align}
where we only expressed explicitly the kernels involving the random fields. The argument exposed in  
App.~\ref{app:rotation-noise} allows one to recast the last term in a form in which the product of 
random fields $H_\rho(t) H_\sigma(t')$ is replaced by its average, $2D \delta_{\rho\sigma} \delta(t-t')$, contracting the indices 
of the factors $Q$ and cutting one of the time integrals. In short, one only needs  
the equal-time responses of the polar coordinates with respect to variations of the random fields. These read, using 
$\int_{t_0}^t\ud{t'} \delta(t'-t) = \Theta(0) = \alpha$,
\begin{align}
 \frac{\delta \theta(t)}{\delta H_\theta(t)} &=  \frac{\eta\gamma_0^2}{1+\eta^2\gamma_0^2} \int_{t_0}^t\ud{t'} \delta(t'-t) = \frac{\alpha\eta\gamma_0^2}{1+\eta^2\gamma_0^2}  \,, 
\label{eq:equal-time1} \\
 \frac{\delta \theta(t)}{\delta H_\phi(t)}  & \
=
  \frac{\gamma_0}{1+\eta^2\gamma_0^2} 
 \int_{t_0}^t {\rm d} t' \ \delta(t-t') =  \frac{\alpha\gamma_0}{1+\eta^2\gamma_0^2} 
\; , 
 \\
\frac{\delta \phi(t)}{\delta H_\theta(t)} &= -\frac{\gamma_0}{1+\eta^2\gamma_0^2} \int_{t_0}^t\ud{t'} \frac{1}{\sin\theta(t')}  \delta(t-t') = - \frac{1}{\sin\theta(t)} \frac{\alpha\gamma_0}{1+\eta^2\gamma_0^2} \,,  \\
\frac{\delta \phi(t)}{\delta H_\phi(t)} &=  \frac{\eta\gamma_0^2}{1+\eta^2\gamma_0^2} \int_{t_0}^t\ud{t'} \frac{1}{\sin\theta(t')}  \delta(t-t') = 
\frac{1}{\sin\theta(t)}  \frac{\alpha\eta\gamma_0^2}{1+\eta^2\gamma_0^2} \,.
\label{eq:equal-time4}
\end{align}
The  first terms in Eqs.~(\ref{eq:integre_theta}) and (\ref{eq:integre_phi}) 
give vanishing contributions since their integrands are finite at all times.

Using these results one calculates $L_\rho$ and $Q_{\rho\rho}$:
\begin{eqnarray}
L_\rho(t) &=& \frac{\alpha\gamma_0}{1+\eta^2\gamma_0^2} 
\left[ 2\eta\gamma_0 \delta_{\rho M_\rms} + \eta\gamma_0 \cot\theta \delta_{\rho\theta} + \cot\theta \delta_{\rho\phi}  \right]
\; , \\ 
Q_{\rho\rho}(t,t) &=& \frac{\alpha^2\gamma_0^2}{1+\eta^2\gamma_0^2} \left(1+\frac{1}{\sin^2\theta} \right)
\; , 
\end{eqnarray}
and, after another  lengthy calculation detailed in App.~\ref{app:rotation-noise}, 
we obtain the following expression for $P_{\rm n}^{\rm sph}$ in terms of the random field components 
and the magnetization polar angles:
\begin{align}
P_{\rm n}^{\rm sph}[H_\theta,H_\phi] 
\propto & 
\exp  \int\ud{t}  
\!\!
\left\{ -\frac{1}{4D} H^2_{\mu_\perp} \! (t) 
+ \frac{\alpha\gamma_0}{1+\eta^2\gamma_0^2} \cot\theta(t) \ [\eta\gamma_0 H_\theta(t) + H_\phi(t) ]
\right.
\nonumber\\
& 
\left.
\qquad\qquad\qquad
- \frac{\alpha^2\gamma_0^2 D}{1+\eta^2 \gamma_0^2} \cot^2\theta(t)
\right\}
\; .
\end{align}
It\^o calculus provides a special case in which a conventional Gaussian distribution is 
recovered.
We stress that there is another special case in which this distribution boils down to a 
standard Gaussian distribution (with zero mean and delta correlations) for all discretization 
prescriptions: the case in which  the magnetization is constrained to rotate on the plane $\theta=\pi/2$.

\subsection{Fokker-Planck approach}
Following steps similar to the ones in Sec.~\ref{sec:Fokker-Planck} and in \cite{Fuller63}, now for the $\alpha$-scheme 
spherical equations of motion, one finds the $\alpha$-generic Fokker-Planck equation
\begin{eqnarray}
\partial_t P(\theta, \phi; t) \! \! & \! = \! & \! \!
- \partial_\theta 
\left\{ [f_\theta + D(2\alpha-1) \frac{\gamma_0^2}{1+\eta^2\gamma_0^2}  \cot\theta 
\right.
\nonumber\\
&& 
\left.
\qquad\qquad
+ D \frac{\gamma_0^2}{1+\eta^2\gamma_0^2} \cot\theta ] P(\theta, \phi;t) 
\right\}
\nonumber\\
&&
-\partial_\phi[f_\phi P(\theta, \phi; t)]
\nonumber\\
&&
-\frac{D\gamma_0^2}{1+\eta^2\gamma_0^2} 
\left\{
\partial_\theta^2  P(\theta,\phi;t)
+\partial_\phi^2 [ \frac{1}{\sin^2\theta} P(\theta,\phi;t) ] 
\right\}
\end{eqnarray}
with 
\begin{eqnarray}
f_\theta &=& \frac{D(1-2\alpha)\gamma_0^2}{1+\eta^2\gamma_0^2} \ \cot \theta + \frac{\gamma_0}{1+\eta^2\gamma_0^2}
 (H_{{\rm eff}, \phi} + \eta\gamma_0 H_{{\rm eff},\theta} )
 \; , 
\nonumber\\
f_\phi &=& 
\frac{\gamma_0}{1+\eta^2\gamma_0^2} \frac{1}{\sin\theta} (\eta\gamma_0 H_{{\rm eff}, \phi} -  H_{{\rm eff},\theta} )
\; . 
\end{eqnarray}
As in the Cartesian case one finds that the drift term in $f_\theta$ cancels the following term and all 
explicit 
$\alpha$ dependence disappears form the Fokker-Planck equation.  One can check that in the 
conservative case
\begin{equation}
H_{{\rm eff},\theta} = - (M_\rms \mu_0^{-1}) \ \partial_\theta U
\qquad\qquad
H_{{\rm eff},\phi} = - (M_\rms \mu_0^{-1}) \ \frac{1}{\sin\theta} \partial_\phi U
\end{equation}
the stationary probability density
\begin{equation}
P_{\rm eq}(M_\rms, \theta,\phi) = N \sin\theta \ e^{-\beta V U(\theta,\phi)} 
\label{eq:equil-Pi-sph}
\end{equation}
with $N$ a normalisation constant
is a solution to the Fokker-Planck equation. 
as long as $D$ is given by Eq.~(\ref{eq:D}).

\subsection{Landau generating functional}
\label{sec:contruct_spherical}

The purpose of this Subsection is to derive the generating functional in the spherical coordinate system.
The steps performed in this Section are very similar to the ones performed in Sec.~\ref{sec:functional} when working with 
Cartesian coordinates.

Given an initial condition ${M_\rms}_0$, $\theta_0$ and $\phi_0$ that we collect in the vector notation ${\mathbf \Omega}_0 = \mathbf{\Omega}(t_0)$, 
and a particular realization of the Gaussian and zero-mean 
random variables $[{\mathbf H}]$ in the Cartesian coordinate system,  
there is a unique trajectory of the variables $[M_\rms]$, $[\theta]$ and $[\phi]$, collected in $[{\mathbf \Omega}]$,
that obeys the equations of motion. 
The generating functional  is defined as 
\begin{align}
 \mathcal{Z}[\boldsymbol{\lambda}] = \langle \ 
 \exp \int \Ud{t} \boldsymbol{\lambda}(t) \cdot \mathbf{\Omega}_{\mathbf H}(t) \ \rangle
  \; ,
\end{align}
where $ \langle \cdots \rangle $ denotes the average over initial conditions and random field realizations.
$\boldsymbol{\lambda}$ is a source that couples linearly to the fluctuating magnetization configuration 
$\mathbf{\Omega}_{\mathbf H}(t)$.

Similarly to the Cartesian case, we construct the MSRJD representation of the generating functional $\mathcal{Z}[\boldsymbol{\lambda}]$
by imposing the equation of motion with a functional delta-function
 \begin{eqnarray}
&&
{\cal Z}[\boldsymbol{\lambda}] = \int {\cal D}[\mathbf{H}] \ P_{\rm n}[\mathbf{H}] \
\int {\cal D}[\mathbf{\Omega}] \
P_{\rm i}[\mathbf{\Omega}(t_0), {\mathbf {H}}_{\rm eff}(t_0)] 
\nonumber\\
&& 
\qquad\quad
  \ \times {\prod_{n=1}^N [M_\rms^2\sin\theta_{n}]^{-1}} 
\delta\Big{[} \mbox{\textbf{Eq}}^{\rm sph}[\mathbf{\Omega},\mathbf{H}] \Big{]}
\nonumber\\
&&
\qquad\qquad
\; \; \times \
\ |{\cal J}^{\rm sph}[\mathbf{\Omega},\mathbf{H}]| 
\ \exp  \int \ud{t} {\boldsymbol{\lambda}}(t) \cdot \mathbf{\Omega}(t) 
\;.
\label{eq:generatingZ24}
\end{eqnarray}
$P_{\rm n}[\mathbf{H}]$ is the probability distribution of the random field in Cartesian coordinates, 
that we still take to be Gaussian with zero mean, delta correlated and variance $2D$.
For the moment we leave the initial probability density $P_{\rm i}$ general. A particular case 
is the one in which it is given by the equilibrium weight (\ref{eq:equil-Pi-sph}).
The measure over the spherical coordinates,  ${\cal D}[\mathbf{\Omega}]$,  is defined in App.~\ref{app:conventions} and includes a summation over the initial conditions at time $t_0$. 
The geometric factor $\prod_{n=1}^N |M_\rms^2 \sin \theta_n |^{-1}$ accompanies the $\delta$ function in the 
spherical coordinate system, see also App.~\ref{app:deltafunction}.
The Jacobian ${\cal J}^{\rm sph}[\mathbf{\Omega},\mathbf{H}]$ is
\begin{eqnarray}\label{eq:defjac}
\mathcal{J}^{\rm sph}[\mathbf{\Omega},\mathbf{H}] \equiv
\det_{\mu\nu;uv}
\begin{array}{c}
\displaystyle
\frac{\delta {\rm{Eq}}^{\rm sph}_{\mu}[\mathbf{\Omega},\mathbf{H}](u)}{\delta \Omega_{\nu}(v)}
\end{array}
\end{eqnarray}	
with the coordinate indices $\mu,\nu=M_\rms,\theta,\phi$ and the times $u,v$.

At this point we have the freedom to write the equation of motion 
in the Landau or Gilbert formulation. The advantage of the former lies in the fact that the time derivatives are well 
separated from the other terms, thus simplifying the analysis. We choose to use a modified Landau formulation that we compactly write as follows:
\begin{align}
&
{\rm Eq}^{\rm sph}_{L,M_\rms}  = \rmd_t M_\rms = 0
\; , 
\nonumber\\
& 
{\rm Eq}^{\rm sph}_{L,\theta} = {\rm D}_t^{(\alpha)}(\theta) - 
\frac{\gamma_0}{1+\eta^2\gamma_0^2} \left[   H_{{\rm eff},\phi} + H_\phi + \eta\gamma_0  (H_{{\rm eff}, \theta} +H_\theta)\right]   = 0 
\; , 
\nonumber\\
& 
{\rm Eq}^{\rm sph}_{L,\phi} = {\rm D}_t^{(\alpha)}(\phi) 
- \frac{\gamma_0}{1+\eta^2\gamma_0^2} \frac{1}{\sin\theta}   \left[\eta\gamma_0 (H_{{\rm eff}, \phi}+H_\phi ) -   (H_{{\rm eff},\theta} +H_\theta ) \right] = 0 
\; .
\nonumber
\end{align}
This form is convenient since, as the derivatives are separated from the rest,
it is relatively simple to compute the  Jacobian (as opposed to what has to be done in 
the Gilbert formulation  that we develop in App.~\ref{sec:construct_spherical_Gilbert}).

The operator in the determinant can be worked out explicitly as explained in App.~\ref{app:MSRDJ-Jacobian-Gus}
\begin{eqnarray}
&& 
{\cal J}_L^{\rm sph} 
= 
\exp 
\left\{
\frac{\alpha\gamma_0}{1+\eta^2\gamma_0^2} \int\Ud{t} 
\left[ 
\frac{D(1-\alpha)\gamma_0}{\sin^2\theta} 
\right.
\right.
\nonumber\\
&& 
\left.
\left.
\qquad\qquad
- \partial_\theta 
\left[
H_{{\rm eff},\phi} + H_{\phi} + \eta\gamma_0   (H_{{\rm eff},\theta} + H_\theta) 
\right]  
\right.
\right.
\nonumber\\
&&
\left.
\left.
\qquad\qquad
+ \frac{1}{\sin\theta} \partial_\phi 
\left[
H_{{\rm eff},\theta} + H_\theta 
-
\eta\gamma_0 (H_{{\rm eff},\phi} + H_\phi)
\right] 
\right]
\right\}
\; . 
\label{eq:Jacobian-Landau-spherical}
\end{eqnarray}
As found in the Cartesian calculation, it does not depend on the parallel component
of the  field, that in spherical coordinates means that ${\cal J}_L^{\rm sph}$ is independent of $H_{{\rm eff},M_\rms}+H_{M_\rms}$.

We next introduce an adimensional  Lagrange
multiplier $[\rmi\hat{\mathbf{\Omega}}]$ to exponentiate the functional delta:
\begin{displaymath}
\int \uD{[\rmi\hat{\mathbf \Omega}]} \exp \left\{ - \int \! \ud{t} 
\left(
\rmi\hat\Omega_{M_\rms} \mbox{{Eq}}^{\rm sph}_{M_\rms}[{\mathbf \Omega}]
+
\rmi\hat\Omega_\theta \mbox{{Eq}}^{\rm sph}_\theta[{\mathbf \Omega}, {\mathbf H}]
+
\rmi\hat\Omega_\phi \mbox{{Eq}}^{\rm sph}_\phi[{\mathbf \Omega}, {\mathbf H}]
\right)
\right\}
\; .  
\end{displaymath}
We identify all the terms 
in the integrand of the exponent that involve the random field ${\mathbf H}$:
\begin{eqnarray*}
&& 
-\frac{1}{4D} H_i^2 
+ \frac{\gamma_0}{1+\eta^2\gamma_0^2}
\left( 
\rmi\hat\Omega_\phi \frac{1}{\sin\theta} \left( \eta\gamma_0 R_{\phi i}  - R_{\theta i} \right)
+
\rmi\hat\Omega_\theta \left( R_{\phi i} + \eta\gamma_0 R_{\theta i} \right)
\right) H_i 
\nonumber\\
&& 
\qquad
+ \frac{\alpha\gamma_0}{1+\eta^2\gamma_0^2}
\left( -\partial_\theta R_{\phi i} 
- \eta\gamma_0 \partial_\theta R_{\theta i} 
-\frac{\eta\gamma_0}{\sin\theta} \partial_\phi R_{\phi i}  
+ \frac{1}{\sin\theta} \partial_\phi R_{\theta i}
\right) H_i\;.
\qquad
\end{eqnarray*}
The quadratic term in $H_i$ comes from its probability distribution. The first set of 
linear terms comes from imposing the equations of motion with the Dirac delta function.
The third group of terms comes from the Jacobian. Contrary to the Cartesian case, 
the latter will yield non-trivial contributions to the action.
After integration and a number of simplifications that use the explicit expression of the rotation matrix $R$
one finds that these terms give rise to
\begin{eqnarray*}
\frac{D \gamma_0^2}{1+\eta^2\gamma_0^2}  \left[
\frac{(\rmi\hat\Omega_\phi)^2}{\sin^2\theta} + (\rmi\hat\Omega_\theta)^2 
+ 2\alpha \rmi\hat\Omega_\theta \cot\theta
+ \frac{\alpha^2}{\sin^2\theta}
\right]
\end{eqnarray*}
(apart from an irrelevant additive constant). The last term is of the same form as the first term in ${\cal J}_L^{\rm sph}$
and we will combine them together when writing $\widetilde S^{\rm sph}_{{\rm jac}}$ below.

We now perform the change of fields
\begin{align}
 \rmi\hat\Omega_\phi \mapsto \sin\theta \, \rmi\hat\Omega_\phi \;,
\end{align}
which comes with a Jacobian
\begin{align}
\prod_{n=0}^{N-1} 
 |
 \sin\bar\theta_n
 |\;,  \label{eq:detedete4}
\end{align}
where we were careful to evaluate the factors on the intermediate points $\bar \theta_n \equiv \alpha \theta_{n+1} + (1-\alpha) \theta_n$. Notice indeed that the discretization matters here since there is no trivial continuous limit of this expression.
See also the discussion in Sect.~\ref{subsec:rules}. The product above can be re-writen as
\begin{align}
 \prod_{n=0}^{N-1} 
 |
 \sin\bar\theta_n
 |
 = 
 \rme^{(1-\alpha)  \ln \left| \frac{\sin\theta_0}{\sin\theta_N} \right| }
 \prod_{n=1}^{N} 
 |
 \sin\theta_n
 |
 \;, \label{eq:detedetee4}
\end{align}
where we used the development
\begin{align}
\sin\bar\theta_n 
=
\alpha \sin \theta_{n+1} + (1-\alpha) \sin \theta_n\;,
\end{align}
and the fact that we do not need to consider higher order terms because they vanish from Eq.~(\ref{eq:detedete4}) once the limit $\delta t \to 0$ is considered. The product $\prod_{n=1}^{N} 
 | \sin\theta_n |$ in Eq.~(\ref{eq:detedetee4}) cancels exactely the geometric one accompanying the delta functions in Eq.~(\ref{eq:generatingZ24}).

We now put all these results together to write the generating functional
\begin{equation}
 \mathcal{Z}[\boldsymbol{\lambda}] = 
 \int \uD[\boldsymbol{\Omega}] {\cal D}[\hat{\boldsymbol{\Omega}}] \ 
 \exp\left( S_L^{\rm sph}[\boldsymbol{\Omega},\hat{\boldsymbol{\Omega}}] + \int\ud{t}\boldsymbol{\lambda} \cdot \boldsymbol{\Omega} \right)\;,
\end{equation}
with the measure ${\cal D}[\boldsymbol{\Omega}] \equiv \lim\limits_{N\to\infty} \prod_{n=0}^N \rmd {M_\rms}_n \rmd \theta_n \rmd \phi_n \, {M_\rms}_n^2 |\sin\theta_n|$
(that includes the initial time $t_0$),
the full action
 \begin{equation}
 S_L^{\rm sph}= S^{\rm sph}_{L,{\rm det}} + \widetilde S^{\rm sph}_{L,{\rm diss}} + \widetilde S^{\rm sph}_{L,{\rm jac}} 
 \end{equation}
 and
 \begin{eqnarray}
 S^{\rm sph}_{L,{\rm det}}  \! & \! = \! & \!
 \ln P_{\rm i}[{\mathbf \Omega}_0, {\mathbf H}_{\rm eff}(t_0)] - 
  \int {\rm d}t \
  \left\{
  \rmi\hat\Omega_{M_\rms} {\rm d}_t M_\rms
 \right.
\nonumber\\
&& 
\;
+ \rmi\hat\Omega_\theta 
\left[ {\rm D}_t^{(\alpha)}(\theta) - \frac{\gamma_0}{1+\eta^2\gamma_0^2} H_{{\rm eff}, \phi}  \right]
\nonumber\\
&& 
\left.
\;
+ {\rmi\hat\Omega_\phi} \left[ \sin\theta {\rm D}_t^{(\alpha)}(\phi)  +  \frac{\gamma_0}{1+\eta^2\gamma_0^2} H_{{\rm eff}, \theta}  \right]
\right\} ,
\qquad
\label{eq:action-det-sph-L}
\\
\widetilde  S^{\rm sph}_{L,{\rm diss}} \! & \! = \! &   \!
 \frac{D \gamma_0^2}{1+\eta^2\gamma_0^2} 
 \int {\rm d}t 
 \left[ 
   {(\rmi\hat\Omega_\phi)^2} +  (\rmi\hat\Omega_\theta)^2 
 + 
 2\alpha \rmi\hat\Omega_\theta \cot\theta 
\right]
\nonumber\\
&&
\;\;
+ \frac{\eta\gamma^2_0}{1+\eta^2\gamma_0^2}
\int \rmd t  \left(\rmi \hat \Omega_\theta  H_{{\rm eff}, \theta} + \frac{\rmi \hat \Omega_\phi}{\sin\theta} H_{{\rm eff}, \phi}
\right)
,
\label{eq:action-diss-sph}
\end{eqnarray}
\begin{eqnarray}
\widetilde S^{\rm sph}_{L,{\rm jac}} \! & \! = \! &\!
(1-\alpha) \ln \left| \frac{\sin\theta_0}{\sin\theta_N} \right|   \nonumber \\
&&
+ \frac{\alpha\gamma_0}{1+\eta^2\gamma_0^2} \int\Ud{t} 
\left[ 
\frac{D\gamma_0}{\sin^2\theta} 
- \partial_\theta 
\left(
H_{{\rm eff},\phi}  + \eta\gamma_0   H_{{\rm eff},\theta}  
\right) 
\right.
\nonumber\\
&&
\left.
\qquad\qquad\qquad\qquad
+ \frac{1}{\sin\theta} \partial_\phi 
\left(
H_{{\rm eff},\theta}
-
\eta\gamma_0 H_{{\rm eff},\phi} 
\right) 
\right]
.
 \end{eqnarray} 

The expressions above can be modified to obtain a slightly more compact, and eventually more convenient, form. 
$\widetilde S_{L,{\rm diss}}^{\rm sph}$ includes a linear term in $\rmi \hat\Omega_\theta$ that can be replaced with the help of 
the identity
\begin{eqnarray}
&&
\int d\phi \ \exp\left[ (\rmi\sigma \phi)^2 - \rmi\phi b - \rmi\phi a  \right]
\nonumber\\
&& 
\qquad\qquad
=
\int d\phi \ \exp\left[ (\rmi\sigma \phi)^2 - \rmi\phi b - \frac{a^2}{4\sigma^2}  - \frac{ab}{2\sigma^2} \right]
\; . 
\label{eq:identity-Gus}
\end{eqnarray}
We apply it to the functional integration over $\rmi \hat \Omega_\theta$
by choosing 
\begin{eqnarray}
&&
\sigma^2 =
\frac{D\gamma_0^2}{1+\eta^2\gamma_0^2}
\; ,
\nonumber\\
&&
a=-\frac{2D\gamma_0^2\alpha}{1+\eta^2\gamma_0^2} \ \cot\theta
\; , 
\nonumber\\
&&
b = 
{\rm D}_t^{(\alpha)}(\theta) -
 \frac{\gamma_0}{1+\eta^2\gamma_0^2} 
(H_{{\rm eff},\phi}+\eta\gamma_0 H_{{\rm eff}, \theta})
\; . 
\nonumber
\end{eqnarray}
The integration generates the terms 
\begin{align}
-\frac{a^2}{4\sigma^2} - \frac{ab}{2\sigma^2} 
= &
-\frac{\alpha^2 D \gamma_0^2}{1+\eta^2\gamma_0^2} \  \cot^2\theta
\nonumber\\
&
+ \alpha\cot\theta \
[{\rm D}_t^{(\alpha)}(\theta) - \frac{\gamma_0}{1+\eta^2\gamma_0^2} (H_{{\rm eff},\phi}+\eta\gamma_0 H_{{\rm eff}, \theta})]
\; . 
\end{align}
We rewrite the full action as $S= S^{\rm sph}_{L,{\rm det}} +  S^{\rm sph}_{L,{\rm diss}} +  S^{\rm sph}_{\rm jac}$, with 
$S^{\rm sph}_{L,{\rm det}}$ given in Eq.~(\ref{eq:action-det-sph-L}) that we repeat here to ease the reading of the 
final result, 
 \begin{eqnarray}
  S^{\rm sph}_{L,{\rm det}}  \! & \! = \! & \!
 \ln P_{\rm i}[{\mathbf \Omega}_0, {\mathbf H}_{\rm eff}(t_0)] - 
  \int {\rm d}t \
  \left\{
  \rmi\hat\Omega_{M_\rms} {\rm d}_t M_\rms
 \right.
\nonumber\\
&& 
\;
+ \rmi\hat\Omega_\theta 
\left[ {\rm D}_t^{(\alpha)}(\theta) 
- \frac{\gamma_0}{1+\eta^2\gamma_0^2} H_{{\rm eff}, \phi}  \right]
\nonumber\\
&& 
\left.
\;
+ {\rmi\hat\Omega_\phi} \left[ \sin\theta {\rm D}_t^{(\alpha)}(\phi)  +   \frac{\gamma_0}{1+\eta^2\gamma_0^2}
H_{{\rm eff}, \theta})  \right]
\right\} ,
\qquad
\label{eq:action-det-sph-L2}
\\
 S^{\rm sph}_{L,{\rm diss}} \! & \! = \! &   \!
 \frac{\gamma_0}{1+\eta^2\gamma_0^2} 
 \int {\rm d}t 
 \left[ 
 D\gamma_0   \left( \rmi\hat\Omega_\phi  \right)^2 +  D\gamma_0 (\rmi\hat\Omega_\theta)^2 
\right.
\nonumber\\
&&
\left.
\;\;\;\;\qquad\qquad + \eta\gamma_0
 \left(\rmi \hat \Omega_\theta  H_{{\rm eff}, \theta} + {\rmi \hat \Omega_\phi} H_{{\rm eff}, \phi}
\right)
\right]
,
\label{eq:action-diss-sph-L}
\\
 S^{\rm sph}_{L,{\rm jac}} \! & \! = \! &\! (1-\alpha) \ln \left| \frac{\sin\theta_0}{\sin\theta_N} \right|  + \alpha \int \rmd t \ \cot\theta \ {\rm D}_t^{(\alpha)}(\theta)
\nonumber\\
&&
\;\; +
\frac{\alpha\gamma_0}{1+\eta^2\gamma_0^2} \int\Ud{t} 
\!
\left[ 
\frac{(1-\alpha)D\gamma_0}{\sin^2\theta} - \cot \theta (H_{{\rm eff},\phi} + \eta\gamma_0 H_{{\rm eff},\theta})
\right.
\nonumber\\
&&
\qquad\qquad\qquad\qquad
- \partial_\theta 
\left(
H_{{\rm eff},\phi}  + \eta\gamma_0   H_{{\rm eff},\theta}  
\right) 
\nonumber\\
&&
\left.
\qquad\qquad\qquad\qquad
+ \frac{1}{\sin\theta} \partial_\phi 
\left(
H_{{\rm eff},\theta}
-
\eta\gamma_0 H_{{\rm eff},\phi} 
\right) 
\right]
.
\label{eq:esferica-jac}
 \end{eqnarray} 
This completes the construction of the generating functional for the Landau formulation of the 
dynamics in the spherical coordinate system.

The construction of the Gilbert action reported in App.~\ref{sec:construct_spherical_Gilbert} leads to
\begin{eqnarray}
S^{\rm sph}_{G,{\rm det}}  \!\! & \!\! = \! \! & \!
 \ln P_{\rm i}[{\mathbf \Omega}_0, {\mathbf H}_{\rm eff}(t_0)] 
 - 
  \int {\rm d}t 
  \left\{ \rmi\hat\Omega_{M_\rms}  {\rm d}_t M_\rms
  + \rmi\hat\Omega_\theta 
[ {\rm D}_t^{(\alpha)}(\theta) - \gamma_0 H_{{\rm eff}, \phi}   
]
\right.
\nonumber\\
&& 
\left.
\!\! 
- \rmi\hat\Omega_\phi 
[ \sin\theta {\rm D}_t^{(\alpha)}(\phi)  + \gamma_0 H_{{\rm eff}, \theta}
]
\right\}
,
\label{eq:action-det-sph-Gilbert-maintext}\\
 S^{\rm sph}_{G,{\rm diss}} \!\! & \!\! = \!\! &   \!\!
 \int {\rm d}t 
 \left[ 
   D \gamma_0^2 (\rmi\hat\Omega_\phi)^2 + D \gamma_0^2 (\rmi\hat\Omega_\theta)^2 
   - \rmi\hat\Omega_\theta \eta\gamma_0 \sin\theta {\rm D}_t^{(\alpha)}(\phi)
\right.
\nonumber\\
&&
\left.
\qquad\quad 
- \rmi\hat\Omega_\phi  \eta\gamma_0 {\rm D}_t^{(\alpha)}(\theta) 
\right]
\;,
\label{eq:action-diss-sph-Gilbert-maintext}
\\
S^{\rm sph}_{G,{\rm jac}} \!\! & \!\! = \!\! &   \!
S^{\rm sph}_{L,{\rm jac}} 
\label{eq:action-jac-sph-Gilbert-maintext}
 \end{eqnarray} 
 and $S_G^{\rm sph} = S_{G, {\rm det}}^{\rm sph} +  S_{G, {\rm diss}}^{\rm sph} + S_{G, {\rm jac}}^{\rm sph}$.
One can easily check that one can go from the Landau to the Gilbert formalism and {\it vice versa} 
within the path-integral {\it via} a change of variables of the auxiliary fields, similarly to what we discussed 
in the Cartesian coordinate system around Eqs.~(\ref{eq:Trans_L2G}) and (\ref{eq:Trans_G2L}):
\begin{eqnarray}
\rmi\hat \Omega^L_\theta &=&  \rmi \hat \Omega^G_\theta + \eta\gamma_0  \ \rmi\hat\Omega^G_\phi
\; , 
\\ 
\rmi\hat\Omega^L_\phi &=& \eta\gamma_0 \ \rmi \hat\Omega^G_\theta - \rmi \hat\Omega^G_\phi
\; ,
\end{eqnarray}
with inverse 
\begin{eqnarray}
\rmi \hat \Omega_\theta^G &=& \frac{1}{1+\eta^2\gamma_0^2} \left[ \rmi\hat \Omega_\theta^L + \eta\gamma_0 \ {\rmi \hat \Omega_\phi^L} \right]
\; , 
\nonumber\\ 
\rmi \hat \Omega_\phi^G &=& \frac{1}{1+\eta^2\gamma_0^2} \left[\eta\gamma_0 \ \rmi \hat \Omega_\theta^L - {\rmi \hat \Omega_\phi^L} \right]
\; . 
\end{eqnarray}

\section{Conclusions}
\label{sec:recap}

In this manuscript we revisited the stochastic approach to the dynamics of a magnetic moment under
the effect of thermal noise, dissipation, magnetic field of potential origin and, also, non-potential forces such 
as spin-torque terms. We used the stochastic Landau-Lifshitz-Gilbert (sLLG) equation as a phenomenological 
description of the dynamics and we constructed a functional generating functional for physical observables.

We found rather confusing statements on the influence (or not) of the discretization 
scheme used to define the stochastic dynamics in discrete time in the literature~\cite{Berkov2002}. Our first goal was to insist upon the fact 
that unless the Stratonovitch prescription is used to define the sLLG, a drift term has to be added to the equation of motion. 
The drift term ensures both the conservation of the magnetization modulus and the approach to Boltzmann equilibrium under conservative 
magnetic fields. 

We also formulated the problem in the spherical coordinate system. Although this is the most natural framework to work
in, due to the explicit conservation of the modulus of the magnetization, it has been the source of many confusing 
statements in the literature. For instance, it is stated in~\cite{Martinez-etal} that  the 
random field in the spherical coordinate system is additive. In Sec.~7.3.1 in~\cite{Langevin-Coffey} it is written that 
the spherical components of the random field is a Gaussian with zero mean. In this paper we showed that the polar 
coordinate field acquires a non-vanishing average. We clarified these issues not only in the standard It\^o and 
Stratonovich schemes but also in the general $\alpha$ prescription. This is an important result  for the 
correct numerical study of the magnetization dynamics.

We then derived the drift term to be added to the equation for the polar angle. We showed that the evolution 
dictated by this $\alpha$ prescription stochastic equations leads to the equilibrium Gibbs-Boltzmann distribution
independently of $\alpha$. 

Next, we focused on the construction of the generating functional.
We stressed that physical results should be independent of the framework used to write the path-integral,
this being the Landau {\it vs.} Gilbert formulation of the 
dynamics, the $\alpha$-prescription, or whether we use 
Cartesian or spherical coordinates.

The equivalence between the Landau and Gilbert formulations at the level of the equations
of motion was carefully discussed in several textbooks on this subject~\cite{Bertotti-etal}. We showed
explicitly how this equivalence is realized in the  path-integral formalism.

We proved the independence of  the $\alpha$-prescription in Sec.~\ref{sec:Fokker-Planck} 
within the framework of the Fokker-Planck equation. The $\alpha$-invariance of physical results 
can also be shown within the generating functional formalism but, as the action depends explicitly
on $\alpha$, this feature is less trivial to show in this set-up.
One way to prove invariance is to use an underlying BRST symmetry~\cite{arenas2010}.
Another possibility is to construct a perturbative expansion and to show invariance in this way~\cite{Honkonen}.
In both cases, an interplay between the contributions of all parts in the action, including the 
Jacobian, are necessary to establish invariance. 

At the level of the stochastic equations of motion, one can go from Cartesian to spherical 
expressions by 
using the transformation rules for the change of basis and the generalized chain-rule for the 
time-derivative. In the Fokker-Planck formalism, a change of variables also allows one to relate
Cartesian and spherical approaches. 
However, at the level of the path-integral, the equivalence is more subtle.
As it is well-known from the results in~\cite{Edwards64,Jevicki76,Tirapegui82,Alfaro92}, a non-linear change of variables 
in the path-integral generates non-trivial extra terms in the action (beyond the formal change of variables and the corresponding Jacobian). These are found also in 
this particular case. We have not discussed this issue in further detail in this manuscript since the 
more adequate scheme to do it is the BRST formalism~\cite{Alfaro92} that we will develop elsewhere.

Our work can be extended in different directions. 
For simplicity, we presented the path-integral for a single magnetic moment.   The  sLLG equation can be easily 
generalized to the case of a space-dependent magnetization by introducing a Ginzburg-Landau free-energy 
functional~\cite{Bertotti-etal,Langevin-Coffey,Martinez-etal}. The generalization of the 
generating functional construction to this case is straightforward. It will be useful to treat micromagnetism~\cite{Ralph2008} and, 
in particular,  the dynamics of magnetic domains.  

A field in which the path-integral formulation of the stochastic dynamics has been specially successful 
is the one of systems with quenched randomness. As known since the work in~\cite{deDominicis76},
the average over quenched disorder is simple to perform within this functional framework and allows the analytic 
treatment of many interesting phenomena~\cite{LesHouches}.


\appendix
\numberwithin{equation}{section}

\section{Path integral measure}
\label{app:discrete}

The time interval $t\in [t_0,\mathcal{T}]$ is divided in $N$ discrete time intervals, $t_n \equiv t_0 + n \Delta t$ with $n=0, \ldots, N$ and increment 
$\Delta t \equiv (\mathcal{T}-t_0)/N$. The continuous time
limit is performed by sending $N$ to infinity while keeping ${\cal T}-t_0$ finite.

We define the path integral over functions defined on the time interval $[t_0,\mathcal{T}]$ as
\begin{align}
 \int\uD{[{\mathbf x}]} \equiv \lim\limits_{N\to\infty} \prod_{n=0}^N \int \Ud{{\mathbf x}_n}
\; . 
\end{align}

When integrating over the magnetization, a 3-dimensional field in Cartesian coordinates,
\begin{align}
 \int\uD{[\mathbf{M}]} \equiv \lim\limits_{N\to\infty} \prod_{n=0}^N \int \Ud{\mathbf{M}_n} \;,
\end{align}
the integration on each time-slice is performed over the $\mathbb{R}^3$ space, and the spherical constraint is imposed by the 
equation of motion. More specifically, it is imposed through the $\rmi\hat{M}_\parallel$ sector of the path-integral expression of the 
generating functional, see Sec.~\ref{sec:conservation_modulus}. 

\section{Gilbert Cartesian generating functional}
\label{app:path-integral:Gilbert}

We start from the evolution equation in the $\alpha$-covariant Gilbert formulation in Cartesian coordinates,
\begin{equation}
\mbox{\bf Eq}_G[ {\mathbf M}, {\mathbf H} ]
\equiv {\rm D}_t^{(\alpha)} {\mathbf M} + \gamma_0 \left( {\mathbf H}_{\rm eff} + {\mathbf H} - \frac{\eta}{M_\rms} {\rm D}_t^{(\alpha)} {\mathbf M} \right) = 
{\mathbf 0}
\; , 
\label{eq:Gilbert-alpha-gen}
\end{equation}
and we impose this equation in a path integral over ${\mathbf M}$ as described in 
Sec.~\ref{sec:functional}. 
The Jacobian that ensures that the integration over $\mathbf{M}$ equals one
is given by
\begin{eqnarray}
\mathcal{J}_G[\mathbf{M},\mathbf{H}] \equiv 
\det_{ij;uv} 
\begin{array}{c}
\displaystyle 
\frac{\delta {\rm Eq}_{Gi}(u)}{\delta M_{j}(v)} 
\end{array}
\end{eqnarray}	
with Eq$_G$ given in Eq.~(\ref{eq:Gilbert-alpha-gen}). 
The operator in the determinant can be worked out explicitly and put into the form
\begin{equation}
\frac{\delta \mbox{Eq}_{Gi}(u)}{\delta M_j(v)}
=
X_{ij}(u) {\rm d}_u \delta(u-v) + A_{ij}(v) \delta(u-v)
\end{equation}
with 
\begin{align}
X_{ij}(u) &\equiv  \delta_{ij} + \frac{\eta\gamma_0}{M_\rms} \epsilon_{ijk} M_k(u)
\; , 
\\
A_{ij} &\equiv 
\gamma_0 \epsilon_{ijk} ( {H_{\rm eff}}_k + H_k) 
+ \gamma_0 \epsilon_{ilk} M_l \partial_j {H_{\rm eff}}_k 
- \frac{\eta\gamma_0}{M_\rms} \epsilon_{ijk} {\rm d}_t M_k 
\nonumber\\
& \;\;\;\; + 2D (1-2\alpha) \frac{\gamma_0^2}{1+\eta^2 \gamma_0^2} \delta_{ij} 
\;.
\end{align}
Factorizing  the operator $X_{ik}(u) {\rm d}_u \delta(u-w)$, with inverse $X^{-1}_{kj}(v) \Theta(w-v)$,
\begin{align}
 X^{-1}_{ij}(u) = \frac{1}{1+\eta^2 \gamma_0^2}  \left[\delta_{ij}  - \frac{\eta\gamma_0}{M_\rms} \epsilon_{ijk} M_k(u) 
 +  \frac{\eta^2\gamma_0^2}{M_\rms^2} M_i(u) M_j(u) \right]
 \; , 
\end{align}
we write
\begin{align}
\frac{\delta \mbox{Eq}_{Gi}(u)}{\delta M_j(v)}
 = &  \int \ud{w} \ X_{ik}(u) {\rm d}_u \delta(u-w) 
\nonumber\\
& 
\qquad \times \left[ \delta_{kj} \delta(w-v) + \Theta(w-v) X^{-1}_{kl}(v) A_{lj} (v)  \right]
\; . 
\end{align}
The Jacobian becomes
\begin{align}
 & \mathcal{J}_G[\mathbf{M}, \mathbf{H}] 
= 
 \; \det_{ik;uw}  \left[  X_{ik}(u)  {\rm d}_u\delta(u-w)\right] 
 \nonumber\\ 
& 
 \qquad \times
 \det_{kj;wv}   \left[
 \delta(w-v) \delta_{kj} 
   + \Theta(w-v) X^{-1}_{kr}(v)  A_{rj}(v) \right]
   \; . 
  \end{align}
\label{sec:discussRVsA2}
Notice that the first factor is actually independent of $\mathbf{M}$. Indeed, using 
 $\det_{ik;uw} [X_{ik}(u) {\rm d}_u \delta(u-w) ]= 
 \det_{ij;uv} [X_{ij}(u) \delta(u-v)] \det_{jk;vw} [\delta_{jk}  {\rm d}_v\delta(v-w)]$,
 one easily finds
 $\det_{ij;uv} [X_{ij}(u) {\rm d}_u \delta(u-v) ]
 \propto
 \prod_u \det_{ij} [\delta_{ij} + \frac{\eta\gamma_0}{M_\rms} \epsilon_{ijk} M_k(u)] 
 = \prod_u [1 + \eta^2\gamma_0^2]$, a trivial constant. 
We treat the second determinant, that depends upon $A_{ij}$ and hence $\mathbf{H}$, with the 
identity~(\ref{eq:identity}) to obtain
 \begin{align}
   \mathcal{J}_G[{\mathbf M}, {\mathbf H}] 
 \propto \; &  
 \exp \left( \alpha \int \Ud{t} X^{-1}_{jr}  A_{rj} \right)
 \;. \end{align}
 In this case only the first term in the expansion yields a non-trivial contribution.
Performing the contractions with $X^{-1}_{jr}$, we find
 \begin{eqnarray}
   \mathcal{J}_G[{\mathbf M}, {\mathbf H}]
& \!\!  \!\!
\propto \!\! \!\! &
 \exp \left\{ \frac{\alpha\gamma_0}{1+\eta^2 \gamma_0^2}   
 \int  \Ud{t}  \Big{[}
\frac{2\eta\gamma_0}{M_\rms} \ \mathbf{M} \cdot  ({\mathbf{H}_{\rm eff}} +\mathbf{H})
\right.
\nonumber\\
&&
\;\;
\left.
+ \frac{\eta\gamma_0}{M_\rms} (M_k M_j -\delta_{kj} M_\rms^2)\partial_j {H_{\rm eff}}_k
+ \epsilon_{jlk} M_l \partial_j {H_{\rm eff}}_k
 \Big{]}\!  \right\} 
 \nonumber
 \end{eqnarray}
 where we omitted a constant factor.
This result coincides with the Jacobian in the  Landau formulation, see Eq.~(\ref{eq:Jcart-L}), and
\begin{equation}
{\cal J}_G[{\mathbf M}, {\mathbf H}]
=
{\cal J}_L[{\mathbf M}, {\mathbf H}]
\; .
\end{equation} 
 
Coming back to the generating functional ${\cal Z}[\boldsymbol{\lambda}]$, 
we now exponentiate the delta functional with an auxiliary field $\rmi\hat {\mathbf M}$
that imposes Eq.~(\ref{eq:Gilbert-alpha-gen})  as 
\begin{align}
\delta \Big{[} \mbox{\textbf{Eq}}_G[\mathbf{M}, \mathbf{H}] \Big{]} 
\propto 
\int {\cal D}[\hat{\mathbf{M}}] \ 
\exp  \int \Ud{t} \ \rmi\hat{\mathbf{M}} \cdot \mbox{\textbf{Eq}}_G[\mathbf{M},\mathbf{H}] 
\; . 
\nonumber
\end{align}
From the very structure of the equation it is clear that after decomposing the
auxiliary field in two components $\rmi\hat{\mathbf M}_\perp$ and 
$\rmi\hat {\mathbf M}_\parallel$ perpendicular and parallel to ${\mathbf M}$, respectively, 
one has 
\begin{align}
\delta \Big{[} \mbox{\textbf{Eq}}_G[\mathbf{M}, \mathbf{H}] \Big{]} 
\propto &
\int {\cal D}[\hat{\mathbf{M}}_\perp]  {\cal D}[\hat{\mathbf{M}}_\parallel] \ 
\exp  \int \Ud{t} \ \rmi\hat{\mathbf{M}}_\perp \cdot \mbox{\textbf{Eq}}_G[\mathbf{M},\mathbf{H}] 
\nonumber\\
&
\qquad\qquad
\times
\exp  \int \Ud{t} \ \rmi\hat{\mathbf{M}}_\parallel \cdot {\rm D}_t^{(\alpha)} {\mathbf M}
\; . 
\nonumber
\end{align}

The integration over the Gaussian white noise $\mathbf{H}$ involves the following terms in the exponential:
\begin{displaymath}
 \frac{2\alpha\eta\gamma_0^2}{(1+\eta^2\gamma_0^2) M_\rms}  \mathbf{M} \cdot \mathbf{H} 
+ \gamma_0 \rmi\hat{\mathbf{M}}_\perp \cdot (\mathbf{M} \wedge \mathbf{H}) 
- \frac{\mathbf{H}^2}{2D}
\; . 
\end{displaymath}
The Gaussian integral then yields 
$D [ 4\alpha^2\eta^2\gamma_0^4/(1+\eta^2 \gamma_0^2)^2 +  \gamma_0^2 (\epsilon_{ijk} \rmi {\hat M}_{\perp_i} M_j)^2] $. 
The first term is just a constant 
while the second one is non-trivial. We then recast the generating functional into the form 
\begin{displaymath}
{\cal Z}[\boldsymbol{\lambda}] = 
\int {\cal D}[\mathbf{M}] {\cal D}[\hat{\mathbf{M}}_\perp] {\cal D}[\hat{\mathbf{M}}_\parallel]
\ \exp \left(S_G+\int \Ud{t} \ \boldsymbol{\lambda} \cdot \mathbf{M} \right)
\; , 
\end{displaymath}
where we neglected all trivial constant factors, 
with an action $S_G$ that
reads
\begin{displaymath}
S_G= \widetilde S_{G,{\rm det}} + \widetilde S_{G, {\rm diss}} + S_{{\rm jac}}\;,
\end{displaymath}
and 
\begin{eqnarray}
 && 
 \widetilde S_{G,{\rm det}} = 
  \ln P_\rmi[\mathbf{M}_0, \mathbf{H}_{\rm eff}(t_0)]
  + 
   \int \ud{t} \rmi\hat{\mathbf M}_\parallel \cdot {\rm D}_t^{(\alpha)} {\mathbf M}
  \nonumber\\
  &&\qquad\qquad 
  +
 \int \ud{t} 
 \rmi\hat{\mathbf{M}}_\perp \cdot \left({\rm D}^{(\alpha)}_t{\mathbf{M}} +  \gamma_0 \mathbf{M} \wedge \mathbf{H}_{\rm eff} 
 \right)  
 \; , 
 \;\;\;\;\;\;\;\;\; \\
&& \widetilde S_{G,{\rm diss}} = 
 \int \ud{t}  
\left(\rmi\hat{\mathbf{M}}_\perp \wedge {\mathbf{M}}\right) \cdot 
 \left( D \gamma_0^2 \ \rmi\hat{\mathbf{M}}_\perp \wedge {\mathbf{M}} - \frac{\eta\gamma_0}{M_\rms} {\rm D}^{(\alpha)}_t{\mathbf{M}} \right)
 \; ,
 \label{eq:S_diss_cart}
\qquad
 \\
&& S_{\rm jac} 
=
 \frac{\alpha\gamma_0}{1+\eta^2\gamma_0^2}   
 \int \ud{t} \Big{\{}
 \frac{2 \eta\gamma_0}{M_\rms} \ \mathbf{M} \cdot \mathbf{H}_{\rm eff}
 \nonumber\\
 && 
\qquad\qquad\qquad + \Big{[}\frac{\eta\gamma_0}{M_\rms} (M_i M_j - \delta_{ij} M_\rms^2) +\epsilon_{ijk} M_k) \Big{]} 
\partial_j {H_{\rm eff}}_i
\Big{\}}
 \; . 
 \;\;\;\;\;\;
 \label{eq:Sjac-cart}
\end{eqnarray}

For reasons that should become clear when reading Sec.~\ref{sec:Landau2Gilbert}, 
we perform the change of (dummy) auxiliary 
fields from  $\rmi\hat{\mathbf{M}}_\perp$ to 
$\rmi\hat{\mathbf{M}}_\perp '\equiv \rmi\hat{\mathbf{M}}_\perp \wedge\mathbf{M} \ M_\rms^{-1}$.
 The Jacobian of this change of variables is a constant
 and the action functional now reads $S_G= S_{G,{\rm det}} + S_{G, {\rm diss}} + S_{{\rm jac}}$ with 
\begin{align}
S_{G, {\rm det}} &= 
 \ln P_\rmi[\mathbf{M}_0, \mathbf{H}_{\rm eff}(t_0)] 
 +
  \int \ud{t} \rmi\hat{\mathbf M}_\parallel \cdot {\rm D}_t^{(\alpha)} {\mathbf M}
\nonumber\\
& \qquad
 +
 \int\! \ud{t}
 \rmi\hat{\mathbf{M}}_\perp \cdot \! 
 \left( \frac{1}{M_\rms}  {\rm D}^{(\alpha)}_t{\mathbf{M}} \wedge {\mathbf M}  
 + \gamma_0 M_\rms \ \mathbf{H}_{\rm eff}   \right) 
\label{eq:S_det_cart-redef}
 \ , \\
S_{G, {\rm diss}} &= \int \ud{t}   \rmi\hat{\mathbf{M}} _\perp \cdot 
 \left( D \gamma_0^2 M_\rms^2 \ \rmi\hat{\mathbf{M}}_\perp 
 - \eta\gamma_0 {\rm D}^{(\alpha)}_t{\mathbf{M}} \right)
 \; ,
 \label{eq:S_diss_cart-redef}
\end{align}
where we used the identity  
$\rmi\hat{\mathbf{M}}_\perp \cdot {\rm D}_t^{(\alpha)}{\mathbf{M}} = - (\rmi\hat{\mathbf{M}}_\perp \wedge \mathbf{M}) \cdot (\mathbf{M} \wedge {\rm D}_t^{(\alpha)} {\mathbf{M}}) M_\rms^{-2}$ 
and we dropped the prime: $\rmi\hat{\mathbf{M}}_\perp' \mapsto \rmi\hat{\mathbf{M}}_\perp$. $S_{\rm jac}$ is unchanged. In 
Sec.~\ref{sec:Landau2Gilbert} we presented the action functional given by the sum of the terms 
in Eqs.~(\ref{eq:Sjac-cart}), (\ref{eq:S_det_cart-redef}) and (\ref{eq:S_diss_cart-redef}) and we showed that it can be taken to the Landau form 
by a suitable change of the auxiliary field. We prove in this way that the functional representations of the Landau and 
Gilbert formulation of the stochastic dynamics are totally equivalent.

\section{Spherical coordinate conventions}
\label{app:conventions}

We are using the spherical coordinate system in which $\theta$ the polar angle, $\theta\in[0,\pi]$, and $\phi$ the azimuthal angle, $\phi \in [0,2\pi]$. The local orthogonal unit vectors are ${\mathbf e}_\mu = (\mathbf{e}_{M_\rms}, \mathbf{e}_\theta, \mathbf{e}_\phi)$.
The link to the Cartesian basis is given by
\begin{eqnarray*}
M_x=M_\rms \, \sin\theta \, \cos\phi \; , \qquad\quad
M_y=M_\rms \, \sin\theta \, \sin\phi \; , \qquad\quad
M_z=M_\rms \, \cos\theta \; .
\end{eqnarray*}
We use Latin indices to label Cartesian coordinates ($i=x,y,z$) while
Greek indices refer to the local basis ($\mu = M_\rms,\theta,\phi$).

The rotation matrix linking Cartesian to local coordinates, ${\mathbf e}_\mu = R_{\mu i } {\mathbf e}_i$, 
is
\begin{equation}
 R_{\mu i}(\mathbf{M}) =  R_{\mu i}(\theta,\phi) =
\begin{array}{cc}
\left[
\begin{array}{ccc}
\sin\theta\cos\phi &  \sin\theta\sin\phi & \cos\theta \\
\cos\theta\cos\phi &  \cos\theta\sin\phi & -\sin\theta \\
-\sin\phi &  \cos\phi & 0
\end{array}
\right]
\end{array}
\label{eq:rotation-matrix}
\end{equation}
with $R_{\alpha i} R_{\beta i } = \delta_{\alpha\beta}$.
Notice that $\mathrm{det} R = 1$ and $R^{-1} = \ ^t\!R$:
\begin{equation}
R^{-1}_{i\mu}(\mathbf{M}) =  R^{-1}_{i \mu}(\theta,\phi) =
\begin{array}{cc}
\left[
\begin{array}{ccc}
\sin\theta\cos\phi &  \cos\theta\cos\phi & -\sin\phi \\
\sin\theta\sin\phi &  \cos\theta\sin\phi & \cos\phi \\
\cos\theta &  -\sin\theta & 0
\end{array}
\right]
\; .
\end{array}
\label{eq:inverse-rotation-matrix}
\end{equation}
and $R^{-1}_{i\alpha} R^{-1}_{i\beta} = \delta_{\alpha\beta}$.

The following properties are useful
\begin{eqnarray*}
&& 
R_{\mu j} \frac{\partial R^{-1}_{j\rho} }{\partial \theta} 
= 
-  \delta_{\mu M_\rms} \delta_{\rho \theta}  + \delta_{\mu\theta} \delta_{\rho M_\rms} 
\; , 
\nonumber\\
&&
R_{\mu j} \frac{\partial R^{-1}_{j\rho} }{\partial \phi} = 
-  \sin \theta \delta_{\mu M_\rms} \delta_{\rho \phi}  -\cos \theta \delta_{\mu\theta} \delta_{\rho \phi} 
+ \sin\theta  \delta_{\mu \phi} \delta_{\rho M_\rms}  + \cos \theta \delta_{\mu\phi} \delta_{\rho \theta} 
\; . 
\qquad\;\;\;\;\;
\nonumber
\end{eqnarray*}

\label{app:deltafunction}

The delta function is not a scalar in the sense that it transforms non-trivially under coordinate transformations. This can be simply seen by considering the property
\begin{displaymath}
1 = \int\ud{\mathbf{x}} \delta^3(\mathbf{x}-\overline{\mathbf{x}})\;,
\end{displaymath}
which after a coordinate change to the  spherical basis reads
\begin{displaymath}
1 = \int\ud{\mathbf{\Omega}}  \delta^3(\mathbf{x}(M_\rms,\theta,\phi)-\mathbf{x}(\overline M_\rms,\overline \theta,\overline\phi))\;,
\end{displaymath}
The measure is $\ud{\mathbf{\Omega}} =  M_\rms^2 \sin\theta \ \rmd M_\rms\, \rmd \theta\, \rmd \phi$.  The integrals run over 
$M_\rms \geq 0$, $\theta\in[0,\pi]$ and $\phi\in [0, 2\pi]$. $\mathbf{x}(M_\rms,\theta,\phi)$ and $\mathbf{x}(\overline M_\rms,\overline\theta,\overline\phi)$ 
are the expressions for  $\mathbf{x}$ and $\overline{\mathbf{x}}$ in terms of the spherical coordinates. Using the identity
\begin{displaymath}
 \delta(\mathbf{x}(M_\rms,\theta,\phi)-\mathbf{x}(\overline M_\rms,\overline\theta,\overline\phi)) = \left|\det_{i\mu} \frac{\partial \Omega_\mu}{\partial x_i}\right| \delta(M_\rms-\overline M_\rms) 
 \delta(\theta-\overline\theta) \delta(\phi-\overline\phi)\;,
\end{displaymath}
we get 
\begin{eqnarray} 
1 &=& \int \ud{\mathbf{\Omega}} \frac{1}{M_\rms^2 \sin \theta} \delta(M_\rms-\overline M_\rms) \delta(\theta-\overline\theta) \delta(\phi-\overline\phi) 
\nonumber\\
&=& \int\ \rmd M_\rms \rmd \theta \rmd \phi \ \delta(M_\rms- \overline M_\rms) \delta(\theta - \overline\theta) \delta(\phi-\overline\phi) 
\; .
\label{eq:delta-spherical-Jacobian}
\end{eqnarray}

\label{app:metric}

If $x_i$ are the Cartesian coordinates of the vector $\mathbf{x}$ and $x_\mu$ the coordinates of the same vector 
in another coordinate system, the Jacobian matrix of the coordinate change is
\begin{align}
 J_{i\mu} \equiv \frac{\partial x_i}{\partial x_\mu}
 \; , 
 \label{eq:Jacobian-matrix}
\end{align}
and the Jacobian of the transformation is 
$
\mathcal{J} \equiv \det_{i\mu} J_{i\mu}
$.

We relate the derivatives with respect to Cartesian coordinates to those with 
respect to spherical coordinates as 
\begin{displaymath}
\partial_i = \frac{\partial x_\mu}{\partial x_i} \ \partial_\mu = J^{-1}_{\mu i} \ \partial_\mu
\; . 
\end{displaymath}

For the vector ${\mathbf M}$ transformed to spherical coordinates, the Jacobian matrix (\ref{eq:Jacobian-matrix}) reads
\begin{align}
 J_{i\nu} = \left[
  \begin{array}{ccc}
           \cos\phi\sin\theta & M_\rms \cos\phi\cos\theta & -M_\rms \sin\phi\sin\theta \\
           \sin\phi\sin\theta & M_\rms \sin\phi\cos\theta & M_\rms \cos\phi\sin\theta \\
           \cos\theta & -M_\rms \sin\theta & 0
          \end{array}
\right]
\end{align}
for $i=x,y,z$ and $\mu=M_\rms,\theta,\phi$.

With spherical coordinates, the integration measure is understood as
\begin{align}
 \int\uD{[\mathbf{\Omega}]} \equiv  \lim\limits_{N\to\infty} \prod_{n=0}^N \int \Ud{M_{\rms\,n}} \Ud{\theta_n} \Ud{\phi_n} M_{\rms\, n}^2 \, |\sin\theta_n| \;.
\end{align}
 
\section{Chain rule in spherical coordinates}
\label{sec:chain-rule-sph}

The matrices introduced in App.~\ref{app:metric} and the properties listed above allow one to derive the 
chain rule in spherical coordinates from the one in Cartesian coordinates,
\begin{align} \label{eq:cr1}
\rmd_t
 & =
  {\rmd_t} M_i  \ \partial_i + 
 \frac{  D(1-2\alpha) \gamma_0^2}{1+\eta^2\gamma_0^2}  \left(M_\rms^2 \delta_{ij} - M_i  M_j \right) \, \partial_i \partial_j \;.
\end{align}
The first term can be re-written as
\begin{align}
 {\rmd_t} M_i  \ \partial_i  & = {\rmd_t} M_i  \ J^{-1}_{\sigma i} \frac{\partial}{\partial \Omega_\sigma} 
 =  \left[ -\frac{  2D(1-2\alpha) \gamma_0^2}{1+\eta^2\gamma_0^2} M_i + g_{ij} \overline H_j  \right] \ J^{-1}_{\sigma i} \frac{\partial}{\partial \Omega_\sigma} \nonumber \\
 & =   -\frac{  2D(1-2\alpha) \gamma_0^2}{1+\eta^2\gamma_0^2} M_\rms \partial_{M_\rms}  +  R_{i\mu} g_{\mu\nu} \overline H_\nu  \ J^{-1}_{\sigma i} \frac{\partial}{\partial \Omega_\sigma} \nonumber \\
 & =   -\frac{  2D(1-2\alpha) \gamma_0^2}{1+\eta^2\gamma_0^2} M_\rms \partial_{M_\rms}
 + \frac{1}{M_\rms}
 \left(
 g_{\theta \nu_\perp} \overline H_{\nu_\perp}  \partial_\theta 
 + \frac{1}{\sin\theta} g_{\phi \nu_\perp} \overline H_{\nu_\perp}  \partial_\phi
 \right) \;,
 \nonumber
\end{align}
where in the second line we introduced the equation of motion in Cartesian coordinates, 
in the third line we used $M_i \partial_i  = M_\rms \partial_{M_\rms}$ and $g_{ij} \overline H_j = R_{i\mu} g_{\mu\nu} \overline H_\nu$, 
and in the last line 
we used the fact that $g_{M_\rms \mu} = g_{\mu M_\rms} = 0$. To shorten the notation we called $\overline {\mathbf H} = {\mathbf H}_{\rm eff} + {\mathbf H}$. 

In order to treat the second term of Eq.~(\ref{eq:cr1}), we notice that
\begin{align}
 (M_\rms^2 \delta_{ij}  - M_i M_j  )  \partial_i \partial_j
 &=  M_\rms^2 \nabla^2 -  M_i M_j  \,  \partial_i \partial_j   \;, \nonumber
\end{align}
with $\nabla^2$  the Laplacian operator in spherical coordinates 
\begin{align}
 \nabla^2  = \frac{1}{M^2_\rms} \left( M_\rms^2 \partial^2_{M_\rms} + 2 M_\rms \partial_{M_\rms}    +  \cot\theta\partial_\theta +  \partial^2_\theta + \frac{1}{\sin^2\theta}\partial^2_\phi \right) \;.
\end{align}
We also have
\begin{align}
  M_i M_j  \,  \partial_i \partial_j    
  &=  M_\rms \ M_i  \frac{\partial}{\partial M_\rms} \partial_{i}  
     = M_\rms \ M_i \frac{\partial}{\partial M_\rms} \left[ J^{-1}_{\sigma i} \frac{\partial}{\partial\Omega_\sigma}  \right] \nonumber \\
 & = M_\rms \ M_i  J^{-1}_{\sigma i} \ \frac{\partial^2}{\partial M_\rms \partial\Omega_\sigma} 
 + M_\rms \ M_i  \ \partial_{M_\rms} J^{-1}_{\sigma i} \ \frac{\partial}{\partial\Omega_\sigma} 
 \nonumber \\
  & =  M^2_\rms \ \frac{\partial^2}{\partial M_\rms^2}  = M_\rms^2 \ \partial^2_{M_\rms}\;,
\end{align}
where in the first line we used $M_i \partial_i  = M_\rms \partial_{M_\rms} $ 
and we later used $M_i J^{-1}_{\sigma i} = M_\rms \delta_{\sigma M_\rms}$ and $M_i \partial_{M_\rms} J^{-1}_{\sigma i} = 0$
to obtain the last line. Therefore, the second term in Eq.~(\ref{eq:cr1}) reads
\begin{align}
  \frac{  D(1-2\alpha) \gamma_0^2}{1+\eta^2\gamma_0^2} M_\rms^2 \left(  \nabla^2 - \partial_{M_\rms}^2 \right)\;.
\end{align}
Altogether the chain rule in spherical coordinates is given by 
\begin{align} \label{eq:cr2}
\rmd_t & =  
 \frac{1}{M_\rms}
 \left( 
 g_{\theta \nu_\perp} \overline H_{\nu_\perp} \partial_\theta 
 + \frac{1}{\sin\theta} g_{\phi \nu_\perp} \overline H_{\nu_\perp} \partial_\phi
 \right)
 \nonumber\\
 &
 \;\;\;\;\;
 + \frac{  D(1-2\alpha) \gamma_0^2}{1+\eta^2\gamma_0^2}
 \left( \cot\theta \ \partial_\theta +  \partial^2_\theta + \frac{1}{\sin^2\theta}\partial^2_\phi \right)\;.
\end{align}
Applying this differential operator to $M_\rms$, $\theta$ and $\phi$ respectively, we obtain
\begin{align}
\rmd_t M_\rms &= 0\;, \\
\rmd_t \theta  &= \frac{1}{M_\rms} g_{\theta \nu_\perp} \overline H_{\nu_\perp}  + \frac{  D(1-2\alpha) \gamma_0^2}{1+\eta^2\gamma_0^2} \cot\theta \;,
\label{eq:equation1}  \\
 \rmd_t \phi &= \frac{1}{M_\rms \sin\theta} g_{\phi \nu_\perp} \overline H_{\nu_\perp} \; .
 \label{eq:equation2}
\end{align}
We now define the covariant derivatives
\begin{align}
{\rm D}^{(\alpha)}_t(\theta) &\equiv \rmd_t \theta-\frac{D(1-2\alpha)\gamma_0^2}{(1+\eta^2\gamma_0^2)} \cot\theta \; , \label{eq:covDtheta2} \\
{\rm D}^{(\alpha)}_t(\phi) &\equiv \rmd_t \phi \;.
\end{align}
and we reintroduce the equations of motion (\ref{eq:equation1}) and (\ref{eq:equation2}) in Eq.~(\ref{eq:cr2}) to re-write the differential operator as 
\begin{align}
 \rmd_t &= 
 {\rm D}^{(\alpha)}_t(\theta)   \partial_\theta  
 + {\rm D}^{(\alpha)}_t(\phi) \partial_\phi
  + \frac{  D(1-2\alpha) \gamma_0^2}{1+\eta^2\gamma_0^2}\left( \cot\theta\partial_\theta +  \partial^2_\theta + \frac{1}{\sin^2\theta}\partial^2_\phi \right) \nonumber \\
  &= 
 \rmd_t \theta  \ \partial_\theta 
 + \rmd_t \phi \ \partial_\phi
  + \frac{  D(1-2\alpha) \gamma_0^2}{1+\eta^2\gamma_0^2}\left(  \partial^2_\theta + \frac{1}{\sin^2\theta}\partial^2_\phi \right)
  \;.
\end{align}
We finally obtain an expression for the chain rule in spherical coordinates that is independent of the external and random fields
\begin{align}
\rmd_t    &=   \  \rmd_t \Omega_{\mu_\perp} \  \partial_{\Omega_{\mu_\perp}}
  + \ \frac{  D(1-2\alpha) \gamma_0^2}{1+\eta^2\gamma_0^2}\left(  \partial^2_\theta + \frac{1}{\sin^2\theta} \ \partial^2_\phi \right)  \;.
\end{align}

\section{Determinants}
\label{sec:determinants}

We will be confronted to the task of calculating the 
determinant of an operator of the form 
\begin{equation}
\delta_{ab} \delta(u-v) + C_{ab}(u,v)
\label{eq:operator-form}
\end{equation} 
where $u$ and $v$ are times and $a$ and $b$ 
are coordinate labels in a generic 
coordinate system. Using the identity
\begin{equation}
\det(1+C) = \exp \mbox{Tr}\ln (1+C) 
\label{eq:identity}
\end{equation}
and expanding $\ln(1+C)$ one has 
\begin{equation}
\det(1+C) 
=
\exp \ \sum_{n=1}^{\infty} \frac{(-1)^{n+1}}{n} 
\int \ \Ud{u} \ 
\left\{  \underbrace{C \circ C \circ ... \circ C}_{n \ \mathrm{times}} \right\}_{\mu\mu}\!(u,u)
\end{equation}
The symbol $\circ$ indicates a matrix product and a time convolution.
Typically, $C$ will be proportional to the Heaviside Theta function,
$C_{ab}(u,v)=\Theta(u-v) A_{ab}(u,v)$, hence causal. In regular 
cases, $A$ does not diverge within the time interval and causality ensures that 
the terms with $n>1$ vanish. This simplification does not necessarily apply to the cases we deal with since the matrix $A$ depends on the
white noise ${\mathbf H}$ and, roughly speaking, two such factors 
together are proportional to a temporal Dirac-delta 
function~\cite{Arnold2000,Lubensky2007}. 
Accordingly, we need to analyze each order in the 
expansion separately to decide which ones yield non-vanishing contributions.

Let us take $A_{ab}(u) = A^1_{ab}(u) + A^2_{abc}(u) H_c(u)$
where $A^1$ and $A^2$ do not depend on the random field. For 
concreteness, let us assume that the field $H_a$ has zero mean and correlations 
$\langle H_a(u) H_b(v) \rangle = 2D \delta_{ab} \delta(u-v)$. These
are the $A$'s we will work with in this manuscript. The first order 
term, $n=1$, in the series is 
\begin{equation}
1\mbox{st} = 
\Theta(0) \int \rmd u \ [A^1_{aa}(u) + A^2_{aac}(u) H_c(u) ] 
\;. 
\end{equation}
The second order term, $n=2$, in the series is 
\begin{eqnarray}
&
2\mbox{nd} = \int \rmd u \int \rmd v \ \Theta(u-v) \Theta(v-u)
 [A^1_{ab}(v) + A^2_{abc}(v) H_c(v) ] 
 \nonumber\\
& 
\qquad\qquad
\times 
  [A^1_{ba}(u) + A^2_{bad}(u) H_d(u) ] 
  \; . 
\end{eqnarray}
Because of the two Theta factors, the only non-vanishing  contribution 
may come from $u=v$ if the integrand diverged at this point.
Let us now assume that one can replace a single random field factor
by its average and two random field factors by their correlations: $H_a(u) \to 
\langle H_a(u) \rangle = 0$ and 
$H_a(u) H_b(v) \to \langle H_a(u) H_b(v) \rangle = 2D \delta_{ab}
\delta(u-v)$. An argument to justify this procedure is given  below.
Thus, 
\begin{align}
2\mbox{nd} 
& =
 \int \rmd u \int \rmd v \ \Theta(u-v) \Theta(v-u) \ 2D\delta_{cd}  
\delta(u-v) \ A^2_{abc}(v) A^2_{bad}(u)
\nonumber\\
& = \Theta^2(0)   \
2D \int \rmd u   
\ A^2_{abc}(u) A^2_{bac}(u)
\end{align}
This term is non-vanishing.
What about higher order terms? Fortunately, they all vanish.
For instance, the third order term is 
\begin{align}
3\mbox{rd} 
= &
\int \rmd u \int \rmd v\int \rmd w \
\Theta(u-v) \Theta (v-w) \Theta(w-u)
  \nonumber\\
  &
   \qquad \times 
    [A^1_{ab}(v) + A^2_{abc}(v) H_c(v) ] 
    \
   [A^1_{bd}(w) + A^2_{bde}(w) H_e(w) ] 
   \nonumber\\
   &
   \qquad \times
    [A^1_{da}(u) + A^2_{daf}(u) H_f(u) ] 
    \nonumber\\
    =& 
   \int \rmd u \int \rmd v\int \rmd w \
\Theta(u-v) \Theta (v-w) \Theta(w-u)
  \nonumber\\
  &
   \qquad \times 
    A^1_{ab}(v)  2D \delta_{ef} \delta(w-u) 
    A^2_{bde}(w) 
    A^2_{daf}(u) +\dots
    \nonumber\\
   = &
     \int \rmd u \int \rmd v \
\Theta(u-v) \Theta (v-u) \Theta(0)
  \nonumber\\
  &
   \qquad \times 
    A^1_{ab}(v)  2D 
    A^2_{bde}(w) 
    A^2_{dae}(u) +\dots
    \nonumber\\
    = 
    & \;\; 0
    \; . 
\end{align}
Similarly, one can prove that there are no further contributions to the series. In conclusion,
\begin{equation}
\det(1+C) = \exp \mbox{Tr} \ln (1+C) = \exp [ \ \mbox{Tr}\ C - \frac{1}{2} \mbox{Tr}\ C^2 \ ]
\label{eq:identity-random-field}
\end{equation}

We now  justify heuristically the replacement of the random field and product of two random fields, in the exponentials, by their averages.
Given a generic functional of the random field, $F[{\mathbf H}]$, multiplied by 
an exponential of the kind
\begin{equation}
\rme ^{ -\frac{1}{2} \int \int \rmd t \rmd t' \  Q_{ab}(t,t') H_a(t) H_b(t') }
\end{equation}
with $Q_{ab}(t,t')$ a generic symmetric matrix in the $ab$ indices and the times $t$ and $t'$,
let us consider its average over random field histories
distributed according to a normal Gaussian pdf
\begin{equation}
P_{\rm n}[{\mathbf H}]  \propto \rme^{-\frac{1}{4D} \int \int \rmd t \ H^2_a(t) }
\end{equation}
that we indicate with $\langle \dots \rangle_0$.
 We now evaluate the average as
 \begin{align}
\mbox{Ave} = \langle  F[\mathbf{H}] \ \rme^{- \frac{1}{2} \iint\udd{t}{t'} Q_{ab}(t,t')  \ H_a(t) H_b(t')  }
 \rangle_0
\nonumber
\end{align}
With a Taylor expansion of the exponential,
\begin{align*}
\mbox{Ave} =  & 
\sum_{n=0} \frac{1}{n !\!} \left( - \frac12 \right)^n
\iint\udd{t_1}{t'_1} \ldots \iint\udd{t_n}{t'_n}  Q_{a_1b_1}(t_1,t_1') \ldots Q_{a_nb_n}(t_n,t_n') \\
 & \qquad \times  \langle F[\mathbf{H} ]  \  H_{a_1}(t_1) H_{b_1}(t_1') \ldots H_{a_n}(t_n)H_{b_n}(t_n')
 \rangle_0
\end{align*}
we set the calculation in a way that we can use Wick theorem. 
Since $Q_{ab}(t,t') = 0 $ for all $t\neq t'$, most of the contractions of the fields on the right side of the average vanish, except for 
the ones that set $t_i=t_i'$ for all $i=1,\dots, n$.
Therefore, the only non-vanishing contributions  should be of the form
\begin{align*}
 & 
 \langle F[{\mathbf H}] \rangle_0 \
 \langle  H_{a_1}(t_1) H_{b_1}(t'_1) \rangle_0
 \; \ldots \; \langle  H_{a_n}(t_n) H_{a_n}(t_n')] \rangle_0
 \\
 &  \qquad\qquad = \langle F[{\mathbf H}] \rangle_0
 \ (2D)^n \ \delta_{a_1 b_1} \ldots \delta_{a_n b_n} 
 \ \delta(t_1-t_1') \ldots \delta(t_n-t_n') \;.
\end{align*}
This can be re-exponentiated to recast the average as
\begin{align*}
    & \langle F[\mathbf{H} ]  \ \rme^{- \frac{1}{2} \iint\udd{t}{t'} Q_{ab}(t,t') H_a(t) H_b(t')} \rangle_0 
= \rme^{-D \int\ud{t} Q_{aa}(t,t)} \ \langle F[\mathbf{H}]  \rangle_0 \;.
 \end{align*}
 In short, the result of the calculation is equivalent to the replacement 
 \begin{equation}
 H_a(t) H_b(t') \to \langle H_a(t) H_b(t')\rangle_0 
 \end{equation}
 in the exponential.
 This argument can be easily generalized to the case in which the random field has a non-zero average.

The line of reasoning followed in this Section is close in spirit to the one consider in~\cite{Arnold2000}. 
A different but equivalent approach has been discussed in~\cite{Lubensky2007} (see also~\cite{Tirapegui-comment}). 
We remark, nevertheless, that we have not used the invertibility of the diffusion matrix $g$ as our $g$ 
is actually not invertible.

\section{Random field in spherical coordinates}
\label{app:rotation-noise}

The probability distribution of the random field components, in the Cartesian coordinate system,  is given by
\begin{equation}
 P_\mathrm{n}[H_x,H_y,H_z] \propto \exp\left\{-\frac{1}{4D}
 \int{} \ud{t}  \left[ H_x(t)^2+  H_y(t)^2 +  H_z(t)^2 \right]  \right\}\;.
\end{equation}
The rotation to the spherical coordinate system, 
\begin{align}
 P^{\rm sph}_{\rm n}[H_{M_\rms}, H_\theta, H_\phi] = |\mathcal{J}^{\rm rot}| \ P_\mathrm{n}[R^{-1}_{x\mu} H_\mu,R^{-1}_{y\mu} H_\mu,R^{-1}_{z\mu} H_\mu]\;,
\end{align}
involves the Jacobian
\begin{eqnarray} \label{eq:Jacobian-rot-app}
\mathcal{J}^{\rm rot}
&\!\! \equiv \!\! &
\mathrm{det}_{i\nu,tt'}  \frac{\delta R^{-1}_{i\mu}(t) H_\mu(t)}{\delta H_\nu(t')} 
\; . 
\end{eqnarray}

\subsection{The Jacobian ${\cal J}^{\rm rot}$}

A series of simple operations allow us to factorize the Jacobian of the change of basis from Cartesian to spherical,
${\cal J}^{\rm rot}$ defined in Eq.~(\ref{eq:Jacobian-rot-app}), in two factors:
\begin{eqnarray}
\mathcal{J}^{\rm rot}
& \!\!\! = \!\! & \mathrm{det}_{i\nu,tt'} \left[ R^{-1}_{i\nu}(t) \delta(t-t') + \frac{\delta R^{-1}_{i\mu}(t)}{\delta H_\nu(t')} H_\mu(t) \right]
\nonumber\\
&\!\!\! = \!\! & \mathrm{det}_{i\mu,tt''} \left[ R^{-1}_{i\mu}(t)\delta(t-t'')  \right]
\nonumber\\
&& \qquad\qquad
\times \ \mathrm{det}_{\mu\nu,t''t'}
\left[ \delta_{\mu\nu}\delta(t''-t') + R_{\mu j}(t'') \frac{\delta R^{-1}_{j\rho}(t'')}{\delta H_\nu(t')} H_\rho(t'') \right]\;,
\nonumber
\end{eqnarray}
that we can now compute since the first term is identical to one and the second one takes the form in Eq.~(\ref{eq:operator-form}) with 
\begin{equation}
C_{\mu\nu}(t,t') \equiv  R_{\mu j}(t) \frac{\delta R^{-1}_{j\rho}(t)}{\delta H_\nu(t')} H_\rho(t)
\; . 
\end{equation} 
The factor $\frac{\delta R^{-1}_{j\rho}(t)}{\delta H_\nu(t')}$ is proportional to $\Theta(t-t')$.
Due to the random field dependence in $C$ we need to use the result in (\ref{eq:identity-random-field})
to express the determinant~\cite{Arnold2000,Lubensky2007}. 
The first term in the sum, $n=1$, is
\begin{eqnarray}
\ln \mathcal{J}^{\rm rot}_1 = \exp \int\Ud{t} R_{\mu j}(t) \frac{\delta R^{-1}_{j\rho}(t)}{\delta H_\mu(t)} H_\rho(t) 
= \int\Ud{t} L_\rho(t) H_\rho(t)
\; .
\end{eqnarray}
The rotation matrix $R^{-1}_{j\rho}$ is a function of $\theta$ and $\phi$ (not of $M_\rms$) and neither $\theta$ nor $\phi$ 
depend on the radial component of the noise $H_{M_\rms}$. Therefore
\begin{align}
 \frac{\delta R^{-1}_{j\rho}(t)}{\delta H_\nu(t')} \Rightarrow 
 \frac{\partial R^{-1}_{j\rho}(t)}{\partial \Omega_\nu} \frac{\delta\Omega_\nu(t)}{\delta H_\mu(t)}
 =
 \frac{\partial R^{-1}_{j\rho}(t)}{\partial \Omega_{\tau_\perp}} \frac{\delta \Omega_{\tau_\perp}(t)}{\delta H_{\nu_\perp}(t')} 
\end{align}
and 
\begin{eqnarray*}
L_\rho(t) = 
R_{\mu j}(t) \frac{\partial R^{-1}_{j\rho}(t)}{\partial \Omega_{\tau_\perp}} \frac{\delta\Omega_{\tau_\perp}(t)}{\delta H_\mu(t)}
\; . 
\end{eqnarray*}
The second term in the series, $n=2$, reads
\begin{align}
\ln {\cal J}_2^{\rm rot} = & -\frac12 \iint\udd{t}{t'} 
 R_{\mu j}(t) \frac{\delta R^{-1}_{j\rho}(t)}{\delta H_\nu(t')} H_\rho(t)
 R_{\nu k}(t') \frac{\delta R^{-1}_{k\sigma}(t')}{\delta H_\mu(t)} H_\sigma(t')
 \nonumber\\
 = &
 -\frac12 \iint\udd{t}{t'} 
 Q_{\rho\sigma}(t,t') H_\rho(t) H_\sigma(t')
\end{align}
with 
\begin{equation}
Q_{\rho\sigma}(t,t') \equiv R_{\mu j}(t) \frac{\partial R^{-1}_{j\rho}(t)}{\partial \Omega_{\tau_\perp}} \frac{\delta \Omega_{\tau_\perp}(t)}{\delta H_{\nu}(t')} R_{\nu k}(t') \frac{\partial R^{-1}_{k\sigma}(t')}{\partial \Omega_{\kappa_\perp}} \frac{\delta \Omega_{\kappa_\perp}(t')}{\delta H_{\mu}(t)}
\; . 
\end{equation}

\subsection{Random field distribution}\label{app:randomfield}

Let us now collect all terms together in a compact notation such that the probability distribution of the random field reads
\begin{align}
\ln P^{\rm sph}_{\rm n}[H_\mu] =&  -\frac{1}{4D} \int {\rm d} t \ H^2_\mu(t) + \int\ud{t}  L_\mu(t) H_\mu(t)
\nonumber\\
&- \frac{1}{2} \iint\udd{t}{t'} Q_{\rho\sigma}(t,t') H_\rho(t) H_\sigma(t') \label{eq:pnoise1}
\; ,
\end{align}
that vanishes for $t' \neq t$ because of the response functions involved in its expression. Therefore, the integrand of the last integral 
above vanishes for $t\neq t'$ but the integral may still yield a non-
trivial contribution at $t=t'$ due to the presence of the two random field factors (which are delta correlated). 

The quadratic weight can be given a usual form by completing the square between the first two terms under the integral
\begin{eqnarray*}
&& - \frac{1}{4D} H_\mu^2(t) + L_\mu(t) H_\mu(t) = 
- \frac{1}{4D} \left[ H_\mu(t) - 2D L_\mu(t) \right]^2
+ D  \ L^2_\mu(t),
\end{eqnarray*}
and the measure can be recast as 
\begin{eqnarray*}
&&
\ln P^{\rm sph}_{\rm n}[H_\mu] =  \ln P^{{\rm sph},0}_{\rm n }[H_\mu]  + D \int\ud{t}  L_\mu^2(t)
 \nonumber\\
 &&
 \qquad\qquad\qquad 
- \frac{1}{2} \iint\udd{t}{t'} Q_{\rho\sigma}(t,t') \, H_\rho(t) H_\sigma(t')
\end{eqnarray*}
where we singled out the conventional Gaussian part of the measure
 \begin{eqnarray}
 \ln P^{{\rm sph},0}_{\rm n}[H_\mu] \equiv  -\frac{1}{4D} \int {\rm d} t \ [H_\mu(t) - 2D L_\mu(t)]^2 \;.
 \end{eqnarray}

We now rewrite the last term in a way that the noises appear as $H_\rho(t) - 2D L_\rho(t)$. This rewriting introduces 
two new terms, one that is linear in $H_\rho(t) - 2D L_\rho(t)$, another one in which this factor does not 
appear:
\begin{eqnarray}
&&
- \frac{1}{2} \iint\udd{t}{t'} Q_{\rho\sigma}(t,t')  \ H_\rho(t) H_\sigma(t') 
\nonumber\\
&&  \qquad
=
- \frac{1}{2} \iint\udd{t}{t'} Q_{\rho\sigma}(t,t')  \ [H_\rho(t) - 2D L_\rho(t)] [H_\sigma(t') -2D L_\sigma(t')]
\nonumber\\
&& \qquad\;\;\;\;
+
\iint\udd{t}{t'} Q_{\rho\sigma}(t,t')  \ 2D L_\rho(t) [H_\sigma(t') -2D L_\sigma(t')]
\nonumber\\
&&
\qquad\;\;\;\;
- \frac{1}{2} \iint\udd{t}{t'} Q_{\rho\sigma}(t,t')  \ 2D A_\rho(t)2D L_\sigma(t')
\end{eqnarray}
The last term vanishes identically since  $Q_{\rho\sigma}(t,t')$ is equal to zero for 
$t\neq t'$ and the accompanying factors do not diverge. A similar argument can be applied to the 
second term as two factors $H_\rho(t)-2D L_\rho(t)$ are needed to get a divergence in the 
integrand. The noise pdf then reads 
\begin{eqnarray*}
&&
\ln P^{\rm sph}_{\rm n}[H_\mu] =  \ln P^{{\rm sph},0}_{\rm n }[H_\mu]  + D \int\ud{t}  L_\mu^2(t)
\nonumber\\
&&
\qquad\qquad
- \frac{1}{2} \iint\udd{t}{t'} Q_{\rho\sigma}(t,t')  \ [H_\rho(t) -2DL_\rho(t)] [H_\sigma(t') - 2D L_\sigma(t')]
\; . 
\end{eqnarray*}

We conclude that the probability density of the random field is 
\begin{eqnarray*}
&&
\ln P^{\rm sph}_{\rm n}[H_\mu] =  -\frac{1}{4D} \int {\rm d} t \ [H_\mu(t) - 2D L_\mu(t)]^2 + D \int\ud{t}  L_\mu^2(t)
\nonumber\\
&&
\qquad\qquad\qquad 
- D \int  {\rm d} t \ Q_{\rho\rho}(t,t)
\; . 
\end{eqnarray*}

After  a lengthy computation that uses the properties 
mentioned in App.~\ref{app:conventions} and the equal-time responses calculated in the main part of the text, 
Eqs.~(\ref{eq:equal-time1})-(\ref{eq:equal-time4}), one derives
\begin{equation}
L_\mu^2(t) = 
\frac{\alpha^2 \gamma_0^2 }{1+\eta^2\gamma_0^2}  \left[ \frac{4\eta^2\gamma_0^2}{1+\eta^2\gamma_0^2} + \cot^2\theta(t) \right]\;,
\label{eq:Acuadrado}
\end{equation}
and
\begin{align}
&  Q_{\rho\rho}(t,t) 
\nonumber\\
& \qquad
 =
 \left [\left( \frac{\delta \theta}{\delta H_\theta} \right)^2 + 2 \sin\theta \frac{\delta \theta}{\delta H_\phi} \frac{\delta \phi}{\delta H_\theta} + \left( \frac{\delta \phi}{\delta H_\phi} \right)^2 + \cos^2\theta \left( \frac{\delta \phi}{\delta H_\theta} \right)^2 \right]
\nonumber\\
& \qquad
= 
\frac{\alpha^2\gamma_0^2 }{1+\eta^2\gamma_0^2}
\cot^2\theta +\dots
\end{align}
with the dots being just constant terms.
Therefore, apart from irrelevant additive constants
we establish that 
\begin{align*}
 L_\mu^2(t) =Q_{\rho\rho}(t,t) = \frac{\alpha^2 \gamma_0^2}{1+\eta^2\gamma_0^2}  \cot^2\theta(t) \;. 
 \label{eq:Acuadrado}  
\end{align*}
In the end, the measure is  
\begin{equation}
P_{\rm n}^{\rm sph}[H_\mu] \propto \exp \left\{ -\frac{1}{4D} \int\ud{t}  [H_\mu(t) - 2D L_\mu(t)]^2 \right\}
\; . 
\end{equation}
One concludes that the random fields remain delta correlated but they acquire a mean-value in the spherical basis.

It is also quite clear that the radial and angular sectors decouple:
\begin{align}
P_{\rm n}^{\rm sph}[H_\mu] \propto  
P_{\rm n}^{\rm sph}[H_{M_\rms}]  \ P_{\rm n}^{\rm sph}[H_{\mu_\perp}] 
 \; . 
\end{align}
As the equations of motion do not depend on the longitudinal noise the 
first term is irrelevant in the context of the LLG equation. The explicit form of the 
perpendicular sector of the random field distribution is
\begin{align}
P_{\rm n}^{\rm sph}[H_{\mu_\perp}] 
\propto & \
\exp  \int\ud{t}  
\left\{ -\frac{1}{4D} H^2_{\mu_\perp}(t) 
+ \frac{\alpha\gamma_0}{1+\eta^2\gamma_0^2} \cot\theta(t) \ [\eta\gamma_0 H_\theta(t) + H_\phi(t) ]
\right.
\nonumber\\
& 
\left.
\qquad\qquad\qquad
- \frac{\alpha^2\gamma_0^2 D}{1+\eta^2 \gamma_0^2} \cot^2\theta(t)
\right\}
\; .
\label{eq:Pn-random-noise-sph}
\end{align}

\section{The Jacobian ${\cal J}_L^{\rm sph}$}
\label{app:MSRDJ-Jacobian-Gus}

We compute here the Jacobian needed for the construction of the generating functional 
in the spherical Landau formalism used in Sec.~\ref{sec:contruct_spherical}. 
\begin{eqnarray}\label{eq:defjac2}
\mathcal{J}_L^{\rm sph}[\mathbf{\Omega},\mathbf{H}] \equiv
\det_{\mu\nu;uv}
\begin{array}{c}
\displaystyle
\frac{\delta {\rm{Eq}}_{L \mu}[\mathbf{\Omega},\mathbf{H}](u)}{\delta \Omega_{\nu}(v)}
\end{array}
\; , 
\end{eqnarray}	
with the coordinate indices $\mu,\nu=M_\rms,\theta,\phi$ and the times $u,v$. From its definition one 
has
\begin{align}
\mathcal{J}_L^{\rm sph}[\mathbf{\Omega},\mathbf{H}] &
=
\det_{uv} \left[ {\rm d}_u \delta_{u-v}  \right]  \ 
\det_{\mu_\perp\nu_\perp;uv} 
\frac{\delta \mbox{Eq}_{L \mu_\perp}(M_\rms,\theta,\phi;u)}{\delta {\nu_\perp}(v)}
\nonumber\\
& = 
\displaystyle
\det_{uv} \left[ {\rm d}_u \delta_{u-v}  \right]  \ 
\det_{\nu_\perp\nu_\perp;uw} [\delta_{\mu_\perp \nu_\perp} \rmd_u \delta_{u-w} ]
\nonumber\\
& \qquad \times \det_{\nu_\perp\nu_\perp;wv} [\delta_{\mu_\perp \nu_\perp} \delta_{w-v}  + \Theta(w-v) A_{\mu_\perp\nu_\perp}(v) ]
\end{align}
with 
\begin{eqnarray}
&&
A_{\theta\theta}(v) 
= 
 \frac{\gamma_0}{1+\eta^2\gamma_0^2} \left[ \frac{D(1-2\alpha)\gamma_0}{\sin^2\theta} 
- \partial_\theta \left(R_{\phi i} + \eta\gamma_0 R_{\theta i} \right) (H_{{\rm eff},i} + H_i)\right] 
\; , 
\nonumber\\
&&
A_{\theta\phi}(v)  
= 
- \frac{\gamma_0}{1+\eta^2\gamma_0^2} 
\partial_\phi \left( R_{\phi i} + \eta\gamma_0  R_{\theta i} \right) (H_{{\rm eff},i} + H_i) 
\; , 
\nonumber\\
&&
A_{\phi\theta}(v) 
= 
- \frac{\gamma_0}{1+\eta^2\gamma_0^2} 
 \frac{1}{\sin\theta} \left[ \partial_\theta \left( \eta\gamma_0  R_{\phi i} -  R_{\theta i} \right)
 \right.
 \nonumber\\
&&  \qquad\qquad\qquad\qquad\qquad\qquad
 \left.
 - \cot\theta \left(  \eta\gamma_0  R_{\phi i} - R_{\theta i}\right) 
\right] (H_{{\rm eff},i} + H_i) 
\; , 
\nonumber\\
&&
A_{\phi\phi}(v) 
= 
- \frac{\gamma_0}{1+\eta^2\gamma_0^2}
 \frac{1}{\sin\theta} \partial_\phi \left( \eta\gamma_0  R_{\phi i} - R_{\theta i} \right)  (H_{{\rm eff},i} + H_i)
 \; . 
 \nonumber
\end{eqnarray}
The first two factors contribute irrelevant constants. The last one can be treated with the 
identity (\ref{eq:identity}) where the time-dependent $2\times 2$ matrix with entries 
is $C_{\mu_\perp\nu_\perp}(w,v)=\Theta(w-v) A_{\mu_\perp\nu_\perp}(v)$.
The causal character of $C$ cuts the expansion at its second order.
The first contribution is
$\mbox{Tr}\ C = A_{\theta\theta}+A_{\phi\phi}$,
\begin{eqnarray}
\mbox{Tr} \ C &=& 
 \frac{\alpha\gamma_0}{1+\eta^2\gamma_0^2} \int \rmd t \left[ \frac{D(1-2\alpha)\gamma_0}{\sin^2\theta} 
- \partial_\theta \left(R_{\phi i} + \eta\gamma_0 R_{\theta i} \right) (H_{{\rm eff},i} + H_i) 
\right.
\nonumber\\
&& 
\left.
\qquad\qquad\qquad\;\;\; -
 \frac{1}{\sin\theta} \partial_\phi \left( \eta\gamma_0  R_{\phi i} - R_{\theta i} \right) (H_{{\rm eff},i} + H_i)
\right]
\; .
\end{eqnarray}
The second order term is given by
$\mbox{Tr}\ C^2 /2= (A^2_{\theta\theta}+2A_{\theta\phi} A_{\phi\theta} + A^2_{\phi\phi})/2$.
Keeping only the terms that are proportional to two random fields, and using $H_i(t) H_j(t') \to
\langle H_i(t) H_j (t') \rangle = 2 D \delta_{ij} \delta(t-t')$, see App.~\ref{sec:determinants},
\begin{eqnarray}
&& 
\mbox{Tr} \ C^2 =
\frac{2D\alpha^2\gamma_0^2}{(1+\eta^2\gamma_0^2)^2}
\int \rmd t
\ \Big{\{}
[ \partial_\theta \left(R_{\phi i} + \eta\gamma_0 R_{\theta i} \right) ]^2
\qquad
 \nonumber\\
&&
 \qquad\qquad\qquad
 + \frac{1}{\sin^2\theta} [\partial_\phi \left( \eta\gamma_0  R_{\phi i} - R_{\theta i} \right)]^2
 + 2 \ \partial_\phi \left( R_{\phi i} + \eta\gamma_0  R_{\theta i} \right) 
 \nonumber\\
 &&
 \qquad\qquad\qquad\qquad
 \times
 \left.
 \frac{1}{\sin\theta} \left[ \partial_\theta \left( \eta\gamma_0  R_{\phi i} -  R_{\theta i} \right)
 - \cot\theta \left(  \eta\gamma_0  R_{\phi i} - R_{\theta i}\right) 
\right]
\right\}
\qquad
\nonumber\\
&& 
\qquad
= \mbox{cst} - 
\frac{2D\alpha^2\gamma_0^2}{1+\eta^2\gamma_0^2}
\int \rmd t \
 \frac{1}{\sin^2\theta} 
 \; . 
\end{eqnarray}
The two terms together yield
\begin{eqnarray}
{\mathcal J}_L^{\rm sph} 
&=&
 \frac{\alpha\gamma_0}{1+\eta^2\gamma_0^2} \int \rmd t \ \left[ \frac{D(1-2\alpha)\gamma_0}{\sin^2\theta} 
- \partial_\theta \left(R_{\phi i} + \eta\gamma_0 R_{\theta i} \right) (H_{{\rm eff},i} + H_i) 
\right.
\nonumber\\
&& 
\left.
\qquad\qquad\qquad\;\;\; -
 \frac{1}{\sin\theta} \partial_\phi \left( \eta\gamma_0  R_{\phi i} - R_{\theta i} \right) (H_{{\rm eff},i} + H_i)
\right]
\nonumber\\
&&
+
\frac{D\alpha^2\gamma_0^2}{1+\eta^2\gamma_0^2}
\int \rmd t \
 \frac{1}{\sin^2\theta} 
 \;. 
\end{eqnarray}
The first and last terms combine to yield a contribution proportional to $\alpha(1-\alpha)/\sin^2\theta$ and
\begin{eqnarray}
&&
{\mathcal J}_L^{\rm sph} 
=
 \frac{\alpha\gamma_0}{1+\eta^2\gamma_0^2} \int \rmd t \left\{ \frac{D(1-\alpha)\gamma_0}{\sin^2\theta} 
- \partial_\theta [ (H_{{\rm eff},\phi} + H_{\phi}) + \eta\gamma_0 (H_{{\rm eff},\theta} + H_{\theta}) ]
\right.
\nonumber\\
&& 
\left.
\qquad\qquad\qquad\qquad\qquad 
-
 \frac{1}{\sin\theta} 
 \partial_\phi [ \eta\gamma_0  (H_{{\rm eff},\phi} + H_\phi) - (H_{{\rm eff},\theta} + H_\theta) ]
\right\}
\; . 
\end{eqnarray}

\section{Gilbert spherical generating functional}
\label{sec:construct_spherical_Gilbert}

We here construct the  generating functional $\mathcal{Z}[\boldsymbol{\lambda}]$
by imposing the equations of motion in the Gilbert formulation
\begin{align}
\mbox{Eq}^{\rm sph}_{G,M_\rms} &\equiv {\rm d}_t M_\rms = 0 
\; , 
\label{eq:Ms-Gilbert-app}
\\
\mbox{Eq}^{\rm sph}_{G,\theta} &\equiv  {\rm D}^{(\alpha)}_t(\theta) + \eta\gamma_0 \ \sin \theta \ {\rm D}^{(\alpha)}_t(\phi) -  \gamma_0 \left( H_{\eff, \phi} + H_\phi \right) = 0 \; ,
 \label{eq:thetaalpha-Gilbert-app} \\
 \mbox{Eq}^{\rm sph}_{G,\phi}  &\equiv  -  \sin\theta\,  {\rm D}^{(\alpha)}_t(\phi) + \eta\gamma_0 \ 
 {\rm D}^{(\alpha)}_t(\theta) -\gamma_0  \left( H_{\eff, \theta} + H_\theta \right) =0 \; .
  \label{eq:phialpha-Gilbert-app} 
\end{align}
The Jacobian 
\begin{eqnarray}\label{eq:defjac4}
\mathcal{J}_G^{\rm sph}[\mathbf{\Omega},\mathbf{H}] \equiv
\det_{\mu\nu;uv}
\begin{array}{c}
\displaystyle
\frac{\delta {\rm{Eq}}^{\rm sph}_{G\mu}[\mathbf{\Omega},\mathbf{H}](u)}{\delta \Omega_{\nu}(v)}
\end{array}
\; , 
\end{eqnarray}	
with the coordinate indices $\mu,\nu=M_\rms,\theta,\phi$ and the times $u,v$,
reads
\begin{align}
\mathcal{J}_G^{\rm sph} [\mathbf{\Omega},\mathbf{H}] 
&
=
\det_{uv} \left[ {\rm d}_u \delta_{u-v}  \right]
\nonumber\\
& \displaystyle
\;\;\; \;\;\; \times 
\det_{\mu_\perp\nu_\perp;uv} \left[ X_{\mu_\perp\nu_\perp}(u)  {\rm d}_u \delta_{u-v} + A_{\mu_\perp\nu_\perp}(u) \delta_{u-v} \right]
\; . 
\nonumber
\end{align}
The first factor is due to the $M_\rms$ diagonal element and it can only yield a constant contribution. We next focus on the second factor.
\begin{align}
{\cal J}_G^{\rm sph}
\displaystyle
& \propto 
\det_{\mu_\perp\rho_\perp;uw} \left[ X_{\mu_\perp\rho_\perp}(u)  {\rm d}_u \delta_{u-w} \right]
\nonumber\\
& \qquad \times
\det_{\mu_\perp\rho_\perp;wv} \left[ \delta_{\rho_\perp \nu_\perp} \delta(w-v) + \theta(w-v) {X^{\,-1}_{\rho_\perp\sigma_\perp}}(v)  A_{\sigma_\perp\nu_\perp}(v)
\right] 
\nonumber\\
&
\displaystyle
= 
\det_{\mu_\perp \nu_\perp; uv} [X_{\mu_\perp \nu_\perp}(u)\delta (u-v)] 
\det_{\nu_\perp \rho_\perp; vw}[ \delta_{\nu_\perp \rho_\perp} {\rm d}_v\delta(v-w)]
\nonumber\\
& \qquad \times
\det_{\mu_\perp\rho_\perp;wv} 
\left[ \delta_{\rho_\perp \nu_\perp} \delta(w-v) + \theta(w-v) {X^{\,-1}_{\rho_\perp\sigma_\perp}}(v) 
A_{\sigma_\perp\nu_\perp}(v)
\right] 
\nonumber\\
& 
\displaystyle
\propto 
\det_{\mu_\perp \nu_\perp; u,v} [X_{\mu_\perp \nu_\perp}(u)\delta (u-v)] 
\nonumber\\
& \qquad \times
\det_{\mu_\perp\rho_\perp;wv} 
\left[ \delta_{\rho_\perp \nu_\perp} \delta(w-v) +  \theta(w-v) {X^{\,-1}_{\rho_\perp\sigma_\perp}}(v) A_{\sigma_\perp\nu_\perp}(v)
\right] 
\;,
\label{eq:factorized-jac}
\end{align}
with
\begin{align}
 X_{\mu_\perp\nu_\perp}(u) & \equiv \left[
\begin{array}{cc}
 1 & \eta\gamma_0 \sin\theta_u
 \\
  \eta\gamma_0 & -\sin\theta_u 
\end{array}
\right]\;, \\
{ X^{\,-1}_{\mu_\perp\nu_\perp}}(v)
 & \equiv
\displaystyle
\frac{1}{1+\eta^2\gamma_0^2}
\left[
\begin{array}{cc}
 1 &  \eta\gamma_0\\
 \displaystyle
  \frac{\eta\gamma_0}{\sin\theta_v}
  &
  \displaystyle
  -\frac{1}{\sin\theta_v} 
\end{array}
\right]\;,
\end{align}
and
\begin{eqnarray*}
&& 
A_{\theta\theta}= 
\frac{D(1-2\alpha)\gamma_0^2}{1+\eta^2\gamma_0^2} \frac{1}{\sin^2\theta} 
- \gamma_0 
\partial_\theta (H_{{\rm eff},\phi} +H_\phi)
\\
&& \qquad\qquad
+\frac{\eta\gamma_0^2}{1+\eta^2\gamma_0^2} \cot\theta \left[ \eta\gamma_0 (H_{{\rm eff},\phi} +H_\phi)  - (H_{{\rm eff},\theta} +H_\theta) \right]
\; , 
\\
&& 
A_{\theta\phi} = - \gamma_0 \partial_\phi (H_{{\rm eff},\phi} +H_\phi)
\; , 
\\
&& 
A_{\phi\theta} = 
 + \eta\gamma_0 \ \frac{D(1-2\alpha)\gamma_0^2}{1+\eta^2\gamma_0^2} \frac{1}{\sin^2\theta} 
- \gamma_0 
\partial_\theta (H_{{\rm eff},\theta} +H_\theta) \\
&& \qquad\qquad - \frac{\gamma_0}{1+\eta^2\gamma_0^2} \cot\theta \left[ \eta\gamma_0 (H_{{\rm eff},\phi} +H_\phi)  - (H_{{\rm eff},\theta} +H_\theta) \right]
\; , 
\\
&&
A_{\phi\phi} =
- \gamma_0 \partial_\phi (H_{{\rm eff},\theta} +H_\theta)
\; . 
\end{eqnarray*}
where we used the equations of motion~(\ref{eq:thetaalpha-Gilbert-app}) and (\ref{eq:phialpha-Gilbert-app})  to replace the 
occurrences of ${\rm D}_t^{(\alpha)}(\phi)$ by its corresponding expression in terms of the random field. Indeed, making 
explicit the random field dependence of the Jacobian is crucial as we shall see below.

The absolute value of the first factor in Eq.~(\ref{eq:factorized-jac}) is
\begin{align}
& 
\Big{|} 
\det_{\mu_\perp \rho_\perp; uv}  X_{\mu_\perp \rho_\perp}(u) \delta(u-v) 
\Big{|}
=
 \prod_{n=0}^{N-1} 
 |
 \sin\bar\theta_n
 |\;, \label{eq:detedete}
\end{align}
where we were careful to evaluate the determinant factors on the intermediate points $\bar \theta_n \equiv \alpha \theta_{n+1} + (1-\alpha) \theta_n$. Notice indeed that the discretization matters here since there is no trivial continuous limit of this expression.
See also the discussion in Sect.~\ref{subsec:rules}. The product above can be re-writen as
\begin{align}
 \prod_{n=0}^{N-1} 
 |
 \sin\bar\theta_n
 |
 = 
 \rme^{(1-\alpha) \ln \left| \frac{\sin\theta_0}{\sin\theta_N} \right| }
 \prod_{n=1}^{N} 
 |
 \sin\theta_n
 |
 \;, \label{eq:detedetee45}
\end{align}
where we used the development
\begin{align}
\sin\bar\theta_n 
=
\alpha \sin \theta_{n+1} + (1-\alpha) \sin \theta_n\;,
\end{align}
and the fact that we do not need to consider higher order terms because they vanish from Eq.~(\ref{eq:detedete}) once the limit $\delta t \to 0$ is considered. The product $\prod_{n=1}^{N} 
 | \sin\theta_n |$ in Eq.~(\ref{eq:detedetee45}) cancels exactely the geometric one accompanying the delta functions in Eq.~(\ref{eq:generatingZ24}). In the following, we use this to drop it from the expressions.

We treat the third factor in Eq.~(\ref{eq:factorized-jac}) with the identity (\ref{eq:identity}) 
and we use the causality of $C_{\rho_\perp\nu_\perp}(w,v)=\Theta(w-v) X^{-1}_{\rho_\perp\sigma_\perp}(v) A_{\sigma_\perp\nu_\perp}(v)$ 
to keep only the first two terms of the expansion. 
Performing the contractions with $X^{-1}_{\rho_\perp\sigma_\perp}$ 
and dropping a constant term in 
the overall normalization, we obtain
\begin{eqnarray*}
&& 
\mbox{Tr} \ C = 
\alpha  \int\Ud{t} X^{-1}_{\nu_\perp\sigma_\perp}(t) A_{\sigma_\perp\nu_\perp}(t)
\nonumber\\
&& \qquad\;\;
=  \frac{\alpha\gamma_0}{1+\eta^2\gamma_0^2}
\int\Ud{t} 
\left[ 
\frac{D(1-2\alpha)\gamma_0}{\sin^2\theta} 
- \partial_\theta (H_{{\rm eff},\phi} + H_{\phi})  
\right.
\nonumber\\
&&
\qquad\qquad\qquad\qquad\qquad
\left.
- \frac{\eta\gamma_0}{\sin\theta} \partial_\phi ( H_{{\rm eff},\phi} + H_\phi)
-\eta\gamma_0 \partial_\theta ( H_{{\rm eff},\theta} + H_\theta) 
\right.
\nonumber\\
&&
\qquad\qquad\qquad\qquad\qquad
\left.
+  \frac{1}{\sin\theta} \partial_\phi ( H_{{\rm eff},\theta} + H_\theta) 
\right]
\; . 
\end{eqnarray*}
  Note that the random field ${\mathbf H}$ is still present in this expression, as it was in the 
Cartesian framework calculation as well.

The second order term in the expansion is a half of
\begin{displaymath}
\mbox{Tr } C^2 = \iint\Ud{t}\Ud{t'}
  \Theta(t-t') \Theta(t'-t)   
 X^{-1}_{\sigma_\perp \mu_\perp}(t) A_{\mu_\perp \nu_\perp}(t) X^{-1}_{\nu_\perp \rho_\perp}(t')  A_{\rho_\perp \sigma_\perp}(t')
 \end{displaymath}
  that  making the sums over the $\perp$ components and using the explicit form of $X^{-1}$
  reads
\begin{align}
 \mbox{Tr } C^2 &=
  \frac{1}{(1+\eta^2\gamma_0^2)^2}  \iint\Ud{t}\Ud{t'}
\Theta(t-t') \Theta(t'-t) \nonumber \\
& \quad \times  \Big{\{}
    \frac{1}{\sin\theta(t)}  \left[ A_{\phi\phi}- \eta\gamma_0  A_{\theta\phi} \right](t) \frac{1}{\sin\theta(t')} \left[A_{\phi\phi}- \eta\gamma_0  A_{\theta\phi} \right](t')  
      \nonumber\\
   & \quad\quad + 
   2\frac{1}{\sin\theta(t)} \left[   \eta\gamma_0  A_{\phi\phi} + A_{\theta\phi} \right](t) \left[ \eta\gamma_0  A_{\theta\theta} -A_{\phi\theta}\right](t') \nonumber\\
& \quad\quad  
   + \left[A_{\theta\theta} + \eta\gamma_0  A_{\phi\theta}\right](t) \left[A_{\theta\theta} + \eta\gamma_0  A_{\phi\theta}  \right](t')
   \Big{\}} 
   \nonumber
   \;. 
\end{align}
We now replace $A_{\mu_\perp\nu_\perp}$ by their explicit form.
Since the two $\Theta$ functions make the integrand vanish for $t\neq t'$, the non-vanishing contributions can only come 
from divergent equal-time terms. Owing to the delta-correlated nature of the random field, we only keep
the terms that are quadratic in the random field, see App.~\ref{sec:determinants}. We find
\begin{align}
&
\mbox{Tr} \ C^2 
 =
 \frac{1}{(1+\eta^2\gamma_0^2)^2}  \iint\Ud{t}\Ud{t'}
\Theta(t-t') \Theta(t'-t) \
H_i(t) H_j(t')
\nonumber \\
& \quad \times  \left\{
    \frac{\gamma_0}{\sin\theta(t)}  \left[ \left( \partial_\phi R_{\theta i} + \eta\gamma_0 \partial_\theta R_{\phi i} \right) + \frac{\eta\gamma_0}{1+\eta^2\gamma_0^2}\cot\theta\left( \eta\gamma_0 R_{\phi i}- R_{\theta i}\right)\right](t) \right. \nonumber \\
    & \quad \quad \times
    \frac{\gamma_0}{\sin\theta(t')} \left[\left( \partial_\phi R_{\theta j} + \eta\gamma_0 \partial_\theta R_{\phi j} \right) + \frac{\eta\gamma_0}{1+\eta^2\gamma_0^2}\cot\theta\left( \eta\gamma_0 R_{\phi j}- R_{\theta j}\right) \right](t')     \nonumber\\
   & \quad\quad - 
   2\frac{\gamma^2_0}{\sin\theta(t)} \left[ \eta\gamma_0\partial_\phi R_{\theta i} + \partial_\phi R_{\phi i}  \right](t) \nonumber \\
    & \qquad \qquad \times
    \left[ \cot\theta \left(\eta\gamma_0 R_{\phi j} - R_{\theta j} \right) + \left( \partial_\theta R_{\theta j} - \eta\gamma_0 \partial_\theta R_{\phi j} \right) \right](t') \nonumber\\
& \quad\quad    
   + \eta^2\gamma_0^4 \left[ \partial_\theta R_{\theta i} + \partial_\theta R_{\phi i} \right](t) \left[\partial_\theta R_{\theta j} + \partial_\theta R_{\phi j}  \right](t')
   \Big{\}} 
   \; . 
\nonumber
\end{align}
Following the same steps as in App.~\ref{app:randomfield}, we now replace the product of random fields $H_i(t) H_j(t')$ 
by its average over the Gaussian measure, $2D \delta_{ij} \delta(t-t')$, see again App.~\ref{sec:determinants}.
After a tedious but straightforward computation, dropping constant terms, we obtain
\begin{align}
-\frac{1}{2} \mbox{Tr } C^2 =   \frac{D\gamma_0^2\alpha^2}{1+\eta^2\gamma_0^2} \int\ud{t} 
 \frac{1}{\sin^2\theta}
 \; . 
\end{align}
Finally, putting all terms together, the Jacobian is
\begin{eqnarray*}
&& {\cal J}_G^{\rm sph} 
=
 \exp\left\{  (1-\alpha) \ln \left| \frac{\sin\theta(t_0)}{\sin\theta(t_N)} \right| \right\}
\nonumber\\
&&
\qquad
\times
 \exp \left\{\frac{\alpha\gamma_0}{1+\eta^2\gamma_0^2} \int\Ud{t} 
\left[ 
\frac{D(1-2\alpha)\gamma_0}{\sin^2\theta} 
- \partial_\theta (H_{{\rm eff},\phi} + H_{\phi})  
\right.\right.
\nonumber\\
&&
\qquad\qquad\qquad\qquad
- \frac{\eta\gamma_0}{\sin\theta} \partial_\phi ( H_{{\rm eff},\phi} + H_\phi)
-\eta\gamma_0 \partial_\theta ( H_{{\rm eff},\theta} + H_\theta) 
\nonumber\\
&&
\qquad\qquad\qquad\qquad
\left.\left.
+  \frac{1}{\sin\theta} \partial_\phi ( H_{{\rm eff},\theta} + H_\theta) 
+ \frac{\alpha\gamma_0 D}{\sin^2\theta}
\right] \right\}
\; . 
\end{eqnarray*}
As found in the Cartesian calculation and in the spherical construction for the Landau formulation of the dynamics, 
the Jacobian ${\cal J}_G^{\rm sph}$ does not depend on the parallel component
of the effective field, $H_{{\rm eff},M_\rms}+H_{M_\rms}$. Moreover, we find that the Landau and Gilbert 
Jacobian in spherical coordinates coincide.
see Eq.~(\ref{eq:Jacobian-Landau-spherical}) and App.~\ref{app:MSRDJ-Jacobian-Gus} for the Landau calculation.

We next introduce a Lagrange
multiplier $[\rmi\hat{\mathbf{\Omega}}]$ to exponentiate the functional delta:
\begin{displaymath}
\int \uD{[\rmi\hat{\mathbf \Omega}]} \exp \left\{ - \int \! \ud{t} 
\left(
\rmi\hat\Omega_{M_\rms} \mbox{{Eq}}^{\rm sph}_{G,M_\rms}[{\mathbf \Omega}]
+
\rmi\hat\Omega_\phi \mbox{{Eq}}^{\rm sph}_{G,\phi}[{\mathbf \Omega}, {\mathbf H}]
+
\rmi\hat\Omega_\theta \mbox{{Eq}}^{\rm sph}_{G,\theta}[{\mathbf \Omega}, {\mathbf H}]
\right)
\right\}
\; .  
\end{displaymath}
We identify all the terms 
in the integrand of the exponent in the exponential that involve the random field ${\mathbf H}$:
\begin{eqnarray*}
&& 
-\frac{1}{4D} H_i^2 
+ \gamma_0 
\left( 
\rmi\hat\Omega_\theta R_{\phi i} 
+
\rmi\hat\Omega_\phi R_{\theta i} 
\right) H_i 
\nonumber\\
&& 
\qquad
+ \frac{\alpha\gamma_0}{1+\eta^2\gamma_0^2}
\left( -\partial_\theta R_{\phi i} 
-\frac{\eta\gamma_0}{\sin\theta} \partial_\phi R_{\phi i}  
- \eta\gamma_0 \partial_\theta R_{\theta i} 
+ \frac{1}{\sin\theta} \partial_\phi R_{\theta i}
\right) H_i\;.
\qquad
\end{eqnarray*}
After integration and a number of simplifications that use the explicit expression of the rotation matrix $R$
we find that these terms give rise to
\begin{eqnarray*}
D \gamma_0^2 \left[
(\rmi\hat\Omega_\phi)^2 + (\rmi\hat\Omega_\theta)^2 
+ \frac{2\alpha}{1+\eta^2\gamma_0^2} \left( \rmi\hat\Omega_\theta + \eta\gamma_0 \rmi\hat\Omega_\phi \right) \cot\theta
+ \frac{\alpha^2}{1+\eta^2\gamma_0^2} \frac{1}{\sin^2\theta}
\right]
\end{eqnarray*}
(apart from an irrelevant additive constant).
Note that minus this form equals  the terms in the exponential of the transverse random field measure
in spherical coordinates, Eq.~(\ref{eq:Pn-random-noise-sph}), after the identification $\rmi\hat\Omega_\theta \to - H_\phi/(2\gamma_0 D)$ and $\rmi\hat\Omega_\phi \to - H_\theta/(2\gamma_0 D)$
(plus a constant).
We put all these results together to write the generating functional
\begin{equation}
 \mathcal{Z}[\boldsymbol{\lambda}] = \int 
{\cal D}[\boldsymbol{\Omega}] {\cal D}[\hat{\boldsymbol{\Omega}}] \exp\left( S_G^{\rm sph}[\boldsymbol{\Omega},\hat{\boldsymbol{\Omega}}] 
+ \int\ud{t}\boldsymbol{\lambda}(t) \cdot \boldsymbol{\Omega}(t) \right)\;,
\end{equation} 
the full action
 \begin{equation}
 S_G^{\rm sph}= \widetilde S^{\rm sph}_{G,{\rm det}} + \widetilde S^{\rm sph}_{G,{\rm diss}} + \widetilde S^{\rm sph}_{G,{\rm jac}} \;,
 \end{equation}
 and the terms 
 \begin{eqnarray}
 \widetilde S^{\rm sph}_{G,{\rm det}}  \! & \! = \! & \!
 \ln P_{\rm i}[{\mathbf \Omega}_0, {\mathbf H}_{\rm eff}(t_0)] 
 - 
  \int {\rm d}t 
  \left[ \rmi\hat\Omega_{M_\rms}  {\rm d}_t M_\rms
\right.
\nonumber\\
&& 
\left.
\;\;
+ \rmi\hat\Omega_\theta 
( {\rm D}_t^{(\alpha)}(\theta) - \gamma_0 H_{{\rm eff}, \phi} )
- \rmi\hat\Omega_\phi (\sin\theta {\rm D}_t^{(\alpha)}(\phi)  + \gamma_0 H_{{\rm eff}, \theta})
\right]
\;,
\label{eq:action-det-sph-Gilbert-tilde}\\
&&
\nonumber\\
\widetilde S^{\rm sph}_{G,{\rm diss}} \! & \! = \! &   \!
 \int {\rm d}t 
 \left[ 
   D \gamma_0^2 (\rmi\hat\Omega_\phi)^2 + D \gamma_0^2 (\rmi\hat\Omega_\theta)^2 
 -
\rmi\hat\Omega_\theta
\eta\gamma_0 \sin\theta {\rm D}_t^{(\alpha)}(\phi)  
- \rmi\hat\Omega_\phi  \eta\gamma_0 {\rm D}_t^{(\alpha)}(\theta) 
\right.
\nonumber\\
&&
\;\;
\left.
 +  \frac{2\alpha\gamma_0^2 D}{1+\eta^2\gamma_0^2} ( \rmi\hat\Omega_\theta + \eta\gamma_0 \rmi\hat\Omega_\phi) \cot\theta
\right]\;,
\label{eq:action-diss-sph-Gilbert-tilde}
\\
\widetilde S^{\rm sph}_{G,{\rm jac}} \! & \! = \! &\!
 (1-\alpha) \ln \left| \frac{\sin\theta(t_0)}{\sin\theta(t_N)} \right|
 +
\frac{\alpha\gamma_0^2 D}{1+\eta^2\gamma_0^2}  \int\ud{t}  \frac{1}{\sin^2\theta}
\nonumber\\
&&
+ \frac{\alpha\gamma_0}{1+\eta^2\gamma_0^2} 
\int {\rm d}t
\left[ 
- \partial_\theta H_{{\rm eff},\phi}  
- \frac{\eta\gamma_0}{\sin\theta} \partial_\phi H_{{\rm eff},\phi}
\right.
\nonumber\\
&& 
\left.
\qquad\qquad\qquad\qquad\qquad
-\eta\gamma_0 \partial_\theta H_{{\rm eff},\theta} 
+  \frac{1}{\sin\theta} \partial_\phi H_{{\rm eff},\theta}
\right]\;.
 \end{eqnarray} 
 (We canceled the last term in the result from the integration over $H_i$ with one term from ${\cal J}^{\rm sph}_{G,{\rm jac}}$.)

We now use the identity (\ref{eq:identity-Gus}) to bring the action $S^{\rm sph}_{G}$ into a form 
that is closer to the one in the Landau formulation. We apply this identity to the integration over $\rmi \hat \Omega_\phi$ and 
$\rmi \hat \Omega_\theta$ separately with 
\begin{align}
&
\sigma^2= D\gamma_0^2
\nonumber\\
&
a = - \frac{2\alpha \gamma_0^2 D}{1+\eta^2\gamma_0^2} \eta\gamma_0 \cot\theta
\nonumber\\
&
b
= -\sin\theta {\rm D}^{(\alpha)}_t(\phi) - \gamma_0 H_{{\rm eff},\theta} + \eta\gamma_0 {\rm D}^{(\alpha)}_t(\theta)
\nonumber
\end{align}
for $\rmi \hat \Omega_\phi$, and 
\begin{align}
&
\sigma^2 = D \gamma_0^2
\nonumber\\
&
a = - \frac{2\alpha \gamma_0^2 D}{1+\eta^2\gamma_0^2}  \cot\theta
\nonumber\\
&
b = 
 {\rm D}^{(\alpha)}_t(\theta) - \gamma_0 H_{{\rm eff},\phi} + \eta\gamma_0 \sin\theta {\rm D}^{(\alpha)}_t(\phi)
\nonumber
\end{align}
for $\rmi \hat \Omega_\theta$. 
The new terms generated by the identity are
\begin{eqnarray*}
- \frac{a^2}{4\sigma^2} - \frac{ab}{2\sigma^2}
& \!\! = \!\! &
-
\frac{\alpha^2\gamma_0^2 D}{(1+\eta^2\gamma_0^2)^2} \eta^2\gamma^2_0 \cot^2\theta 
\nonumber\\
& &
+
\frac{\alpha}{1+\eta^2\gamma_0^2} \eta\gamma_0 \cot\theta \
[-\sin\theta {\rm D}^{(\alpha)}_t(\phi) - \gamma_0 H_{{\rm eff},\theta} + \eta\gamma_0 {\rm D}^{(\alpha)}_t(\theta)]
\end{eqnarray*}
in the first case, and 
\begin{eqnarray*}
- \frac{a^2}{4\sigma^2} - \frac{ab}{2\sigma^2}
& \!\! = \!\! &
-
\frac{\alpha^2\gamma_0^2 D}{(1+\eta^2\gamma_0^2)^2} \cot^2\theta 
\nonumber\\
& &
+
\frac{\alpha}{1+\eta^2\gamma_0^2}  \cot\theta \
[ {\rm D}^{(\alpha)}_t(\theta) - \gamma_0 H_{{\rm eff},\phi} + \eta\gamma_0 \sin\theta {\rm D}^{(\alpha)}_t(\phi)]
\end{eqnarray*}
in the second case. Adding them up one finds the total contribution
\begin{eqnarray*}
-
\frac{\alpha^2\gamma_0^2 D}{1+\eta^2\gamma_0^2} \cot^2\theta 
+ \alpha \cot\theta {\rm D}^{(\alpha)}_t(\theta)
- \frac{\alpha\gamma_0}{1+\eta^2\gamma_0^2} \cot\theta \
[\eta\gamma_0 H_{{\rm eff},\theta}+H_{{\rm eff},\phi}
]
\; . 
\end{eqnarray*}
We will add the first and last term to $\widetilde S_{G,{\rm jac}}$ to get 
$S_{L,{\rm jac}}^{\rm sph}$, see Eq.~(\ref{eq:esferica-jac}).  In order to put together all terms in 
$1/\sin^2\theta$ we dropped an irrelevant constant. 
We are left with the rest of the contributions  that we rearrange as
\begin{eqnarray}
S^{\rm sph}_{G,{\rm det}}  \!\! & \!\! = \! \! & \!
 \ln P_{\rm i}[{\mathbf \Omega}_0, {\mathbf H}_{\rm eff}(t_0)] 
 - 
  \int {\rm d}t 
  \left[ \rmi\hat\Omega_{M_\rms}  {\rm d}_t M_\rms
  + \rmi\hat\Omega_\theta 
( {\rm D}_t^{(\alpha)}(\theta) - \gamma_0 H_{{\rm eff}, \phi}   
) 
\right.
\nonumber\\
&& 
\left.
\!\! 
- \rmi\hat\Omega_\phi 
( \sin\theta {\rm D}_t^{(\alpha)}(\phi)  + \gamma_0 H_{{\rm eff}, \theta}
)
\right]
,
\label{eq:action-det-sph-Gilbert}\\
 S^{\rm sph}_{G,{\rm diss}} \!\! & \!\! = \!\! &   \!\!
 \int {\rm d}t 
 \left[ 
   D \gamma_0^2 (\rmi\hat\Omega_\phi)^2 + D \gamma_0^2 (\rmi\hat\Omega_\theta)^2 
   - \rmi\hat\Omega_\theta \eta\gamma_0 \sin\theta {\rm D}_t^{(\alpha)}(\phi)
\right.
\nonumber\\
&&
\left.
\qquad\quad 
- \rmi\hat\Omega_\phi  \eta\gamma_0 {\rm D}_t^{(\alpha)}(\theta) 
\right]
\;,
\label{eq:action-diss-sph-Gilbert}
\\
S^{\rm sph}_{G,{\rm jac}} \!\! & \!\! = \!\! &   \!
S^{\rm sph}_{L,{\rm jac}} 
\label{eq:action-jac-sph-Gilbert}
 \end{eqnarray} 
 and $S_G^{\rm sph} = S_{G, {\rm det}}^{\rm sph} +  S_{G, {\rm diss}}^{\rm sph} + S_{G, {\rm jac}}^{\rm sph}$.

\newpage

\noindent
{\rm \bf Acknowledgements.}
We thank H. H\"uffel and F. Rom\`a for very useful discussions.
We acknowledge financial support from ANR-BLAN-0346 (FAMOUS), PICT-2008-0516 (Argentina), 
NSF grants No. DMR-0906943 and DMR-1151810 (USA) and the Funda\c c\~ao  de Amparo \`a Pesquisa do Estado do Rio de Janeiro (FAPERJ) and Conselho Nacional de Desenvolvimento Cient\'\i fico e Tecnol\'ogico (CNPq).
DGB  thanks the ICMP at the University of Illinois at Urbana-Champaign, where part of this work was developed. 
GSL thanks CNRS for an associate researcher position at LPTHE, Jussieu. LFC and GSL thank the ICTP Trieste for hospitality during the first steps of this investigation.
ZGA is a CNPq fellow (Brasil).

\bibliographystyle{phaip}

\bibliography{magn-FDT}

\begin{thebibliography}{10}

\bibitem{Brown1963}
W.~F. Brown,
\newblock Phys. Rev. {\bf 130}, 1677 (1963).

\bibitem{Hillebrands2002}
B.~Hillebrands and K.~Ounadjela, editors,
\newblock {\em Spin dynamics in confined magnetic structures},
\newblock Springer, Berlin, 2002.

\bibitem{Mattis1988}
D.~C. Mattis,
\newblock {\em Theory of magnetism I: statics and dynamics},
\newblock Springer-Verlag, Berlin, 1988.

\bibitem{Mattis1985}
D.~C. Mattis,
\newblock {\em Theory of magnetism II: thermodynamics and statistical
  mechanics},
\newblock Springer-Verlag, Berlin, 1985.

\bibitem{Johnson1985}
M.~Johnson and R.~H. Silsbee,
\newblock Phys. Rev. Lett. {\bf 55}, 1790 (1985).

\bibitem{Brataas12}
A.~Brataas, A.~D. Kent, and H.~Ohno,
\newblock Nat. Mater. {\bf 11}, 372 (2012).

\bibitem{Landau-Lifshitz}
L.~D. Landau and E.~M. Lifshitz,
\newblock Phys. Z. Sowietunion {\bf 8}, 153 (1935).

\bibitem{Gilbert55}
T.~L. Gilbert,
\newblock Phys. Rev. {\bf 100}, 1243 (1955).

\bibitem{Gilbert04}
T.~L. Gilbert,
\newblock IEEE Trans. Mag. {\bf 40}, 3443 (2004).

\bibitem{Bertotti-etal}
G.~Bertotti, I.~Mayergoyz, and C.~Serpico,
\newblock {\em Nonlinear magnetization dynamics in nanosystems},
\newblock Elsevier, Amsterdam, 2009.

\bibitem{Cimrak2008}
I.~Cimr\'ak,
\newblock Arch. Comp. Meth. Eng. {\bf 15}, 277 (2008).

\bibitem{Fuller63}
W.~F. Brown,
\newblock Phys. Rev. {\bf 130}, 1677 (1963).

\bibitem{Aditi2012}
D.~Pinna, A.~Mitra, D.~L. Stein, and A.~D. Kent,
\newblock App. Phys. Lett. {\bf 101}, 262401 (2012).

\bibitem{Slonczewski1996}
J.~Slonczewski,
\newblock J. Magn. Magn. Mat. {\bf 159}, L1 (1996).

\bibitem{Berger1996}
L.~Berger,
\newblock Phys. Rev. B {\bf 54}, 9353 (1996).

\bibitem{Tsoi1998}
M.~Tsoi et~al.,
\newblock Phys. Rev. Lett. {\bf 80}, 4281 (1998).

\bibitem{Myers1999}
E.~B. Myers, D.~C. Ralph, J.~A. Katine, R.~N. Louie, and R.~A. Buhrman,
\newblock Science {\bf 285}, 867 (1999).

\bibitem{Brataas02}
Y.~Tserkovnyak, A.~Brataas, and G.~E.~W. Bauer,
\newblock Phys. Rev. Lett. {\bf 88}, 117601 (2002).

\bibitem{Parcollet05}
O.~Parcollet and X.~Waintal,
\newblock Phys. Rev. B {\bf 73}, 144420 (2006).

\bibitem{Brataas08}
A.~Brataas, Y.~Tserkovnyak, and G.~E.~W. Bauer,
\newblock Phys. Rev. Lett. {\bf 101}, 037207 (2008).

\bibitem{Bode11}
N.~Bode, L.~Arrachea, G.~S. Lozano, T.~S. Nunner, and F.~von Oppen,
\newblock Phys. Rev. B {\bf 85}, 115440 (2012).

\bibitem{Coffey12}
W.~T. Coffey and Y.~P. Kalmykov,
\newblock J. App. Phys. {\bf 112}, 121301 (2012).

\bibitem{Martin73}
P.~C. Martin, E.~Siggia, and H.~A. Rose,
\newblock Phys. Rev. A {\bf 8}, 423 (1973).

\bibitem{Janssen76}
H.~K. Janssen,
\newblock Z. Phys. B {\bf 23}, 377 (1976).

\bibitem{deDominicis76}
C.~de~Dominicis,
\newblock J. Phys. Colloq. {\bf 37}, C1 (1976).

\bibitem{Lubensky2007}
A.~W.~C. Lau and T.~C. Lubensky,
\newblock Phys. Rev. E {\bf 76}, 011123 (2007).

\bibitem{Andreanov05}
A.~Andreanov, G.~Biroli, and A.~Lef\`evre,
\newblock J. Stat. Mech. , P07008 (2006).

\bibitem{Velenich2008}
A.~Velenich, C.~Chamon, L.~F. Cugliandolo, and D.~Kreimer,
\newblock J. Phys. A {\bf 41}, 235002 (2008).

\bibitem{Kim2013}
B.~Kim, K.~Kawasaki, H.~Jacquin, and F.~van Wijland,
\newblock Equilibrium dynamics of the dean-kawasaki model: Mct and beyond,
\newblock arXiv:1307.1359.

\bibitem{Dean97}
D.~S. Dean,
\newblock J. Phys. A {\bf 29}, 613 (1996).

\bibitem{AronLeticia2010}
C.~Aron, G.~Biroli, and L.~F. Cugliandolo,
\newblock J. Stat. Mech. P11018 (2010).

\bibitem{arenas2010}
Z.~{Gonz{\'a}lez~Arenas} and D.~G. Barci,
\newblock Phys. Rev. E {\bf 81}, 051113 (2010).

\bibitem{arenas2012a}
Z.~{Gonz\'alez~Arenas} and D.~G. Barci,
\newblock Phys. Rev. E {\bf 85}, 041122 (2012).

\bibitem{arenas2012}
Z.~{Gonz\'alez~Arenas} and D.~G. Barci,
\newblock J. Stat. Mech. P12005 (2012).

\bibitem{Honkonen}
J.~Honkonen,
\newblock Ito and stratonovich calculuses in stochastic field theory,
\newblock arXiv:1102.1581v2.

\bibitem{Berkov2007}
D.~V. Berkov,
\newblock Magnetization dynamics including thermal fluctuations: basic
  phenomenology, fast demagnetisation processes and transitions over
  high-energy barriers,
\newblock in {\em Handbook of Magnetism and advanced magnetic materials},
  edited by H.~Kronm\"uller and S.~Parkin, J. Wiley \& sons, 2007.

\bibitem{Lakshmanan2013}
M.~Lakshmanan,
\newblock Phil. Trans. Roy. Soc. A {\bf 369}, 1280 (2013).

\bibitem{Kubo66}
R.~Kubo,
\newblock Rep. Prog. Phys. {\bf 29}, 255 (1966).

\bibitem{Langevin-Coffey}
W.~T. Coffey, Y.~P. Kalmykov, and J.~T. Waldron,
\newblock {\em The Langevin equation}, volume~14 of {\em World Scientific
  series in contemporary chemical physics},
\newblock World Scientific, Singapore, 2005.

\bibitem{Martinez-etal}
E.~Mart\'{\i}nez, L.~L\'opez-D\'{\i}az, L.~Torres, and O.~Alejos,
\newblock Physica B {\bf 343}, 252 (2004).

\bibitem{Stiles2006}
M.~D. Stiles and J.~Miltat,
\newblock Spin torque and dynamics,
\newblock in {\em Spin dynamics in confined magnetic structures III}, edited by
  B.~Hillebrands and A.~Thiaville, Springer, Berlin, 2006.

\bibitem{Berkov2008}
D.~V. Berkov and J.~Miltat,
\newblock J. Magn. Magn. Mat. {\bf 320}, 1238 (2008).

\bibitem{Myers2002}
E.~B. Myers et~al.,
\newblock Phys. Rev. Lett. {\bf 89}, 196801 (2002).

\bibitem{Koch2004}
R.~Koch, J.~A. Katine, and J.~Z. Sun,
\newblock Phys. Rev. Lett. {\bf 92}, 088302 (2004).

\bibitem{Li2004}
Z.~Li and S.~Zhang,
\newblock Phys. Rev. B {\bf 69}, 134416 (2004).

\bibitem{Russek2005}
S.~E. Russek, S.~Kaka, W.~H. Rippard, M.~R. Pufall, and T.~J. Silva,
\newblock Phys. Rev. B {\bf 71}, 104425 (2005).

\bibitem{Apalkov2005}
D.~M. Apalkov and P.~B. Visscher,
\newblock J. Magn. Magn. Mater {\bf 286}, 370 (2005).

\bibitem{Xiao2005}
J.~Xiao, A.~Zangwill, and M.~D. Stiles,
\newblock Phys. Rev. B {\bf 72}, 014446 (2005).

\bibitem{Gardiner}
C.~W. Gardiner,
\newblock {\em Handbook of stochastic methods for physics, chemistry and the
  natural sciences},
\newblock Springer-Verlag, Berlin Heidelberg, 1996.

\bibitem{Berkov2002}
D.~V. Berkov and N.~L. Gorn,
\newblock J. Phys.: Cond. Matt. {\bf 14}, L281 (2002).

\bibitem{Hanggi1978}
P.~H\"anggi,
\newblock Helv.\ Phys.\ Acta {\bf 51}, 183 (1978).

\bibitem{Janssen-RG}
H.~K. Janssen,
\newblock {\em Topics in Modern Statistical Physics},
\newblock World Scientific, Singapore, 1992.

\bibitem{Hanggi1982}
P.~H\"anggi and H.~Thomas,
\newblock Phys. Rep. {\bf 88}, 207 (1982).

\bibitem{Klimontovich}
Y.~L. Klimontovich,
\newblock Physics-Uspekhi {\bf 37}, 737 (1994).

\bibitem{Kupferman}
R.~Kupferman, G.~A. Pavliotis, and A.~M. Stuart,
\newblock Phys. Rev. E {\bf 70}, 036120 (2004).

\bibitem{Hanggi-shot}
P.~H\"anggi,
\newblock Z. Physik B {\bf 75}, 275 (1989).

\bibitem{vanKampen}
N.~G. van Kampen,
\newblock {\em {Stochastic Processes in Physics and Chemistry}},
\newblock Elsevier, London, UK, 2007.

\bibitem{Zinn-Justin}
J.~Zinn-Justin,
\newblock {\em Quantum field theory and critical phenomena},
\newblock Oxford University Press, USA, 2002.

\bibitem{GarciaPalacios98}
J.~L. Garc\'{\i}a-Palacios and F.~J. L{\'a}zaro,
\newblock Phys. Rev. B {\bf 58}, 14937 (1998).

\bibitem{deDominicis78}
C.~de~Dominicis,
\newblock Phys. Rev. B {\bf 18}, 4913 (1978).

\bibitem{Tirapegui82}
F.~Langouche, D.~Roekaerts, and E.~Tirapegui,
\newblock {\em Functional integration and semiclassical expansions},
\newblock Kluwer Academic Publishers, Dordsecht, 1982.

\bibitem{Arnold2000}
P.~Arnold,
\newblock Phys. Rev. E {\bf 61}, 6099 (2000).

\bibitem{Edwards64}
S.~F. Edwards and Y.~V. Gulyaev,
\newblock Proc. Roy. Soc. A {\bf 279}, 229 (1964).

\bibitem{Jevicki76}
J.-L. Gervais and A.~Jevicki,
\newblock Nucl. Phys. B {\bf 110}, 93 (1976).

\bibitem{Alfaro92}
J.~Alfaro and P.~H. Damgaard,
\newblock Ann. Phys. {\bf 220}, 188 (1992).

\bibitem{Ralph2008}
D.~C. Ralph and M.~D. Stiles,
\newblock J. Magn. Magn. Mat. {\bf 320}, 1190 (2008).

\bibitem{LesHouches}
L.~F. Cugliandolo,
\newblock Dynamics of glassy systems,
\newblock in {\em Slow Relaxation and non equilibrium dynamics in Condensed
  Matter: Les Houches Session LXXVII}, edited by J.-L. Barrat, J.~Kurchan,
  M.~V. {Feigel'man}, and J.~Dalibard, Elsevier, Amsterdam, 2003.

\bibitem{Tirapegui-comment}
H.~Calisto and E.~Tirapegui,
\newblock Phys. Rev. E {\bf 65}, 038101 (2002).

\end{thebibliography}

\end{document}